\documentclass[showpacs,showkeys,12pt,preprint,preprintnumbers,
nofootinbib,groupedaddress,superscriptaddress,amsmath,amssymb]{revtex4}

\usepackage[dvipdfm]{graphicx}
\usepackage{dcolumn}
\usepackage{bm}
\usepackage{amssymb}
\usepackage{amsmath}
\usepackage{epsfig}
\usepackage{here}

\def\beq{\begin{equation}}
\def\eeq{\end{equation}}
\def\bea{\begin{array}}
\def\eea{\end{array}}
\def\be{\begin{equation}}
\def\ee{\end{equation}}
\def\ba{\begin{eqnarray}}
\def\ea{\end{eqnarray}}

\def\to{\rightarrow}

\def\[{\left[}
\def\]{\right]}
\def\({\left(}
\def\){\right)}

\def\sm0{{\widetilde{m}_0}}

\def\U1em{{U(1)_{\rm em}}}
\def\to{\rightarrow}

\def\sq2{\sqrt{2}}

\def\ee{e^+e^-}

\def\End{\end{document}}

\newcommand{\gsim}{\mbox{ \raisebox{-1.0ex}{$\stackrel{\textstyle >}
{\textstyle \sim}$ }}}
\newcommand{\lsim}{\mbox{ \raisebox{-1.0ex}{$\stackrel{\textstyle <}
{\textstyle \sim}$ }}}

\def\Journal#1#2#3#4{{#1} {\bf #2} (#4) #3}

\def\PLB{{\em Phys. Lett.}  B}

\def\PRD{{\em Phys. Rev.} D}

\def\PTP{\em Prog.~Theor.~Phys.}

\def\fsl#1{\setbox0=\hbox{$#1$}                 
   \dimen0=\wd0                                 
   \setbox1=\hbox{/} \dimen1=\wd1               
   \ifdim\dimen0>\dimen1                        
      \rlap{\hbox to \dimen0{\hfil/\hfil}}      
      #1                                        
   \else                                        
      \rlap{\hbox to \dimen1{\hfil$#1$\hfil}}   
      /                                         
   \fi}

\begin{document}

\title{Models of Yukawa interaction in the two Higgs doublet model,
and their collider phenomenology}%
\author{Mayumi Aoki}
\address{Department of Physics, Tohoku University,
Aramaki, Aoba, Sendai, Miyagi 980-8578, Japan}
\author{Shinya Kanemura}
\address{Department of Physics, University of Toyama,
3190 Gofuku, Toyama 930-8555, Japan}
\author{Koji Tsumura}
\address{The Abdus Salam ICTP of UNESCO and IAEA,
Strada Costiera 11, 34151 Trieste, Italy}
\author{Kei Yagyu}
\address{Department of Physics, University of Toyama,
3190 Gofuku, Toyama 930-8555, Japan}

\begin{abstract}
Possible models of Yukawa interaction are discussed
in the two Higgs doublet model (THDM) under the discrete symmetry
imposed to avoid the flavor changing neutral current at the leading order.
It is known that there are four types of such models corresponding to the
possible different assignment of charges for the discrete symmetry on
quarks and leptons.
We first examine decay properties of Higgs bosons in each type model,
and summarize constraints on the models from current experimental data.
We then shed light on the differences among these models
in collider phenomenology.
In particular, we mainly discuss the so-called type-II THDM and type-X
THDM. The type-II THDM corresponds to the model
with the same Yukawa interaction as the minimal supersymmetric
standard model. On the other hand, in the type-X THDM,
additional Higgs bosons can predominantly decay into leptons.
This scenario may be interesting
because of the motivation for a light charged Higgs boson scenario
such as in the TeV scale model of neutrino, dark matter and baryogenesis.
We study how we can distinguish the type-X THDM from the 
minimal supersymmetric standard model
at the Large Hadron Collider and the International Linear Collider.


\preprint{TU-839, UT-HET 022, IC/2009/007}
\end{abstract}

\maketitle

\setcounter{footnote}{0}
\renewcommand{\thefootnote}{\arabic{footnote}}

\section{Introduction}

%
The CERN LHC, which is a proton-proton collider
with maximal energy $14$ TeV, has started its operation~\cite{Ref:LHC}.
The most important purpose of the LHC is hunting for the Higgs boson,
the last unknown particle in the standard model (SM).
The Higgs sector is introduced for the 
spontaneous breakdown of the electroweak gauge symmetry.
Weak gauge bosons ($W^\pm$ and $Z$) then receive their masses
via the Higgs mechanism, and quarks and charged leptons 
receive their masses through the Yukawa interaction.
Because the SM Higgs sector has not been confirmed yet, the
possibility of its non-minimal form should also be considered in order to
understand the nature of the symmetry breaking.
In fact, it is well known that many candidates of physics beyond the SM
predict extended Higgs sectors.
For example, supersymmetric extensions of the SM have at least two Higgs
doublets~\cite{Ref:HHG}.  Successful models based on dynamical symmetry breaking also
may require the non-minimal form as the Higgs sector in the low energy
effective theory~\cite{Ref:EWSB}.
In addition, some of new physics models that are intended to solve the problems
which the SM cannot explain, such as tiny neutrino masses, essence of dark matter,
and baryogenesis, has been built with the extension of
the electroweak symmetry breaking sector~\cite{Ref:Nu,Ref:DM,Ref:Baryo,Ref:AKS}.

There are two basic experimental constraints which an extended Higgs
sector must respect:
those on the electroweak rho parameter ($\rho$) as well as the
flavour changing neutral current (FCNC).
The measured value of the rho parameter ($\rho_\text{exp} \approx 1$)
suggest that the electroweak sector of the model would
approximately have a global $SU(2)$ symmetry (the custodial symmetry)
which guarantees $\rho = 1$ at tree level~\cite{Ref:rho}.
For this to be satisfied, a possibility of
the structure with (multi-) isospin doublets (plus singlets)
would be natural as the Higgs sector.
In the SM, FCNC phenomena are suppressed due to the electromagnetic gauge
symmetry and due to the Glashow-Iliopoulos-Maiani mechanism~\cite{Ref:GIM},
so that the experimental bounds on the FCNC processes are satisfied.
In models with more than one Higgs doublet, this is not true in general,
because two or more Yukawa matrices
for each fermion cannot be  simultaneously diagonalized.
It is well known that, to avoid such Higgs-boson-associated FCNC interactions,
each fermion should couple to only one of the Higgs doublets.
This can be realized in a natural way by imposing a discrete $Z_2$
symmetry~\cite{Ref:GW}.

In this paper, we discuss phenomenological differences in various models of Yukawa
interactions under the discrete $Z_2$ symmetry in the two Higgs doublet model (THDM).
It has been known that there are four patterns of the Yukawa interaction
depending on the assignment of charges for quarks and
leptons under the $Z_2$ symmetry~\cite{Ref:Barger,Ref:Grossmann,Ref:Akeroyd}.
We refer them as type-I, type-II, type-X and
type-Y THDMs\footnote{The type-X (type-Y) THDM is referred to as type-IV 
(type-III) THDM in Ref.~\cite{Ref:Barger}
and the type-I' (type-II') THDM in Ref.~\cite{Ref:Grossmann}.
Sometimes the most general THDM, in which each fermion couples to both
Higgs doublet fields, is called the type-III~\cite{Ref:TypeIII}.
To avoid confusion we just call them as these type-X and type-Y THDMs.}.
The type in the Yukawa interaction can be related to
the new physics scenarios.
For example, the Higgs sector of the minimal supersymmetric standard model (MSSM)
is the THDM with a supersymmetric relation~\cite{Ref:HHG}
among the parameters of the Higgs sector, whose Yukawa interaction is
of type-II, in which only a Higgs doublet couples
to up-type quarks and the other couples to down-type quarks and charged leptons.
On the other hand, a TeV scale model to try to explain neutrino masses, dark
matter and baryogenesis has been proposed in Ref.~\cite{Ref:AKS}.
In this model the Higgs sector is the two Higgs doublets 
with extra scalar singlets, and  
the Yukawa interaction corresponds to the type-X, in which only a Higgs
doublet couples to quarks, and the other couples to leptons.
Therefore, in order to select the true model from various new physics candidates
that predict THDMs (and their variations with singlets), it is important
to experimentally determine the type of the Yukawa interaction. 

There have been many studies for the phenomenological properties of
the type-II THDM, often in the context of the MSSM~\cite{Ref:HHG}.
On the contrary, there has been fewer studies for the other
types of Yukawa interactions in the THDM.
The purpose of this paper is to clarify phenomenological differences among these
types of Yukawa interactions in the THDM at
the LHC and the International Linear Collider (ILC)~\cite{Ref:ILC}.
We first study the decay rates and the decay branching ratios
of the CP-even ($h$ and $H$) and CP-odd ($A$) neutral Higgs bosons
and the charged Higgs bosons ($H^\pm$) in various types of Yukawa interactions.
It is confirmed that there are large differences in the Higgs boson decays
among these types of Yukawa interactions in the THDM.
In particular, in the case where the CP-even Higgs boson $h$
is approximately SM-like, $H$ and $A$ decay mainly into $\tau^+\tau^-$
in the type-X scenario for the wide range of parameter space,
while they decay mainly into $b\bar b$ in the type-II scenario.
We then summarize constraints on the mass of $H^\pm$ from current
experimental bounds in various types of Yukawa interactions.
In addition to the lower bounds on the mass ($m_{H^\pm}^{}$)
from CERN LEP and Tevatron direct searches~\cite{Ref:LEP,Ref:Tevatron},
$m_{H^\pm}^{}$ can also be constrained by the $B$-meson decay data
such as $B\to X_s\gamma$~\cite{Ref:bsgNLO,Ref:bsgNLO2,Ref:bsgNNLO,Ref:bsgNNLO_THDM} and
$B\to\tau\nu$~\cite{Ref:MK,Ref:BTauNu}, depending on the model of Yukawa
interaction.
The $B\to X_s\gamma$ results give a severe lower bound, 
$m_{H^\pm}^{}\gtrsim 295$ GeV, at the next-to-next-to-leading order (NNLO)
in the (non-supersymmetric) type-II THDM
and the type-Y THDM~\cite{Ref:bsgNNLO,Ref:bsgNNLO_THDM},
but provide no effective bound in
the type-I (type-X) THDM for $\tan\beta \gtrsim 2$, where $\tan\beta$ is the
ratio of the vacuum expectation values (VEVs) of the CP-even Higgs bosons.
We also discuss the experimental bounds on the charged Higgs sector
from purely leptonic observables
$\tau\to\mu\overline{\nu}\nu$~\cite{Ref:TauMuNuNu}
and the muon anomalous magnetic moment~\cite{Ref:g-2MuSM,Ref:g-2MuTHDM}.
We finally discuss the possibility of discriminating between the types of
Yukawa interactions at the LHC and also at the ILC.
We mainly study collider phenomenology in the type-X THDM
in the light extra Higgs boson scenario,
and see differences from the results in the MSSM (the type-II THDM).
We discuss the signal of neutral and charged Higgs bosons at the LHC,
which may be useful to distinguish the type of the Yukawa interaction.
The feasibility of the direct production processes
from gluon fusion $gg\to A$ ($H$) and the associated production
from $pp\to b\bar b A$ ($b \bar b H$) is studied and the difference
in the signal significance of their leptonic decay channels is
evaluated in the type-X THDM and the MSSM.
We also consider the Higgs boson pair production $pp \to AH^\pm, HH^\pm, AH$
and find that the leptonic decay modes are also useful to
explore the type of the Yukawa interaction.
At the ILC, the process of $e^+e^- \to AH$ is useful to examine
the type-X THDM, because the final states
are completely different from the case of the MSSM.

In Sec.~II, we give a brief review of
the types of the Yukawa interactions in the THDM. In Sec.~III,
the decay widths and the branching ratios are evaluated in the
four different types of the Yukawa interactions.
Section~IV is devoted to a discussion of current experimental
constraints on the THDM in each type of the Yukawa interaction.
In Sec.~V, the possibility of discriminating
the type of the Yukawa interaction 
at the LHC and the ILC is discussed.
Conclusions are given in Sec.~VI.
The formulae of the decay rates of the Higgs bosons are listed in the 
Appendix.

\section{Two Higgs doublet models under the $Z_2$ symmetry}

In the THDM with isospin doublet scalar fields $\Phi_1$ and $\Phi_2$
with a hypercharge of $Y=1/2$,
the discrete $Z_2$ symmetry ($\Phi_1 \to \Phi_1$ and $\Phi_2 \to
-\Phi_2$) may be imposed to avoid FCNC at the lowest order~\cite{Ref:GW}.
The most general Yukawa interaction under the $Z_2$ symmetry
can be written as
\begin{align}
{\mathcal L}_\text{yukawa}^\text{THDM} =
&-{\overline Q}_LY_u\widetilde{\Phi}_uu_R^{}
-{\overline Q}_LY_d\Phi_dd_R^{}
-{\overline L}_LY_\ell\Phi_\ell \ell_R^{}+\text{H.c.},
\end{align}
where $\Phi_f$ ($f=u,d$ or $\ell$) is either $\Phi_1$ or $\Phi_2$.
There are four independent $Z_2$ charge assignments
on quarks and charged leptons, as summarized in TABLE~\ref{Tab:type}
~\cite{Ref:Barger,Ref:Grossmann}.
In the type-I THDM, all quarks and charged leptons obtain their masses
from the VEV of $\Phi_2$.
In the type-II THDM, masses of up-type quarks are generated by the VEV
of $\Phi_2$, while those of down-type quarks and charged leptons
are acquired by that of $\Phi_1$. The Higgs sector of the
MSSM is a special THDM whose Yukawa interaction is of type-II.
\begin{table}[t]
\begin{center}
\begin{tabular}{|c||c|c|c|c|c|c|}
\hline & $\Phi_1$ & $\Phi_2$ & $u_R^{}$ & $d_R^{}$ & $\ell_R^{}$ &
 $Q_L$, $L_L$ \\  \hline
Type-I  & $+$ & $-$ & $-$ & $-$ & $-$ & $+$ \\
Type-II & $+$ & $-$ & $-$ & $+$ & $+$ & $+$ \\
Type-X  & $+$ & $-$ & $-$ & $-$ & $+$ & $+$ \\
Type-Y  & $+$ & $-$ & $-$ & $+$ & $-$ & $+$ \\
\hline
\end{tabular}
\end{center}
\caption{Variation in charge assignments of the $Z_2$ symmetry.} \label{Tab:type}
\end{table}
The type-X Yukawa interaction (all quarks couple to $\Phi_2$
while charged leptons couple to $\Phi_1$) 
is used in the model in Ref.~\cite{Ref:AKS}.
The remaining one is referred to as the type-Y THDM.

The most general Higgs potential under the softly broken $Z_2$
symmetry is given by
\begin{align}
V^\text{THDM}
&= m_1^2\Phi_1^\dag\Phi_1+m_2^2\Phi_2^\dag\Phi_2
-m_3^2\left(\Phi_1^\dag\Phi_2+\Phi_2^\dag\Phi_1\right)
+\frac{\lambda_1}2(\Phi_1^\dag\Phi_1)^2
+\frac{\lambda_2}2(\Phi_2^\dag\Phi_2)^2\nonumber \\
&\qquad+\lambda_3(\Phi_1^\dag\Phi_1)(\Phi_2^\dag\Phi_2)
+\lambda_4(\Phi_1^\dag\Phi_2)(\Phi_2^\dag\Phi_1)
+\frac{\lambda_5}2\left[(\Phi_1^\dag\Phi_2)^2
+(\Phi_2^\dag\Phi_1)^2\right], \label{Eq:PotTHDM}
\end{align}
where the parameters $m_3^2$ and $\lambda_5$ are complex, in general.
In this paper, they are taken to be real by assuming CP invariance.
The Higgs doublet fields can be parametrized as
\begin{align}
\Phi_i=\begin{pmatrix}\omega_i^+\\\frac1{\sqrt2}(v_i+h_i-i\,z_i)
\end{pmatrix},
\end{align}
and the mass eigenstates are defined by
\begin{align}
\begin{pmatrix}h_1\\h_2\end{pmatrix}=\text{R}(\alpha)
\begin{pmatrix}H\\h\end{pmatrix},\quad
\begin{pmatrix}z_1\\z_2\end{pmatrix}=\text{R}(\beta)
\begin{pmatrix}z\\A\end{pmatrix},\quad
\begin{pmatrix}\omega_1^+\\\omega_2^+\end{pmatrix}=\text{R}(\beta)
\begin{pmatrix}\omega^+\\H^+\end{pmatrix},
\end{align}
where the rotation matrix is given by
\begin{align}
\text{R}(\theta)=\begin{pmatrix}\cos\theta&-\sin\theta\\
\sin\theta&\cos\theta\end{pmatrix}.
\end{align}
There are five physical Higgs bosons, i.e.,
two CP-even states $h$ and $H$, one CP-odd state $A$, and a pair of
charged states $H^\pm$, and $z$ and $\omega^\pm$ are
Nambu-Goldstone bosons that are eaten as the
longitudinal components of the massive gauge bosons.
The eight parameters $m_1^2$--$m_3^2$ and
$\lambda_1$--$\lambda_5$ in the Higgs sector are replaced
by eight physical parameters: i.e., the VEV
$v=\sqrt{v_1^2+v_2^2}\simeq 246$ GeV,
the mixing angles $\alpha$ and $\beta$ ($\tan\beta=v_2/v_1$),
the Higgs boson masses  $m_h^{}, m_H^{}, m_A^{}, m_{H^\pm}^{}$,
and the soft breaking mass parameter
$M=m_3/\sqrt{\sin\beta\cos\beta}$.
The mixing angle $\alpha$ is defined such that
$h$ is the SM-like Higgs boson when $\sin(\beta-\alpha)=1$.

\begin{table}[tb]
\begin{center}
\begin{tabular}{|c||c|c|c|c|c|c|c|c|c|}
\hline
& $\xi_h^u$ & $\xi_h^d$ & $\xi_h^\ell$
& $\xi_H^u$ & $\xi_H^d$ & $\xi_H^\ell$
& $\xi_A^u$ & $\xi_A^d$ & $\xi_A^\ell$ \\ \hline
Type-I
& $c_\alpha/s_\beta$ & $c_\alpha/s_\beta$ & $c_\alpha/s_\beta$
& $s_\alpha/s_\beta$ & $s_\alpha/s_\beta$ & $s_\alpha/s_\beta$
& $\cot\beta$ & $-\cot\beta$ & $-\cot\beta$ \\
Type-II
& $c_\alpha/s_\beta$ & $-s_\alpha/c_\beta$ & $-s_\alpha/c_\beta$
& $s_\alpha/s_\beta$ & $c_\alpha/c_\beta$ & $c_\alpha/c_\beta$
& $\cot\beta$ & $\tan\beta$ & $\tan\beta$ \\
Type-X
& $c_\alpha/s_\beta$ & $c_\alpha/s_\beta$ & $-s_\alpha/c_\beta$
& $s_\alpha/s_\beta$ & $s_\alpha/s_\beta$ & $c_\alpha/c_\beta$
& $\cot\beta$ & $-\cot\beta$ & $\tan\beta$ \\
Type-Y
& $c_\alpha/s_\beta$ & $-s_\alpha/c_\beta$ & $c_\alpha/s_\beta$
& $s_\alpha/s_\beta$ & $c_\alpha/c_\beta$ & $s_\alpha/s_\beta$
& $\cot\beta$ & $\tan\beta$ & $-\cot\beta$ \\
\hline
\end{tabular}
\end{center}
\caption{The mixing factors in Yukawa interactions in Eq.~\eqref{Eq:Yukawa}} \label{Tab:MixFactor}
\end{table}

The Yukawa interactions are expressed in terms of mass eigenstates
of the Higgs bosons as
\begin{align}
{\mathcal L}_\text{yukawa}^\text{THDM} =
&-\sum_{f=u,d,\ell} \( \frac{m_f}{v}\xi_h^f{\overline
f}fh+\frac{m_f}{v}\xi_H^f{\overline
f}fH-i\frac{m_f}{v}\xi_A^f{\overline f}\gamma_5fA\)\nonumber\\
&-\left\{\frac{\sqrt2V_{ud}}{v}\overline{u}
\left(m_u\xi_A^u\text{P}_L+m_d\xi_A^d\text{P}_R\right)d\,H^+
+\frac{\sqrt2m_\ell\xi_A^\ell}{v}\overline{\nu_L^{}}\ell_R^{}H^+
+\text{H.c.}\right\},\label{Eq:Yukawa}
\end{align}
where $P_{L/R}$ are projection operators for left-/right-handed fermions,
and the factors $\xi^f_\varphi$ are listed in TABLE~\ref{Tab:MixFactor}.

For the successful electroweak symmetry breaking,
a combination of quartic coupling constants should satisfy
the condition of vacuum stability~\cite{Ref:VS2,Ref:VS}.
We also take into account bounds from perturbative unitarity~\cite{Ref:LQT}
to restrict parameters in the Higgs potential~\cite{Ref:PU,Ref:PU2}.
The top and bottom Yukawa coupling constants cannot be taken too large.
The requirement $|Y_{t,b}|^2<\pi$ at the tree level can provide a milder
constraint $0.4\lesssim\tan\beta\lesssim 91$, where $|Y_t|=(\sqrt{2}/v)
m_t \cot\beta$ and $|Y_b|=(\sqrt{2}/v) m_b \tan\beta$.

\section{Decays of Higgs bosons}

\begin{figure}
\begin{center}
\begin{minipage}{0.275\hsize}
\vspace{0.8ex}
\includegraphics[width=4.5cm,angle=-90]{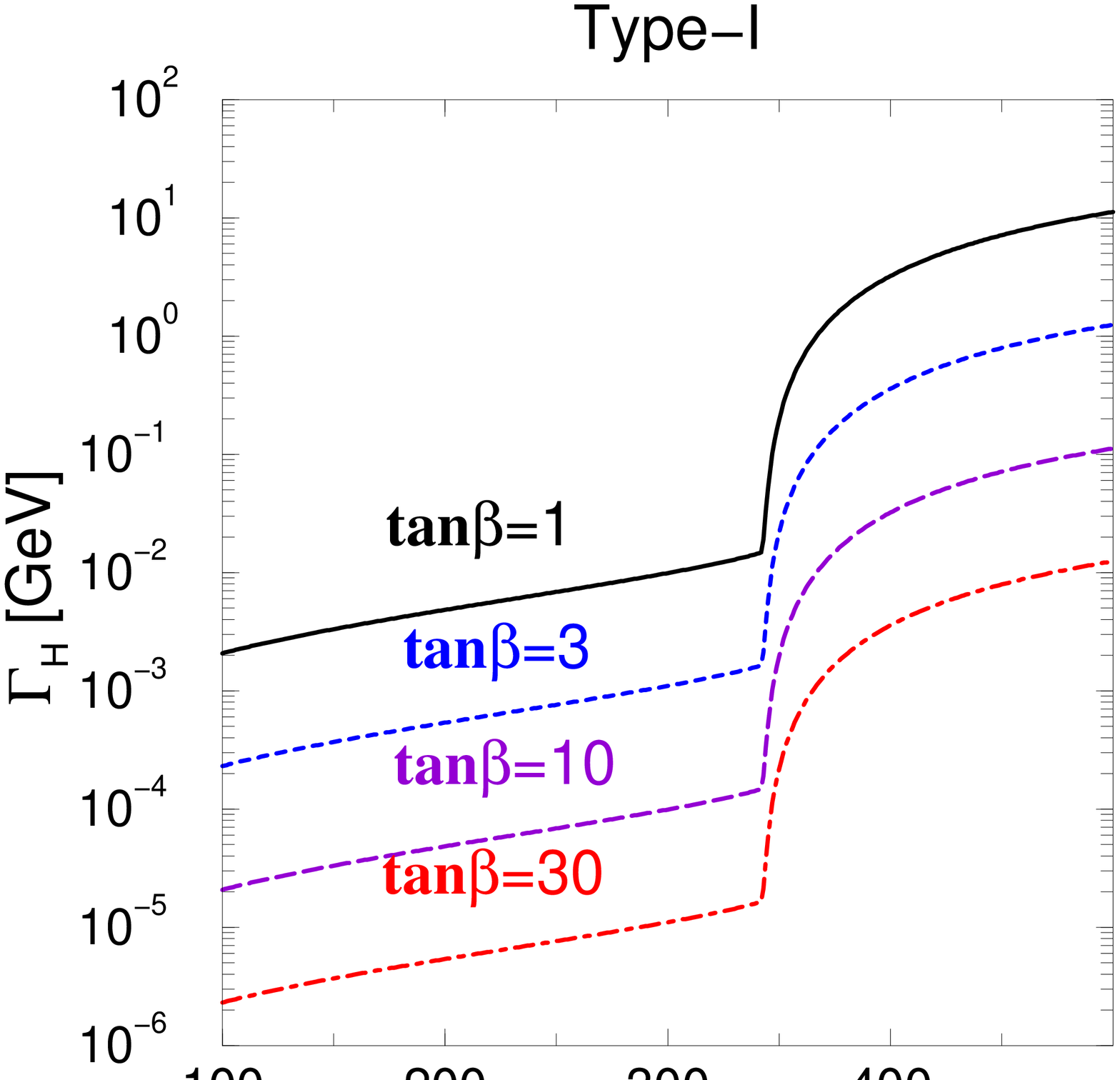}
\end{minipage}
\begin{minipage}{0.23\hsize}
\includegraphics[width=4.4cm,angle=-90]{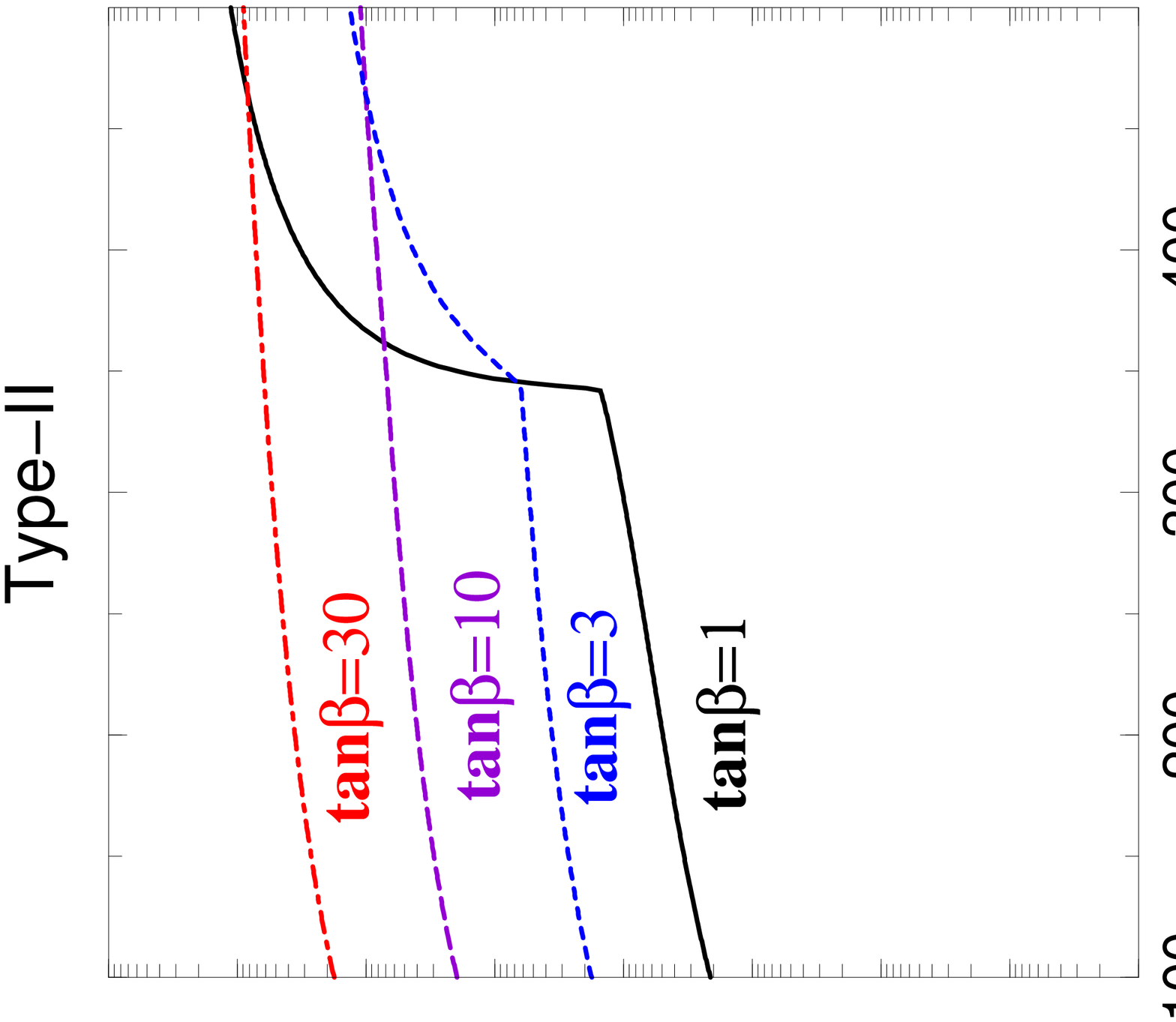}
\end{minipage}
\begin{minipage}{0.23\hsize}
\includegraphics[width=4.4cm,angle=-90]{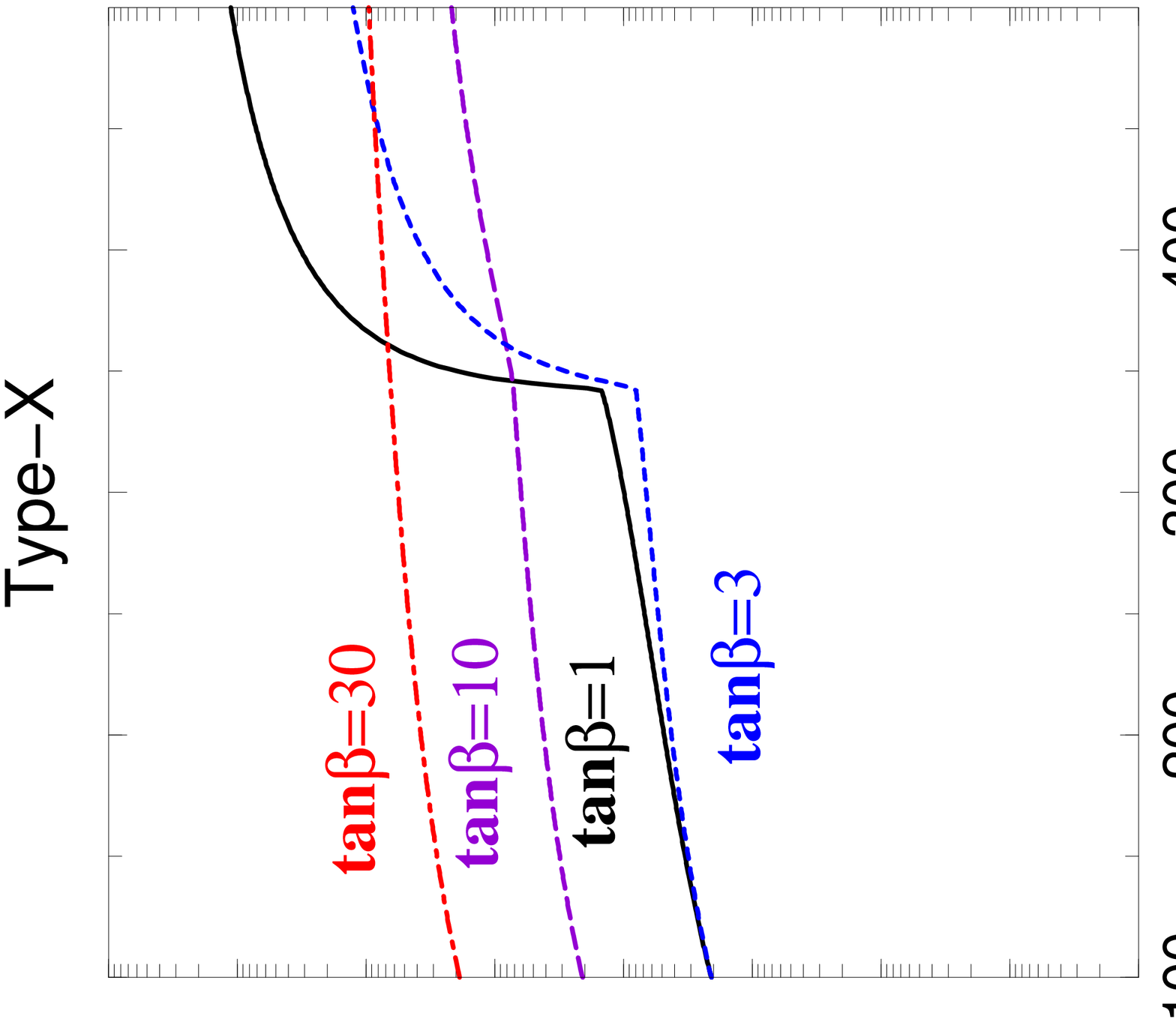}
\end{minipage}
\begin{minipage}{0.23\hsize}
\includegraphics[width=4.4cm,angle=-90]{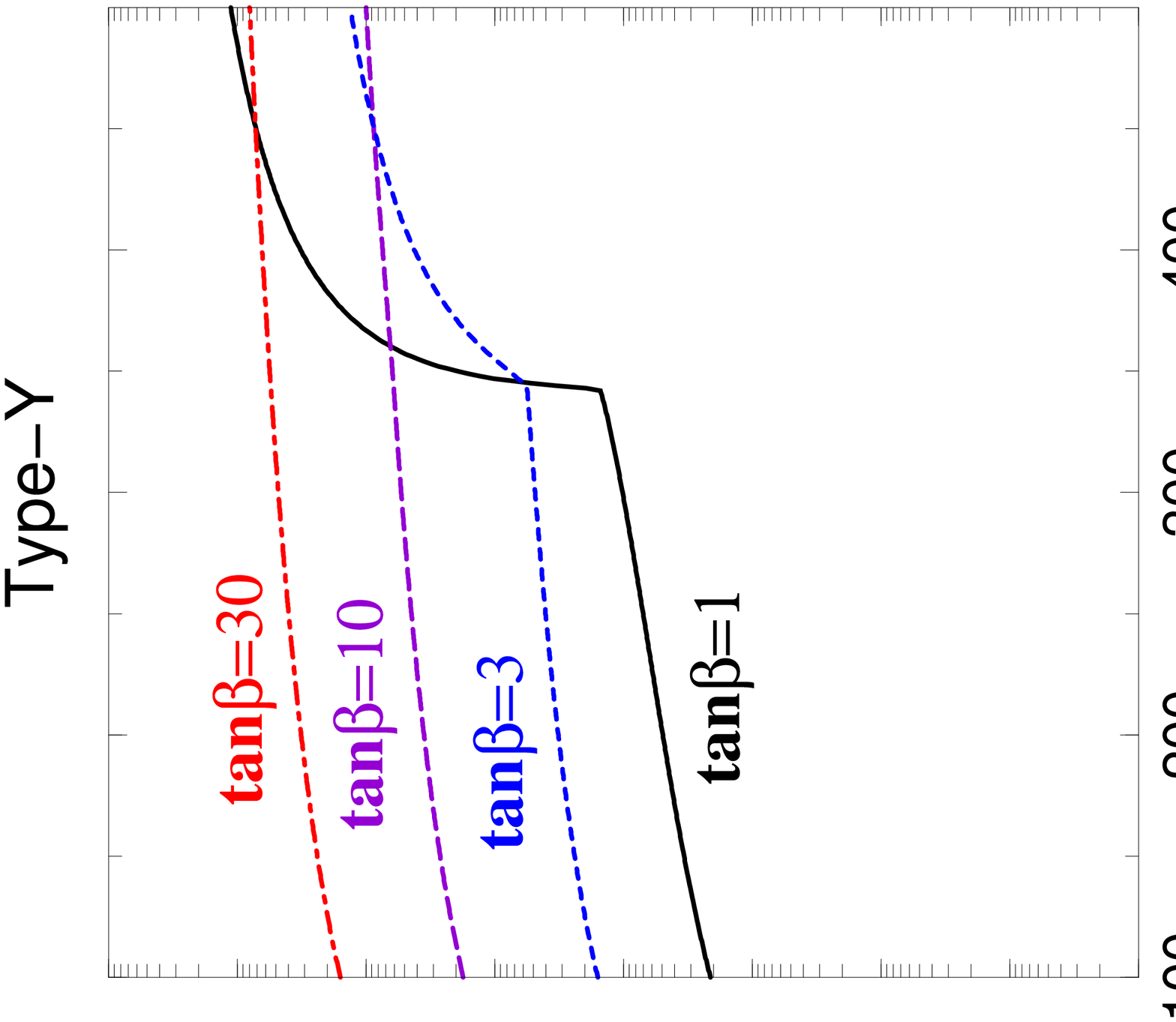}
\end{minipage}
\end{center}
\begin{center}
\begin{minipage}{0.275\hsize}
\vspace{0.8ex}
\includegraphics[width=4.5cm,angle=-90]{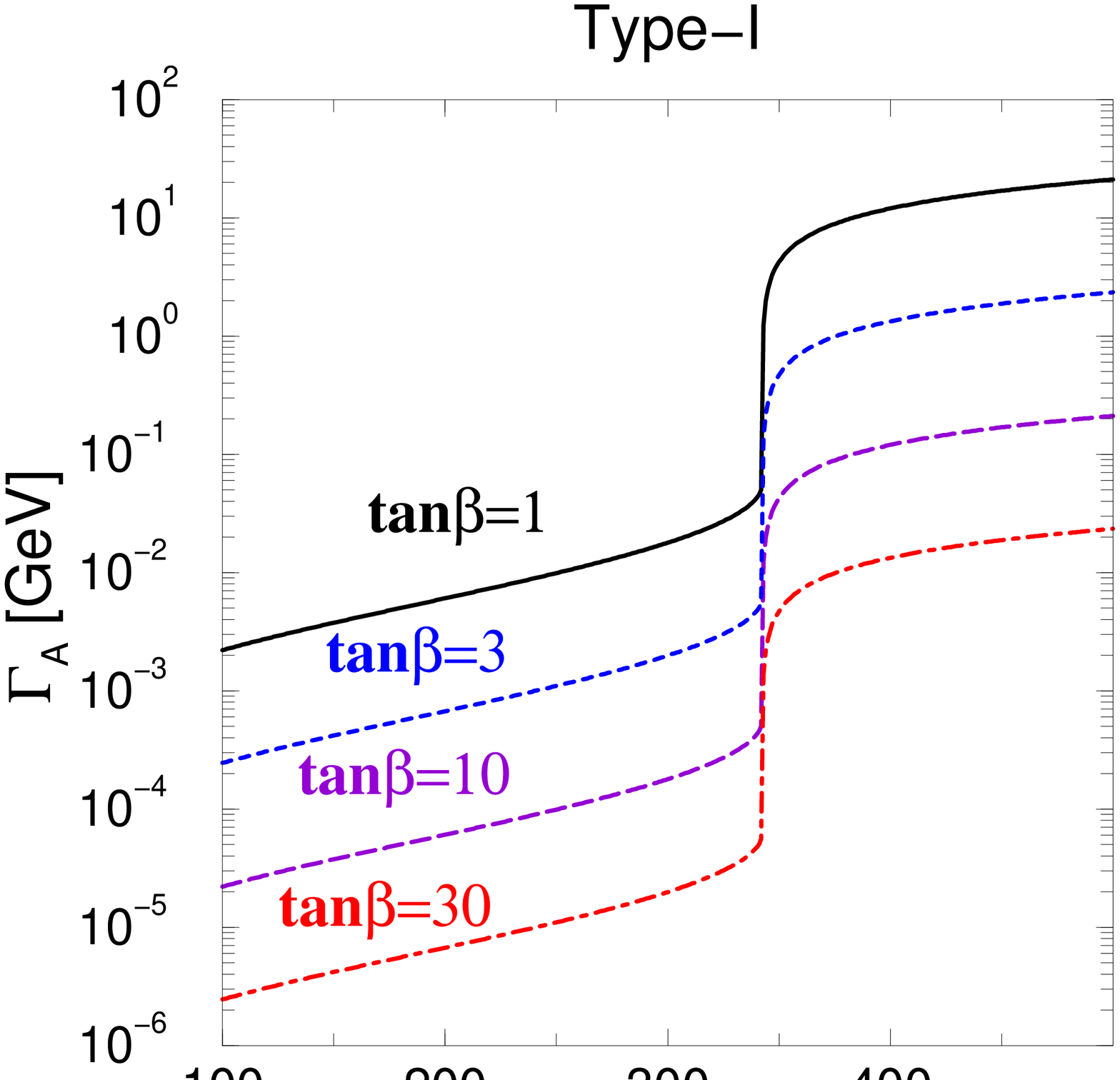}
\end{minipage}
\begin{minipage}{0.23\hsize}
\includegraphics[width=4.4cm,angle=-90]{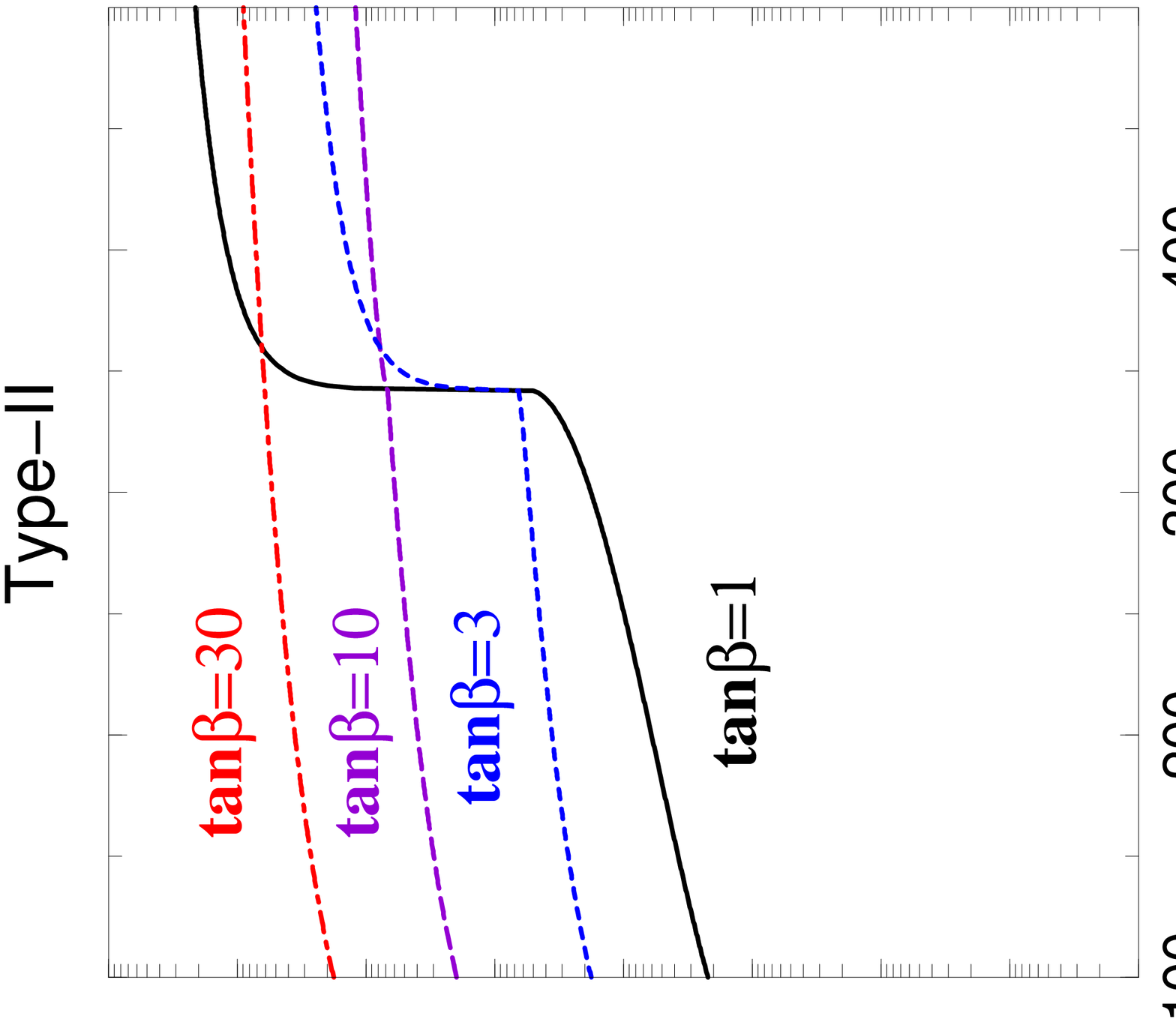}
\end{minipage}
\begin{minipage}{0.23\hsize}
\includegraphics[width=4.4cm,angle=-90]{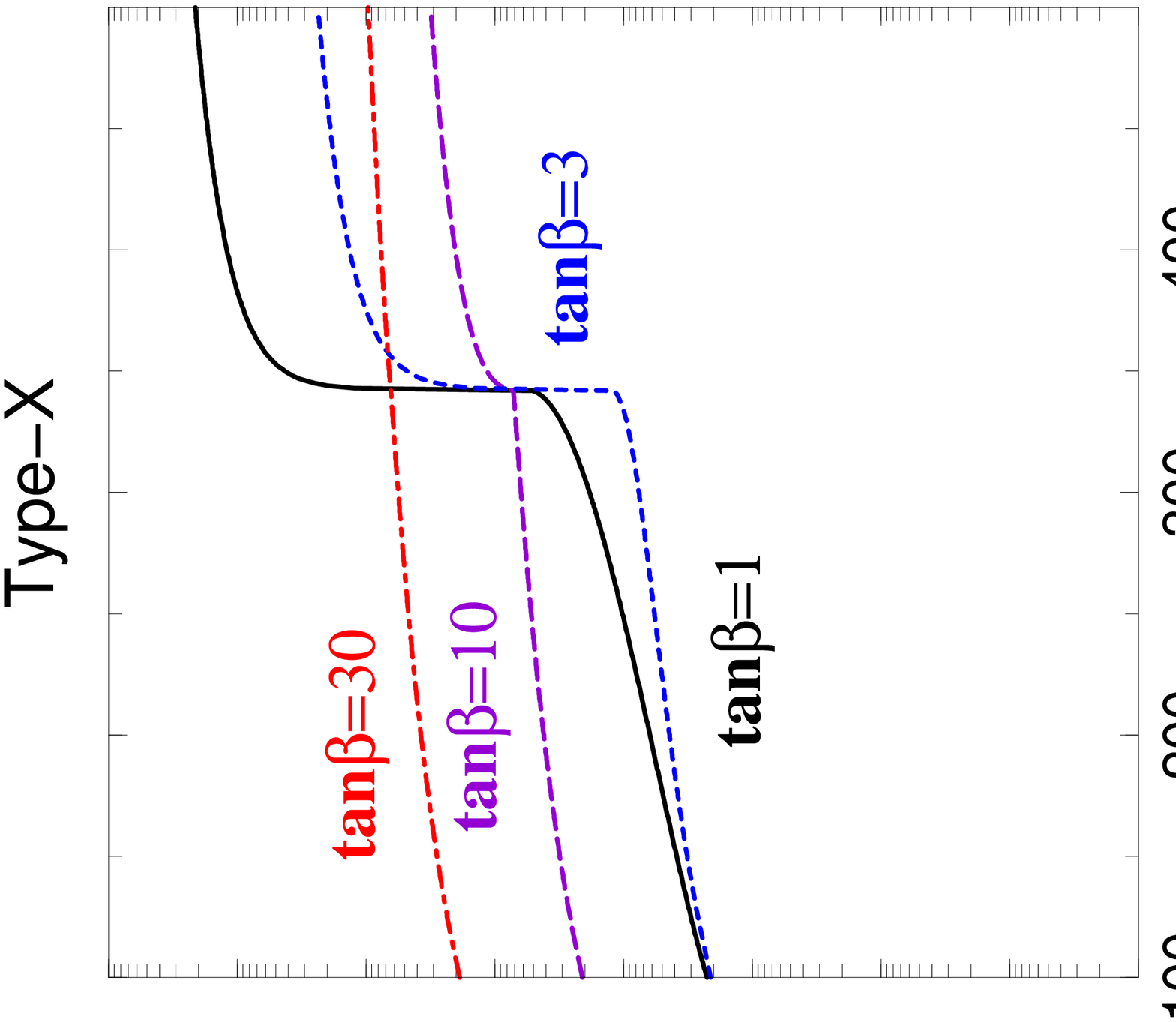}
\end{minipage}
\begin{minipage}{0.23\hsize}
\includegraphics[width=4.4cm,angle=-90]{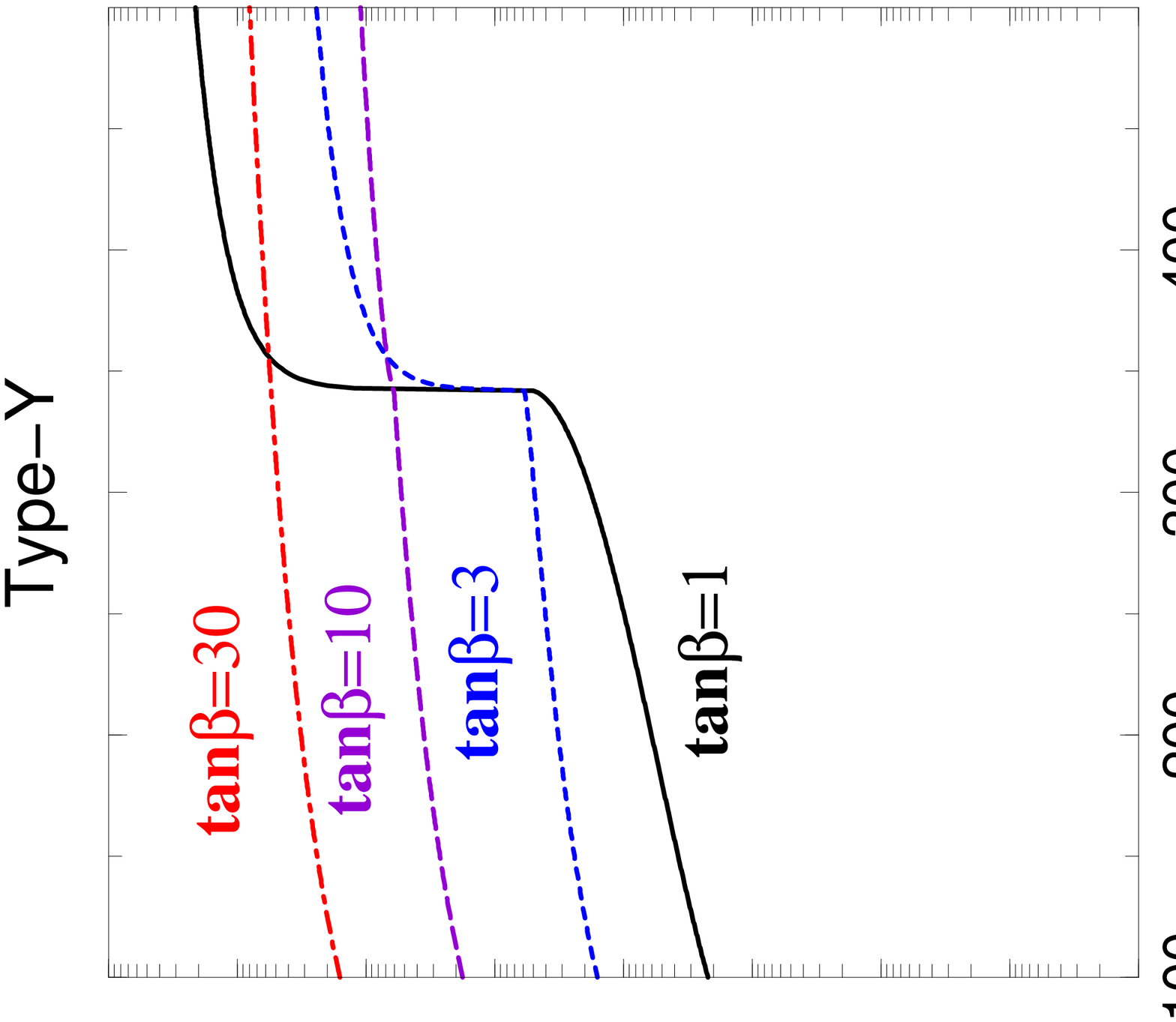}
\end{minipage}
\end{center}
\begin{center}
\begin{minipage}{0.275\hsize}
\vspace{0.8ex}
\includegraphics[width=4.5cm,angle=-90]{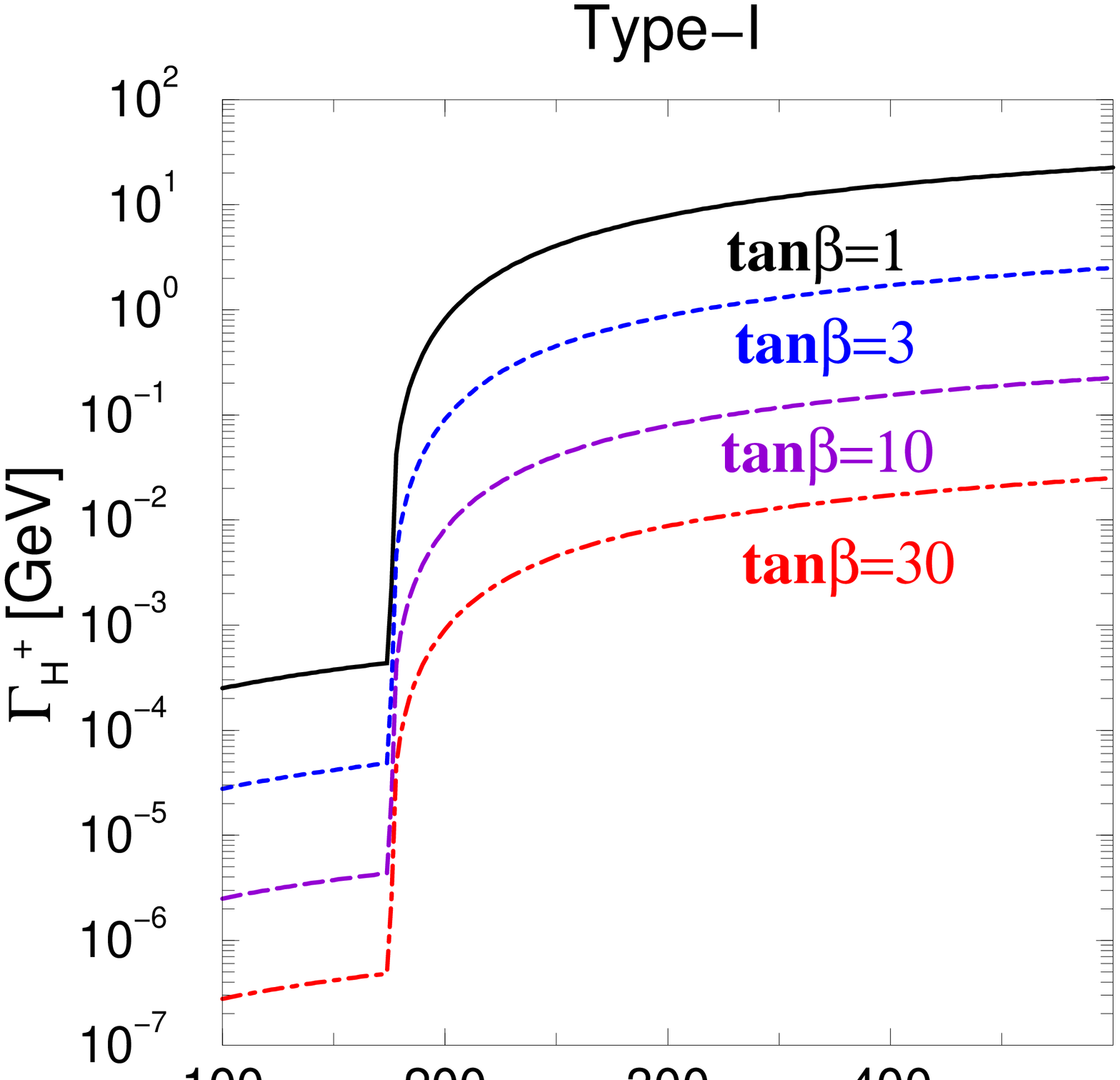}
\end{minipage}
\begin{minipage}{0.23\hsize}
\includegraphics[width=4.4cm,angle=-90]{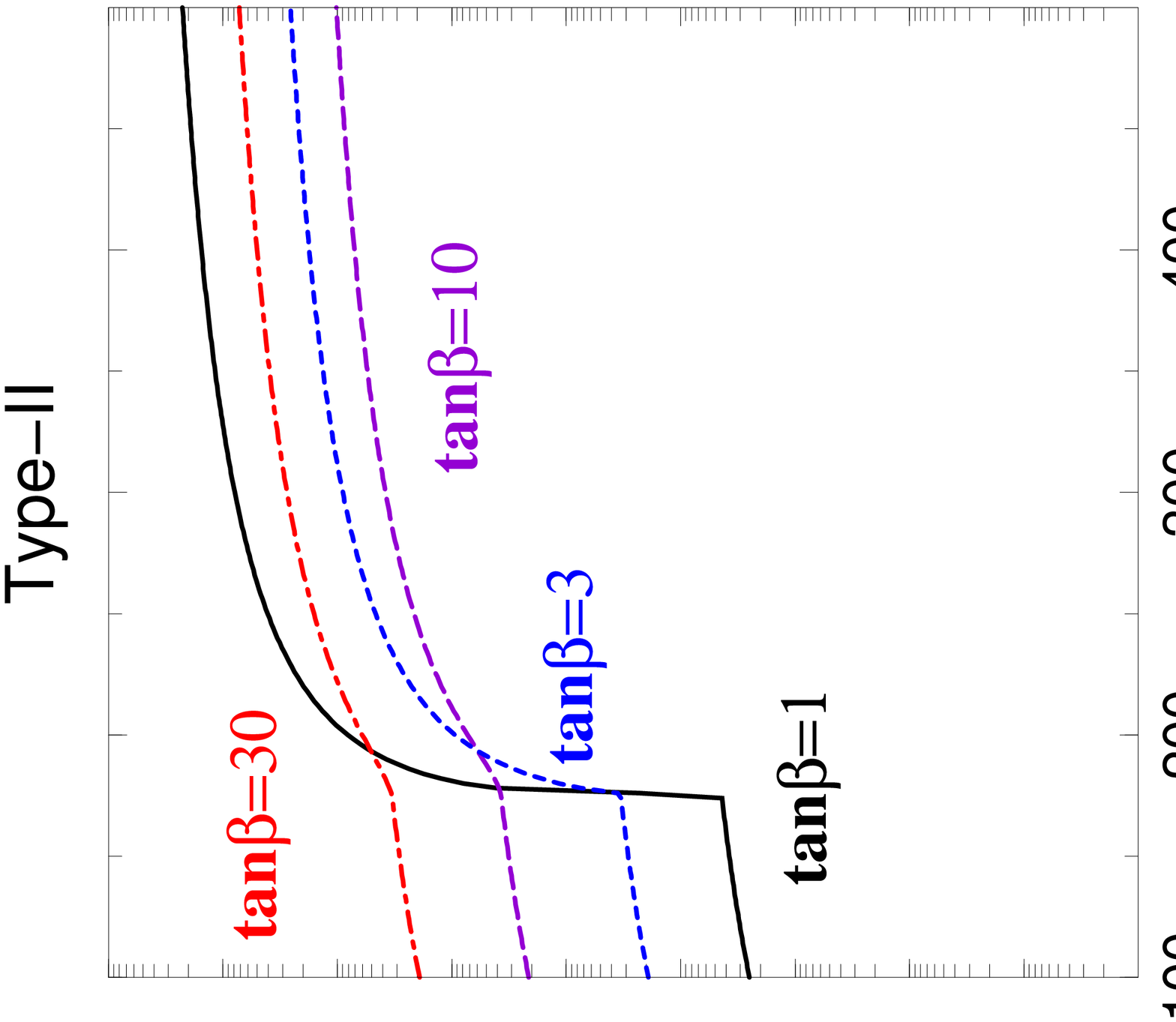}
\end{minipage}
\begin{minipage}{0.23\hsize}
\includegraphics[width=4.4cm,angle=-90]{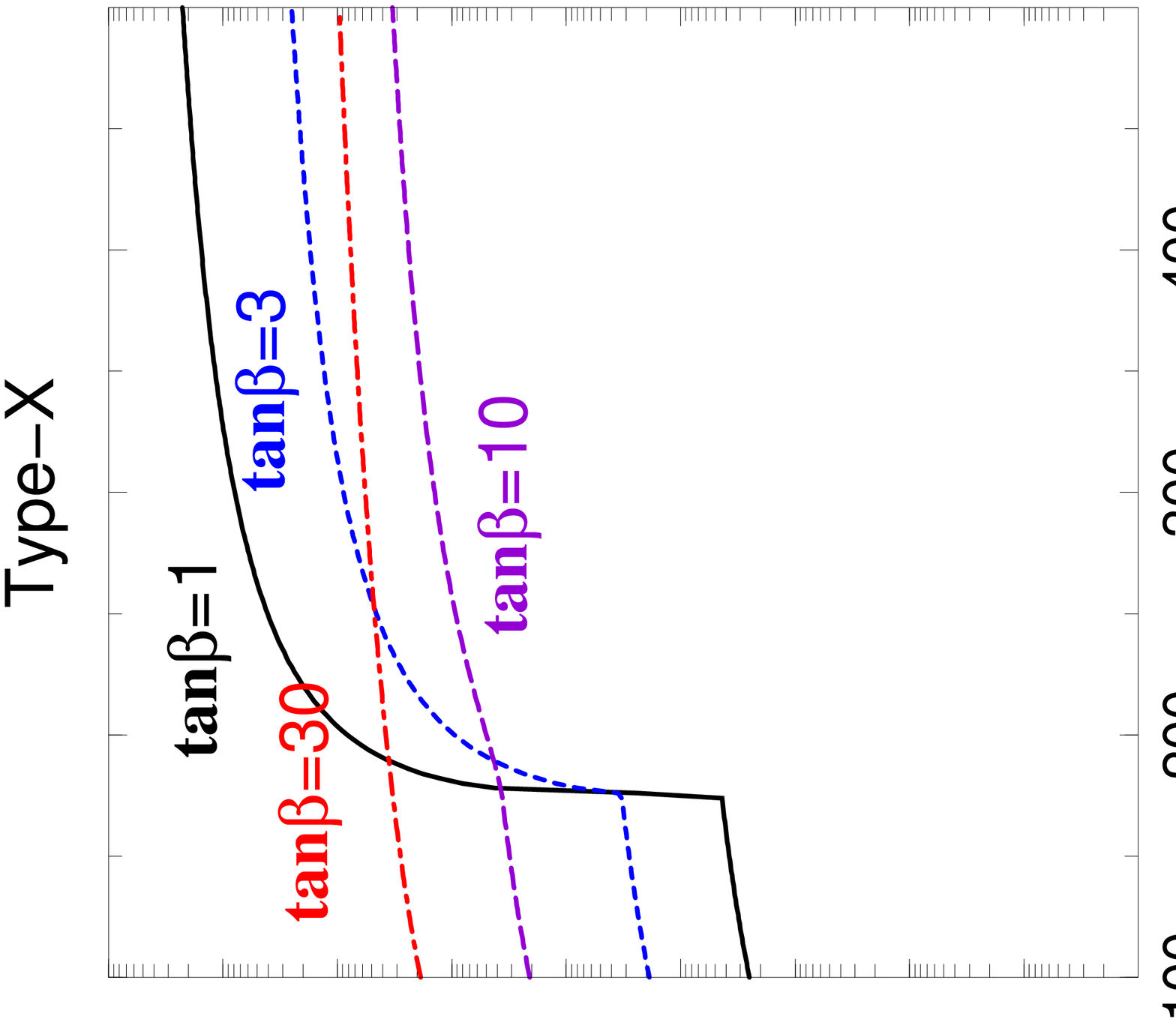}
\end{minipage}
\begin{minipage}{0.23\hsize}
\includegraphics[width=4.4cm,angle=-90]{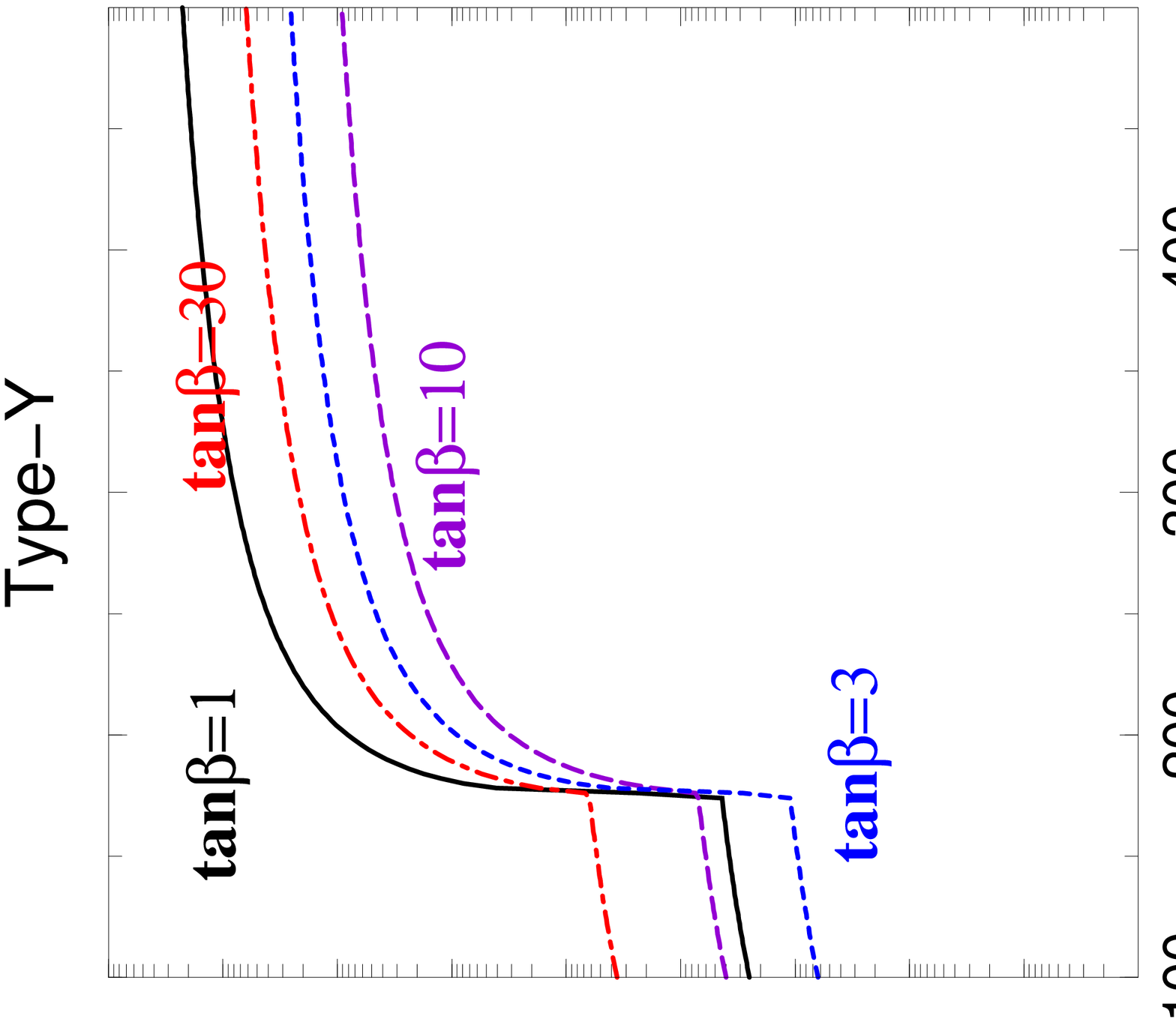}
\end{minipage}
\end{center}
\caption{Total decay widths of $H$, $A$ and $H^\pm$ in the four
 different  types of THDM as a function of the decaying scalar boson mass
for several values of $\tan\beta$ under the assumption $m_\Phi^{}=m_H^{}=m_A^{}=m_{H^\pm}^{}$
and $M=m_\Phi^{}-1$ GeV. The SM-like limit $\sin(\beta-\alpha) =1$ is
taken, where $h$ is the SM-like Higgs boson.}
\label{FIG:width_mass}
\end{figure}

\begin{figure}
\begin{center}
\begin{minipage}{0.285\hsize}
\includegraphics[width=4.4cm,angle=-90]{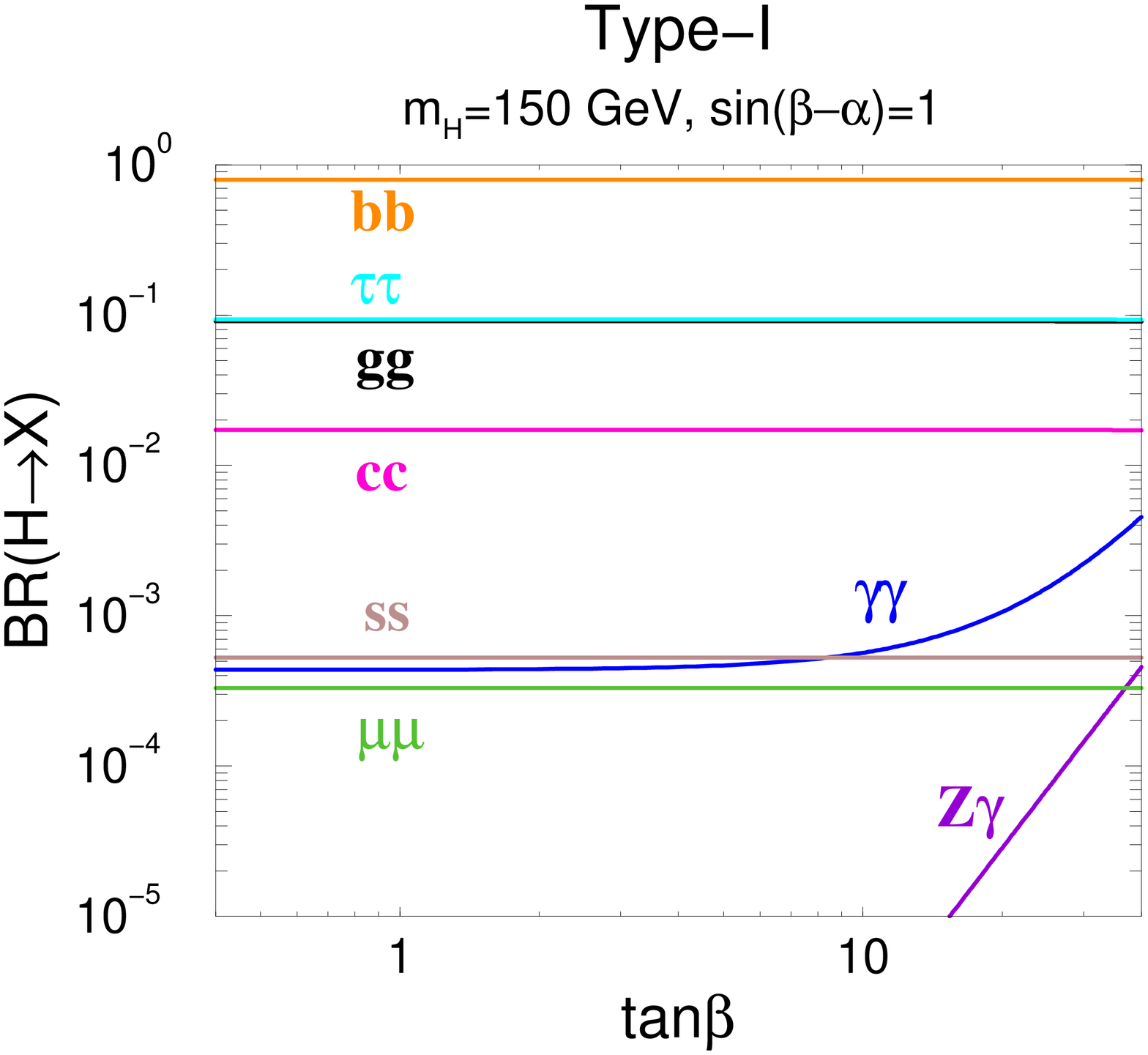}
\end{minipage}
\begin{minipage}{0.23\hsize}
\includegraphics[width=4.4cm,angle=-90]{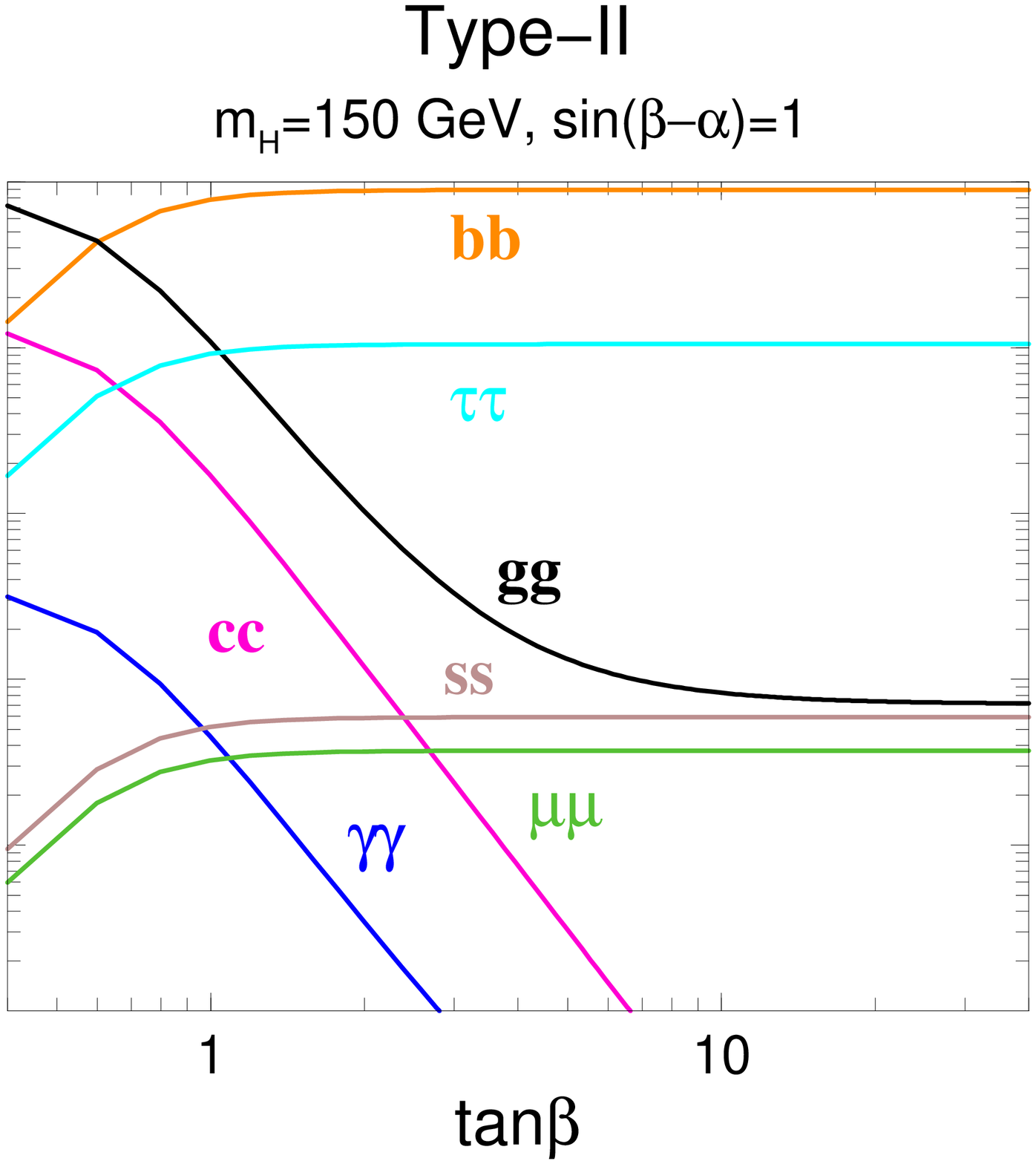}
\end{minipage}
\begin{minipage}{0.23\hsize}
\includegraphics[width=4.4cm,angle=-90]{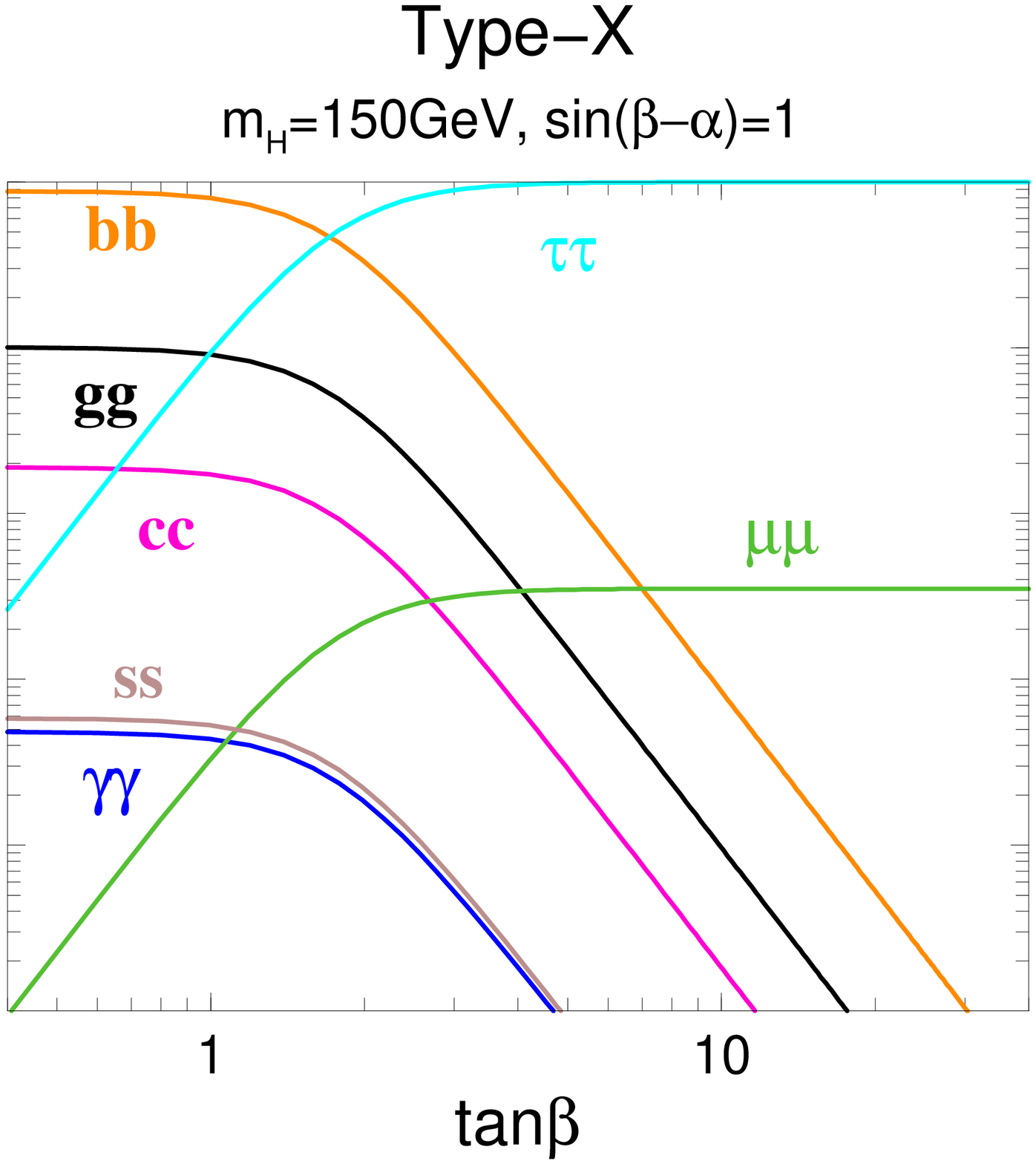}
\end{minipage}
\begin{minipage}{0.23\hsize}
\includegraphics[width=4.4cm,angle=-90]{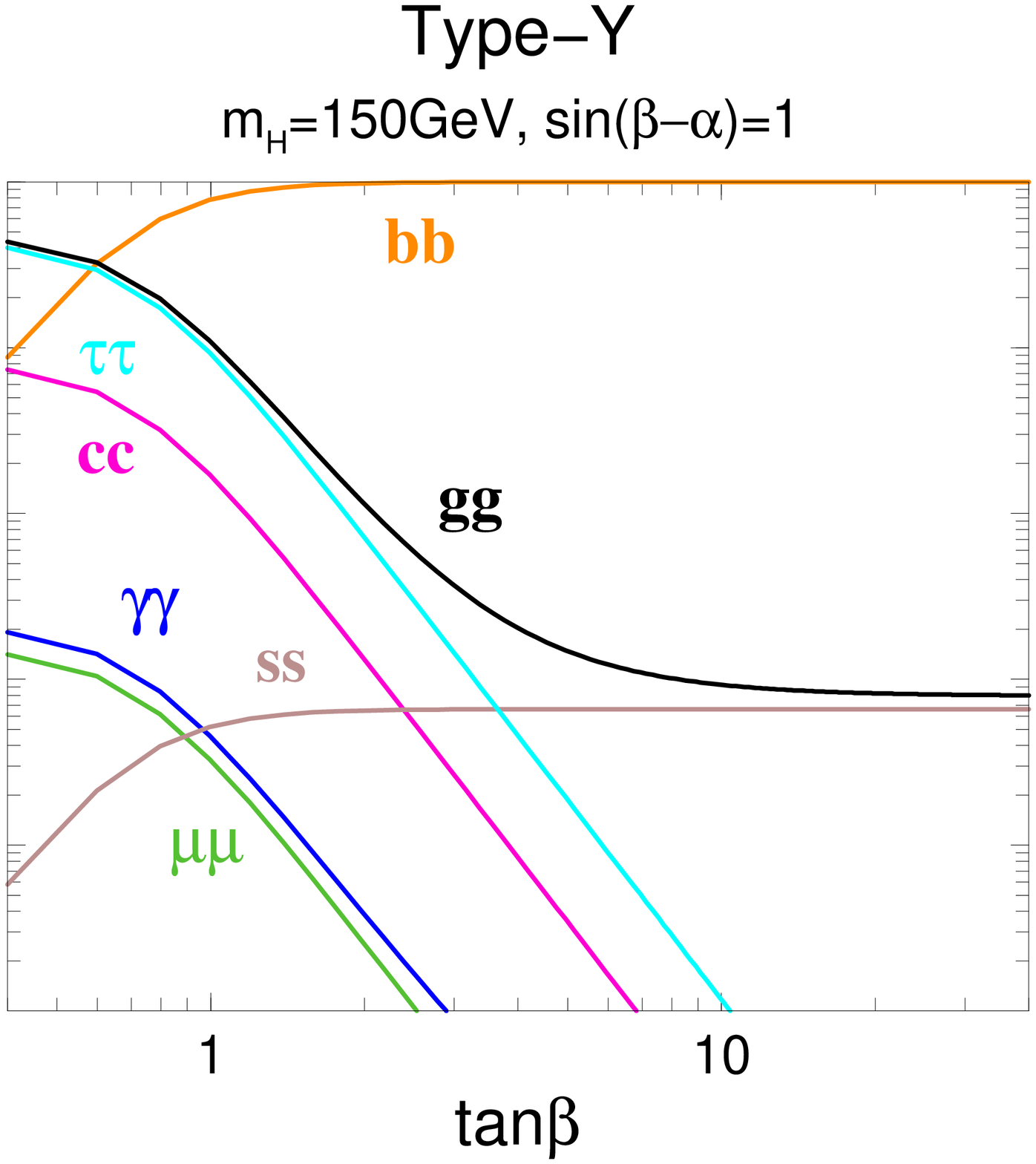}
\end{minipage}
\begin{minipage}{0.285\hsize}
\includegraphics[width=4.4cm,angle=-90]{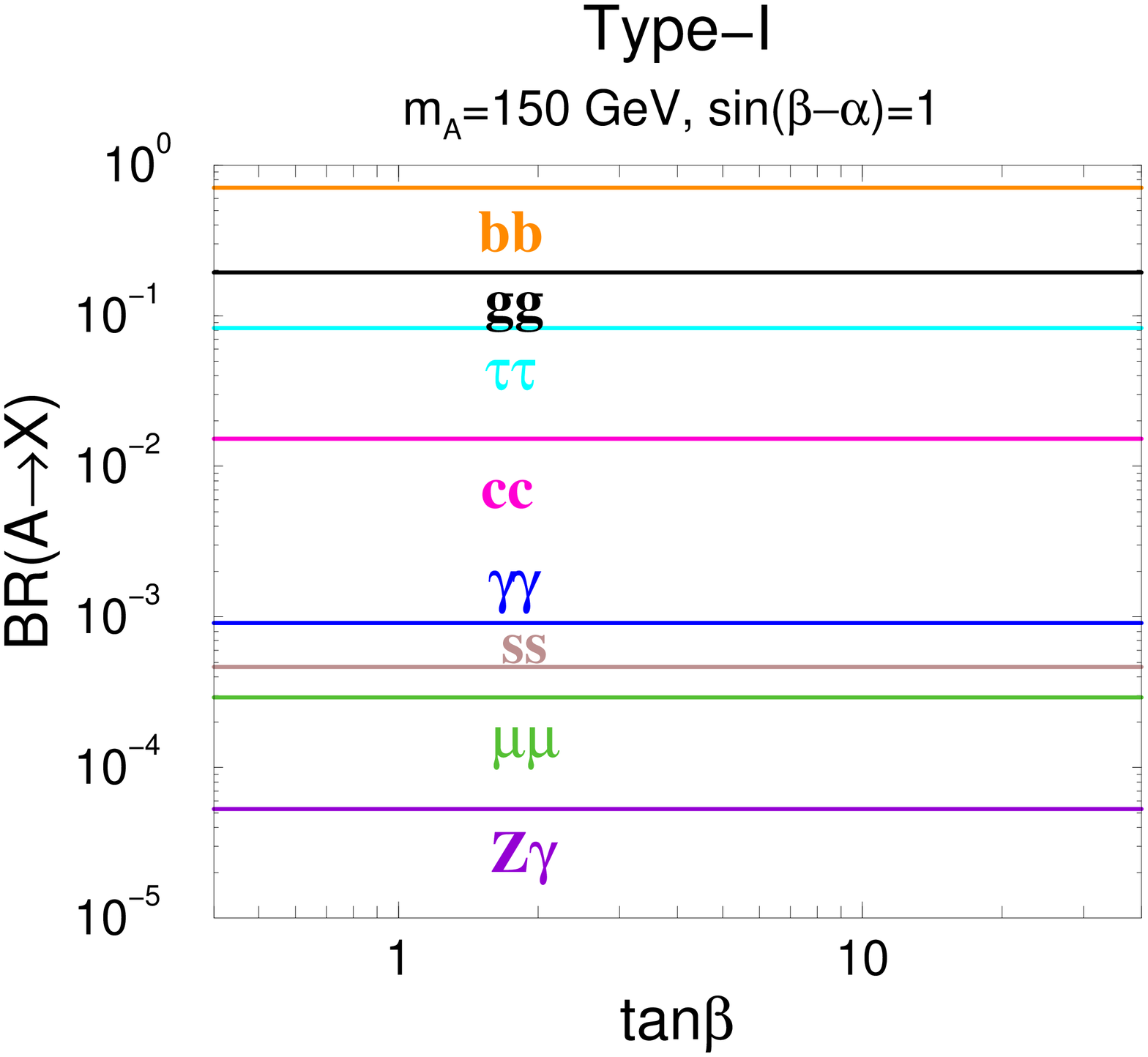}
\end{minipage}
\begin{minipage}{0.23\hsize}
\includegraphics[width=4.4cm,angle=-90]{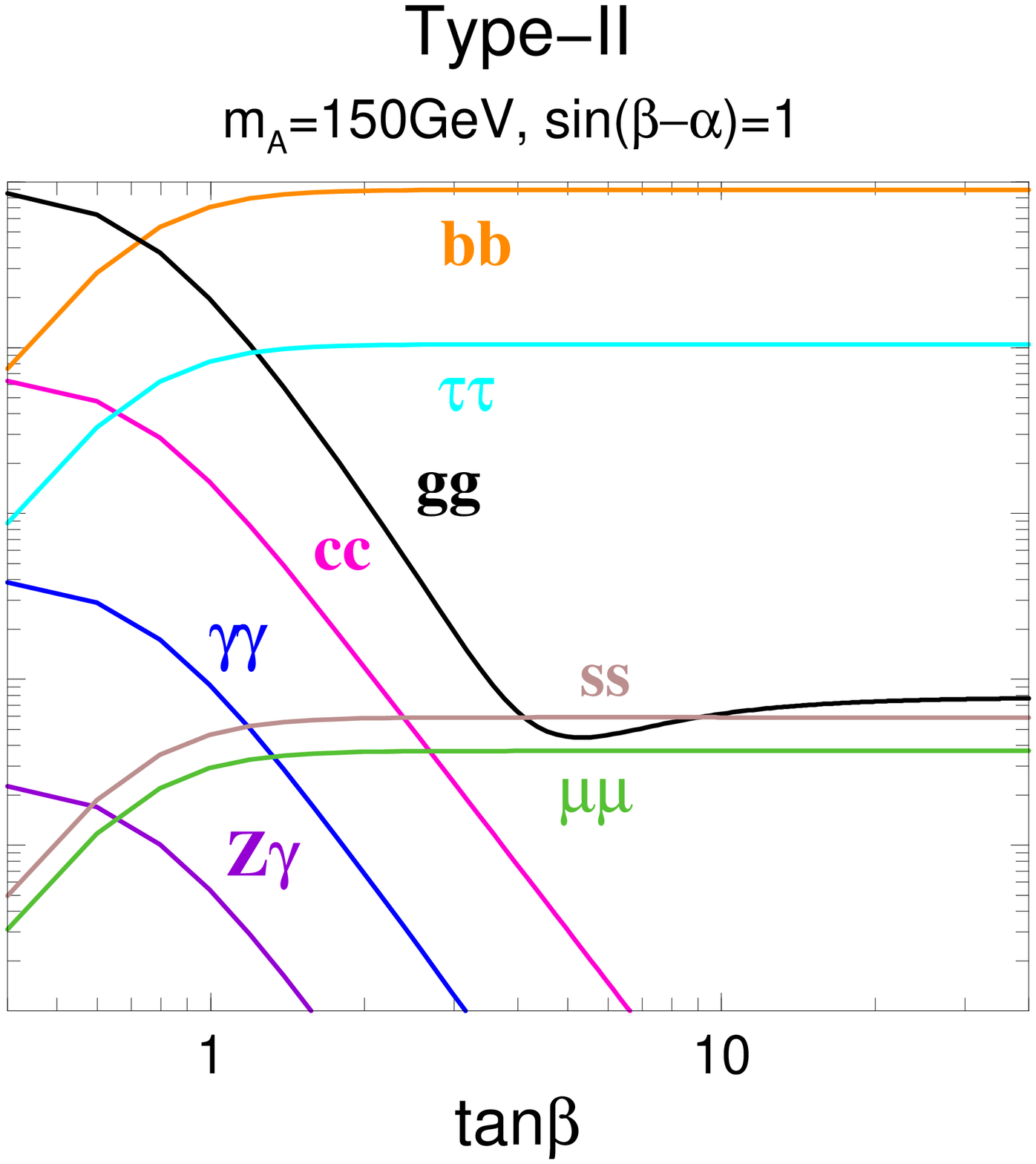}
\end{minipage}
\begin{minipage}{0.23\hsize}
\includegraphics[width=4.4cm,angle=-90]{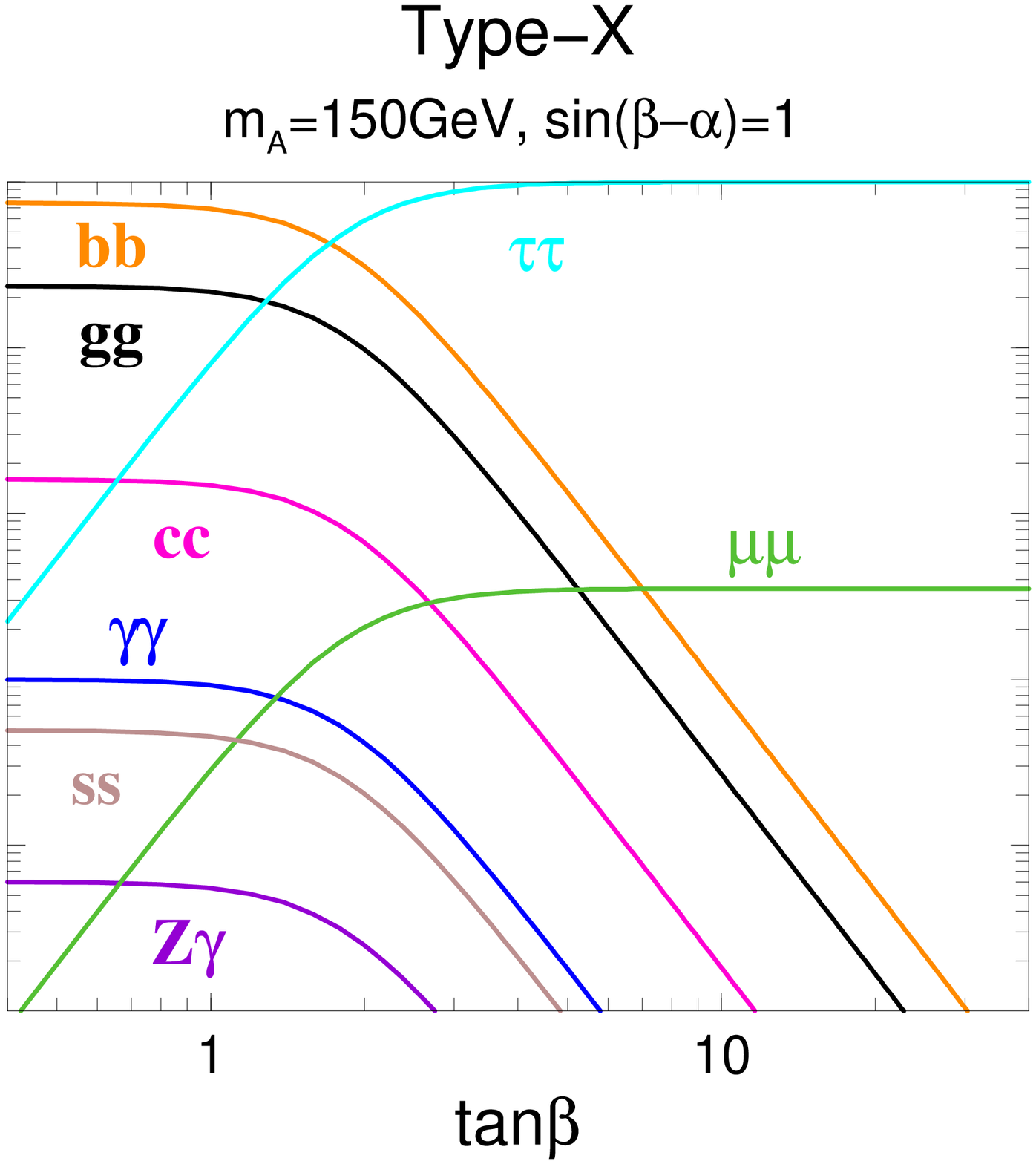}
\end{minipage}
\begin{minipage}{0.23\hsize}
\includegraphics[width=4.4cm,angle=-90]{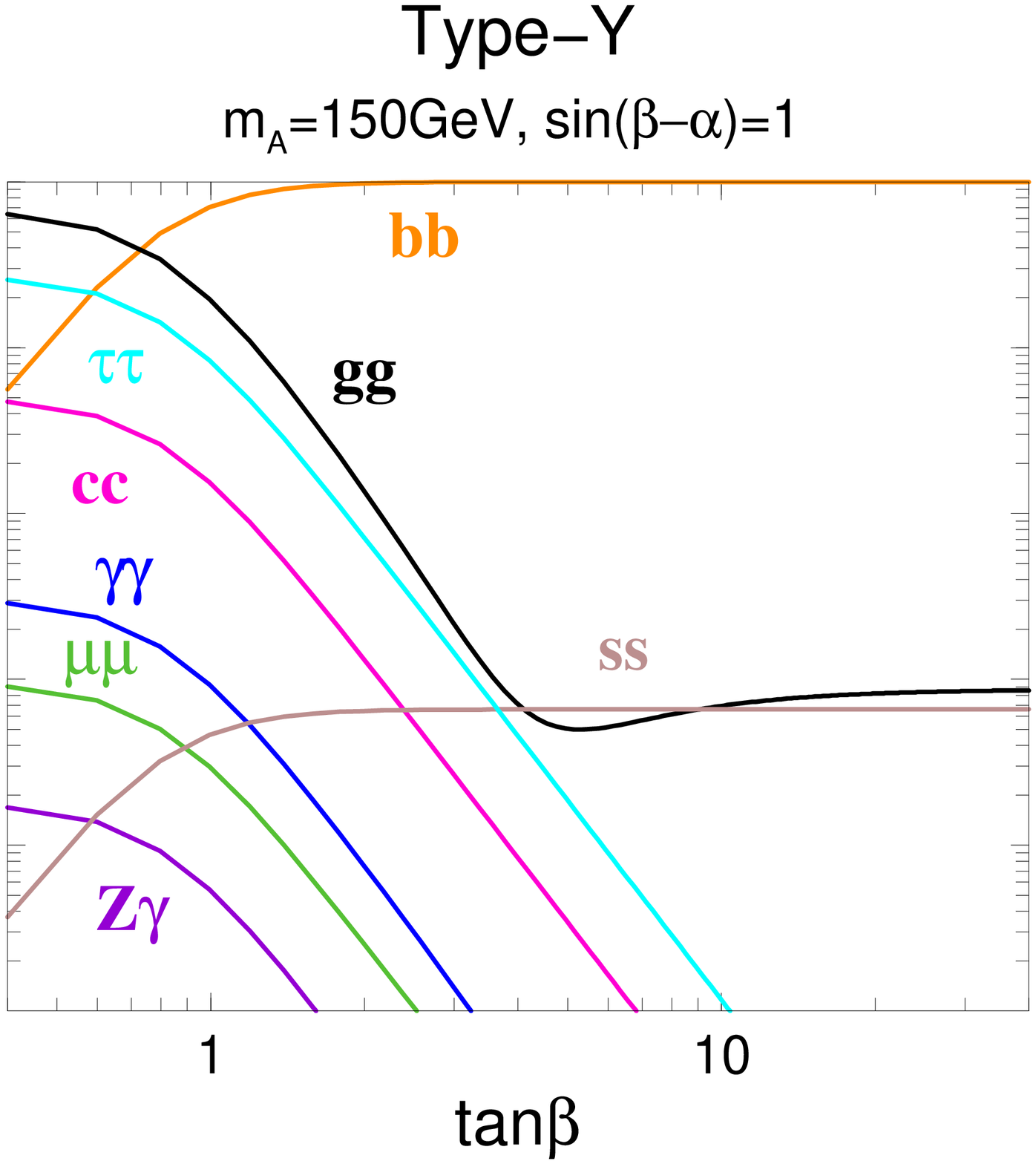}
\end{minipage}
\begin{minipage}{0.285\hsize}
\includegraphics[width=4.4cm,angle=-90]{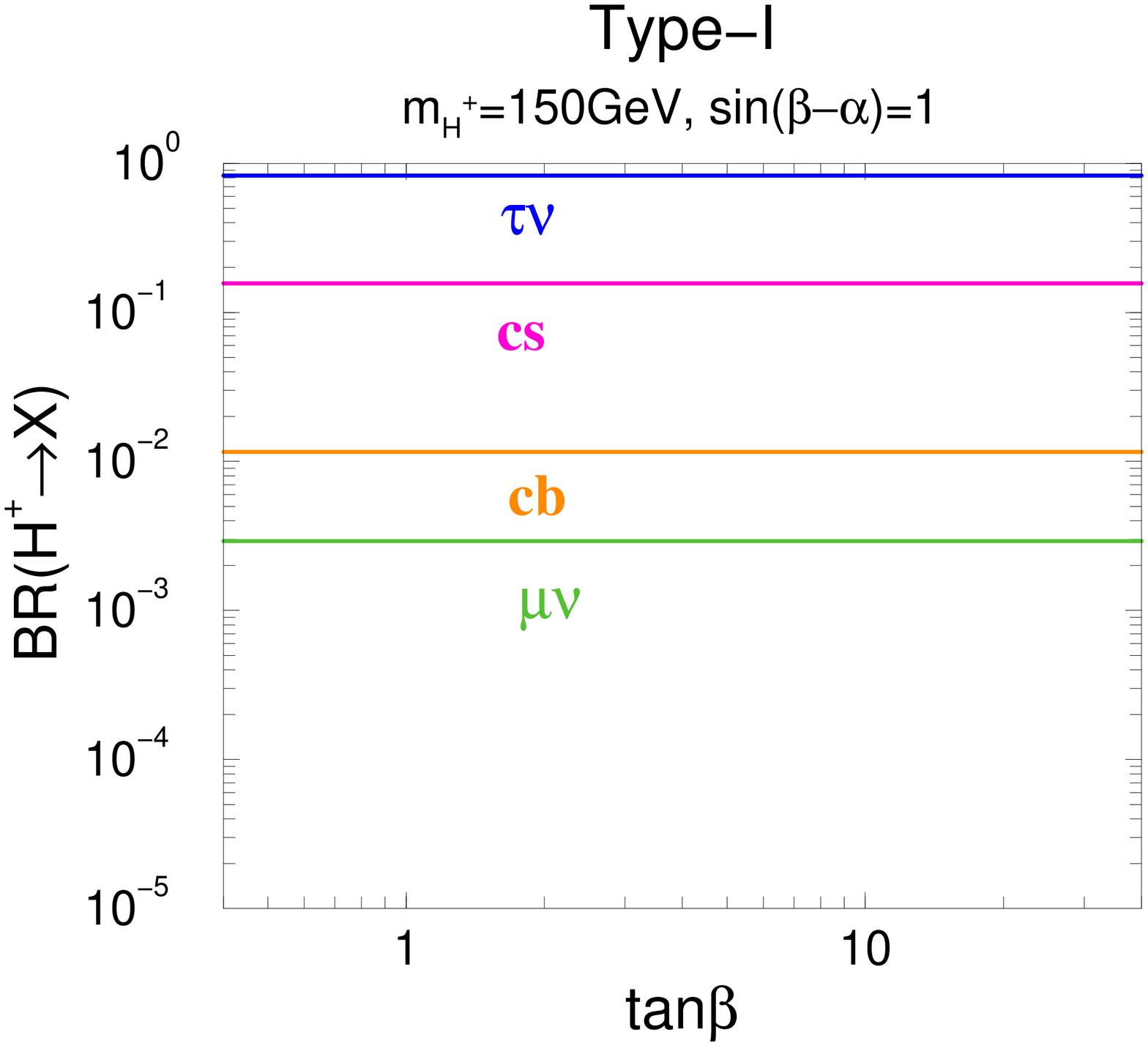}
\end{minipage}
\begin{minipage}{0.23\hsize}
\includegraphics[width=4.4cm,angle=-90]{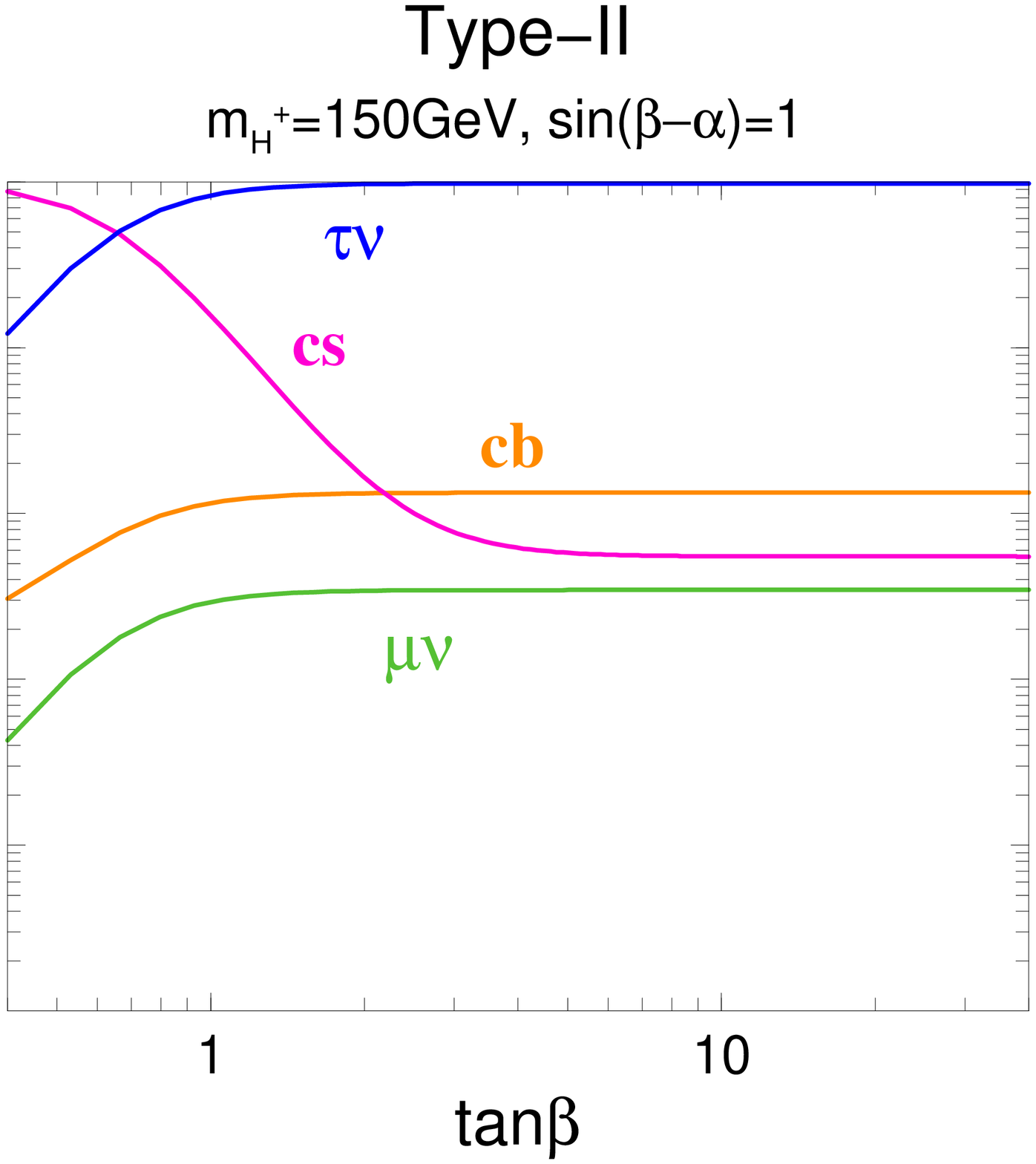}
\end{minipage}
\begin{minipage}{0.23\hsize}
\includegraphics[width=4.4cm,angle=-90]{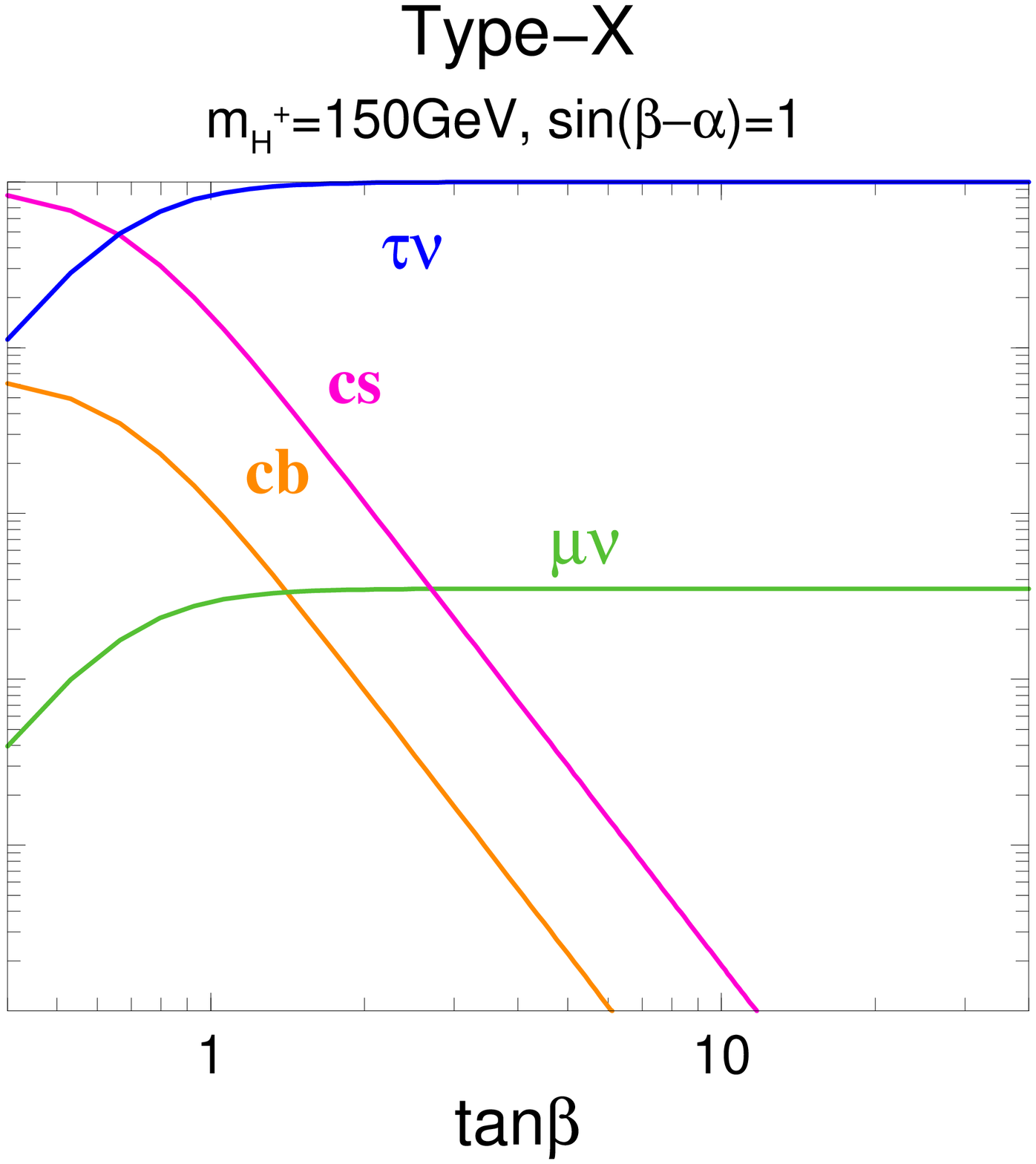}
\end{minipage}
\begin{minipage}{0.23\hsize}
\includegraphics[width=4.4cm,angle=-90]{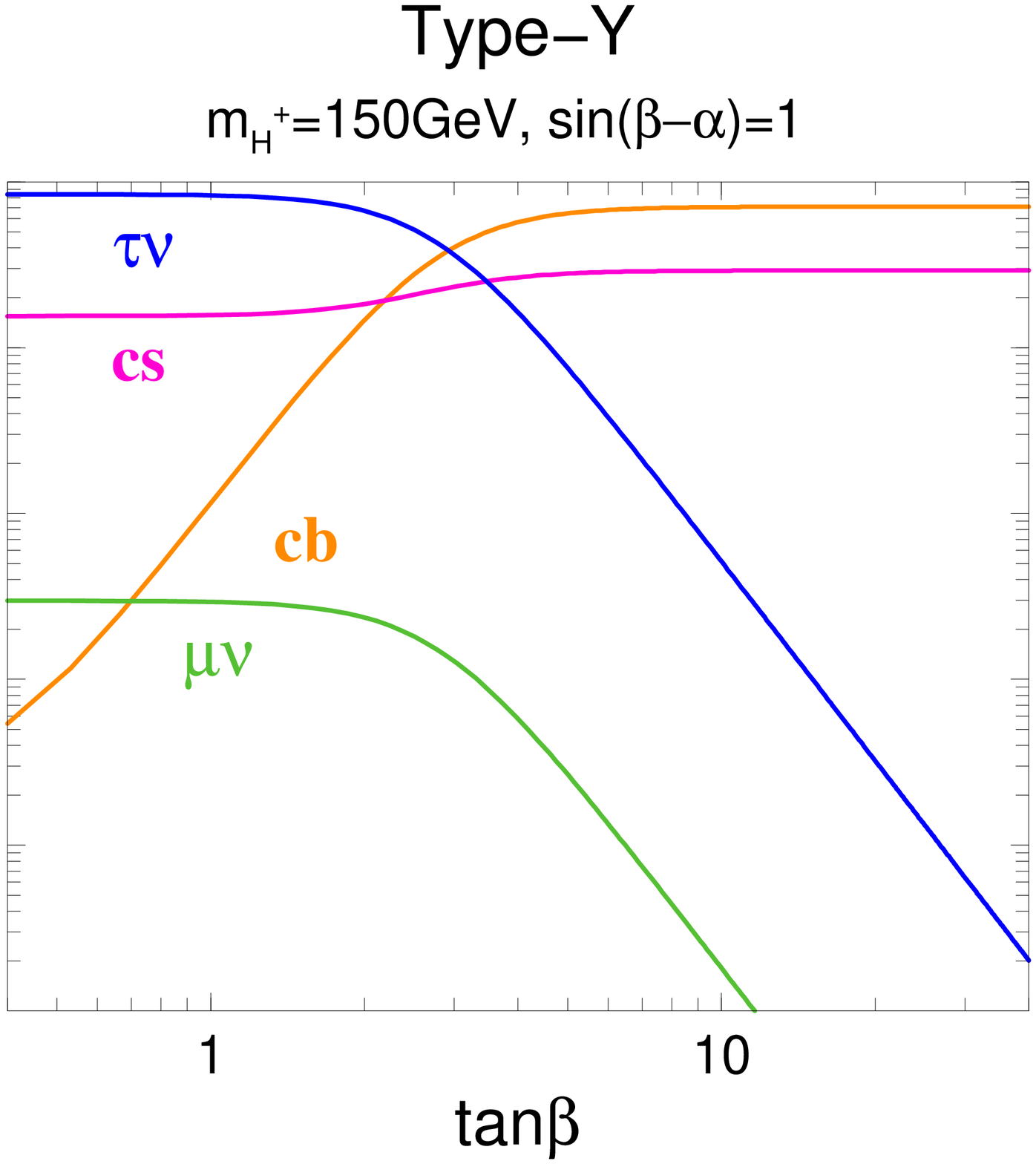}
\end{minipage}
\caption{Decay branching ratios of $H$, $A$ and $H^\pm$
in the four different types of THDM as a function of $\tan\beta$
for $m_H^{}=m_A^{}=m_{H^\pm}^{}=150$ GeV and $M=149$ GeV.
The SM-like limit $\sin(\beta-\alpha) =1$ is taken, where $h$
is the SM-like Higgs boson.}
\label{FIG:br_150}
\end{center}
\end{figure}

\begin{figure}
\begin{center}
\begin{minipage}{0.285\hsize}
\includegraphics[width=4.4cm,angle=-90]{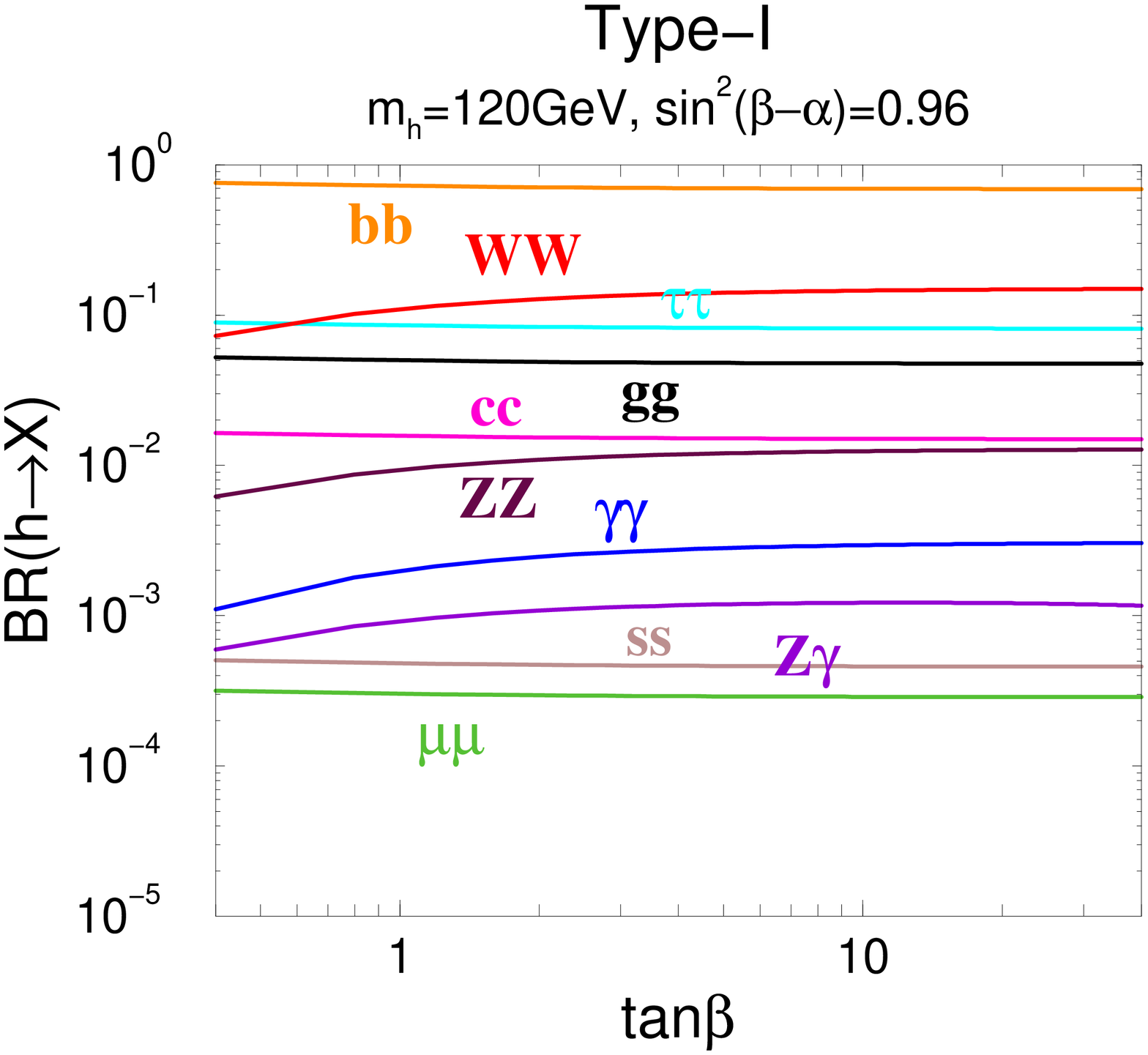}
\end{minipage}
\begin{minipage}{0.23\hsize}
\includegraphics[width=4.4cm,angle=-90]{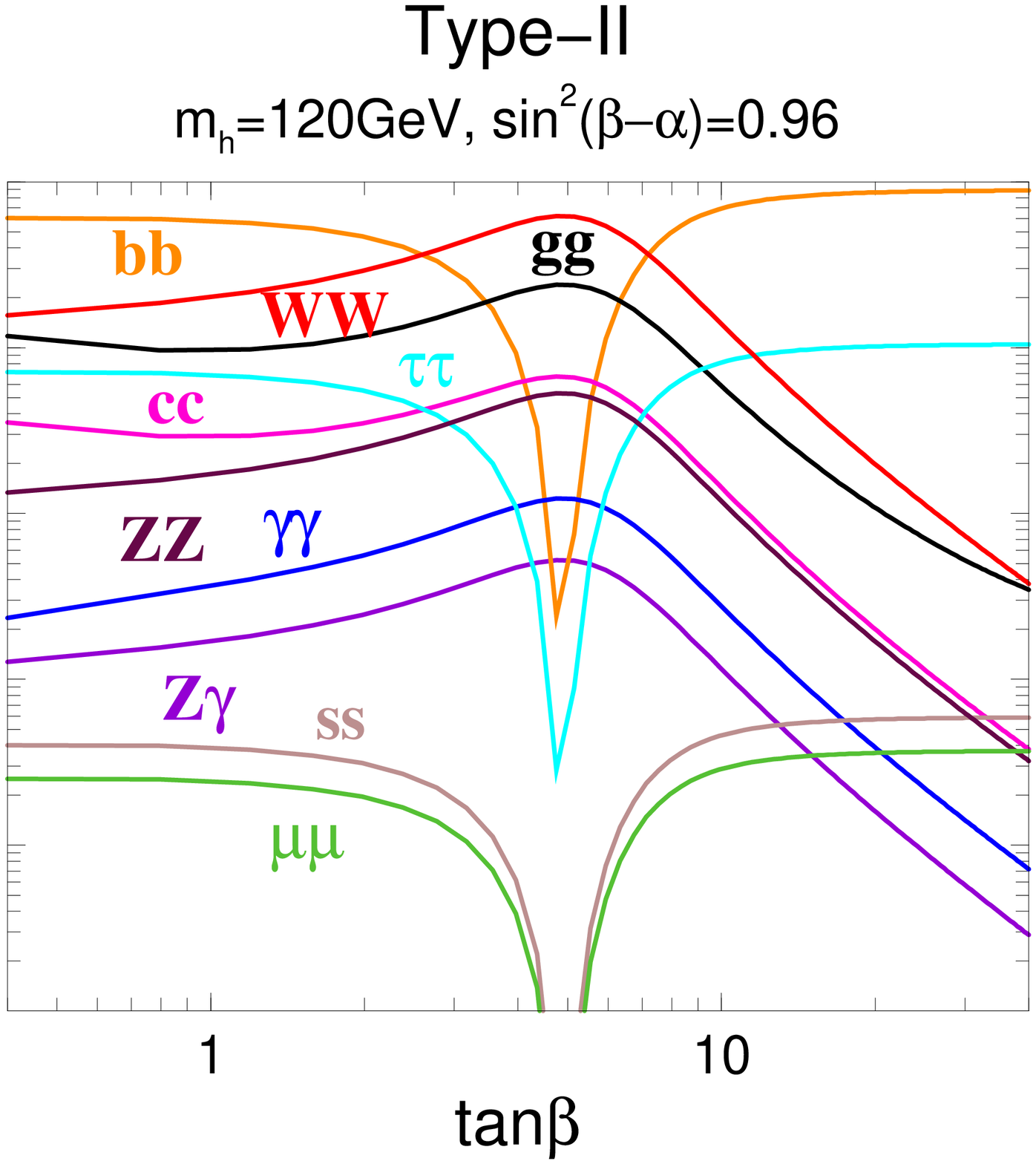}
\end{minipage}
\begin{minipage}{0.23\hsize}
\includegraphics[width=4.4cm,angle=-90]{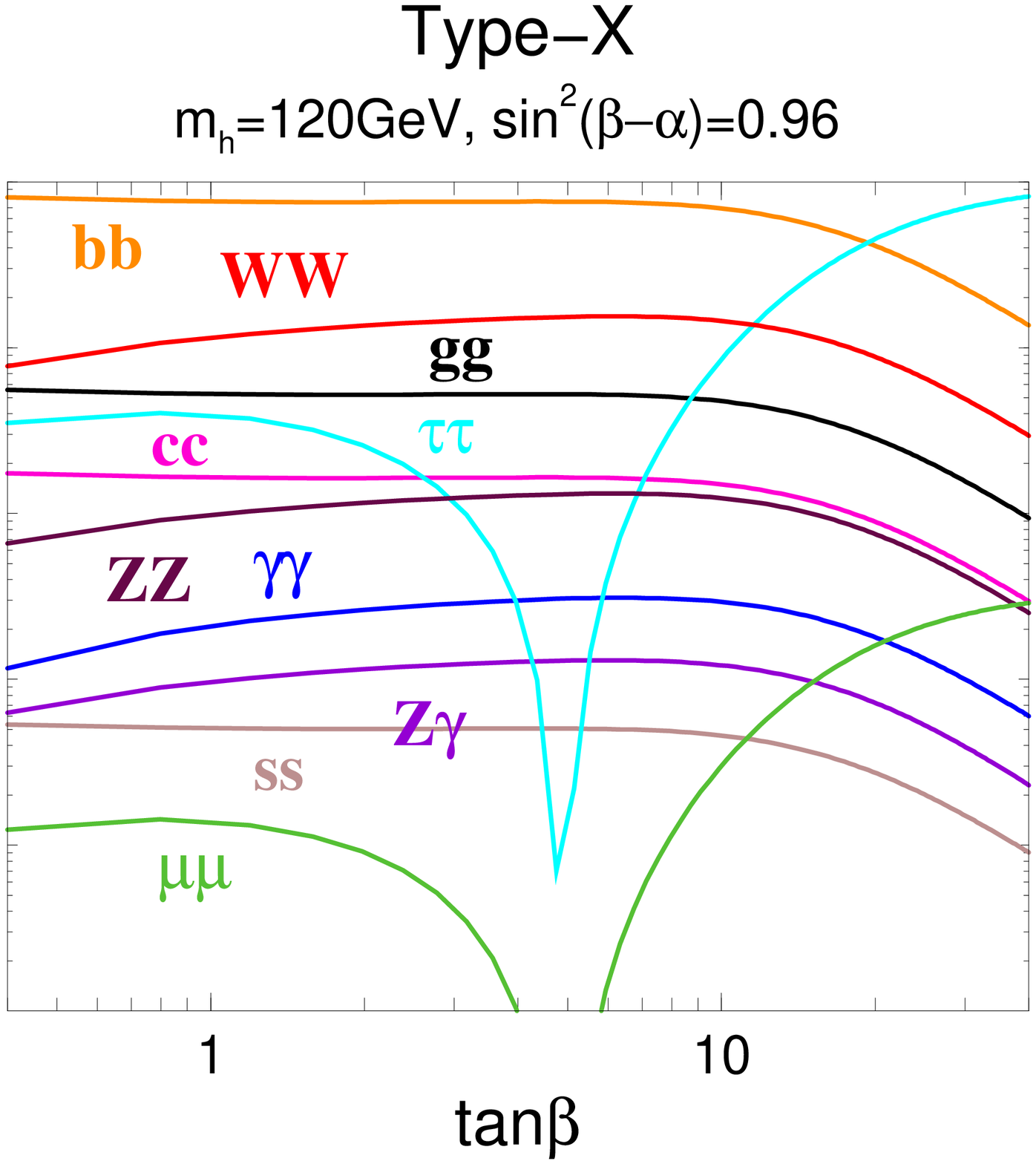}
\end{minipage}
\begin{minipage}{0.23\hsize}
\includegraphics[width=4.4cm,angle=-90]{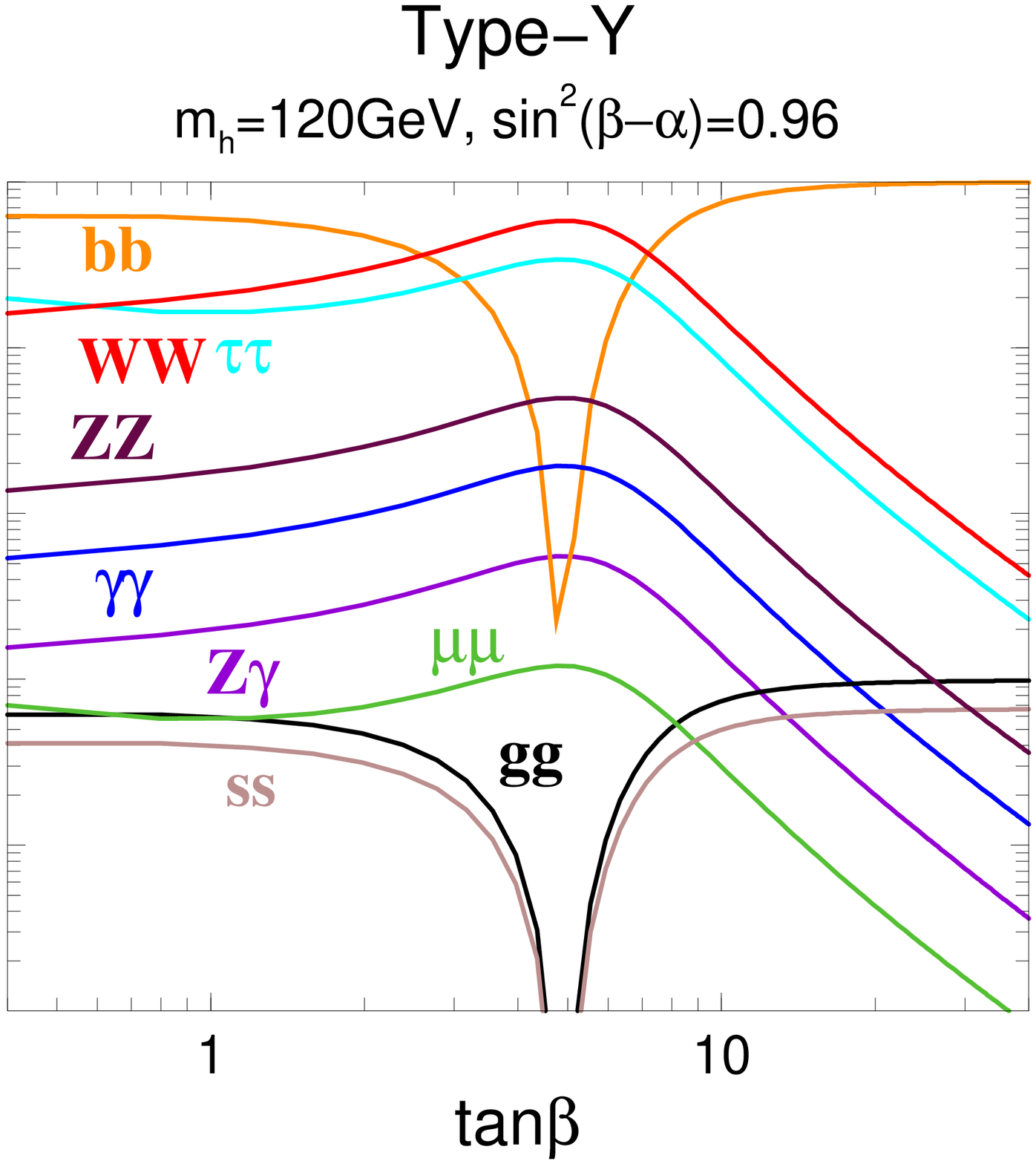}
\end{minipage}
\begin{minipage}{0.285\hsize}
\includegraphics[width=4.4cm,angle=-90]{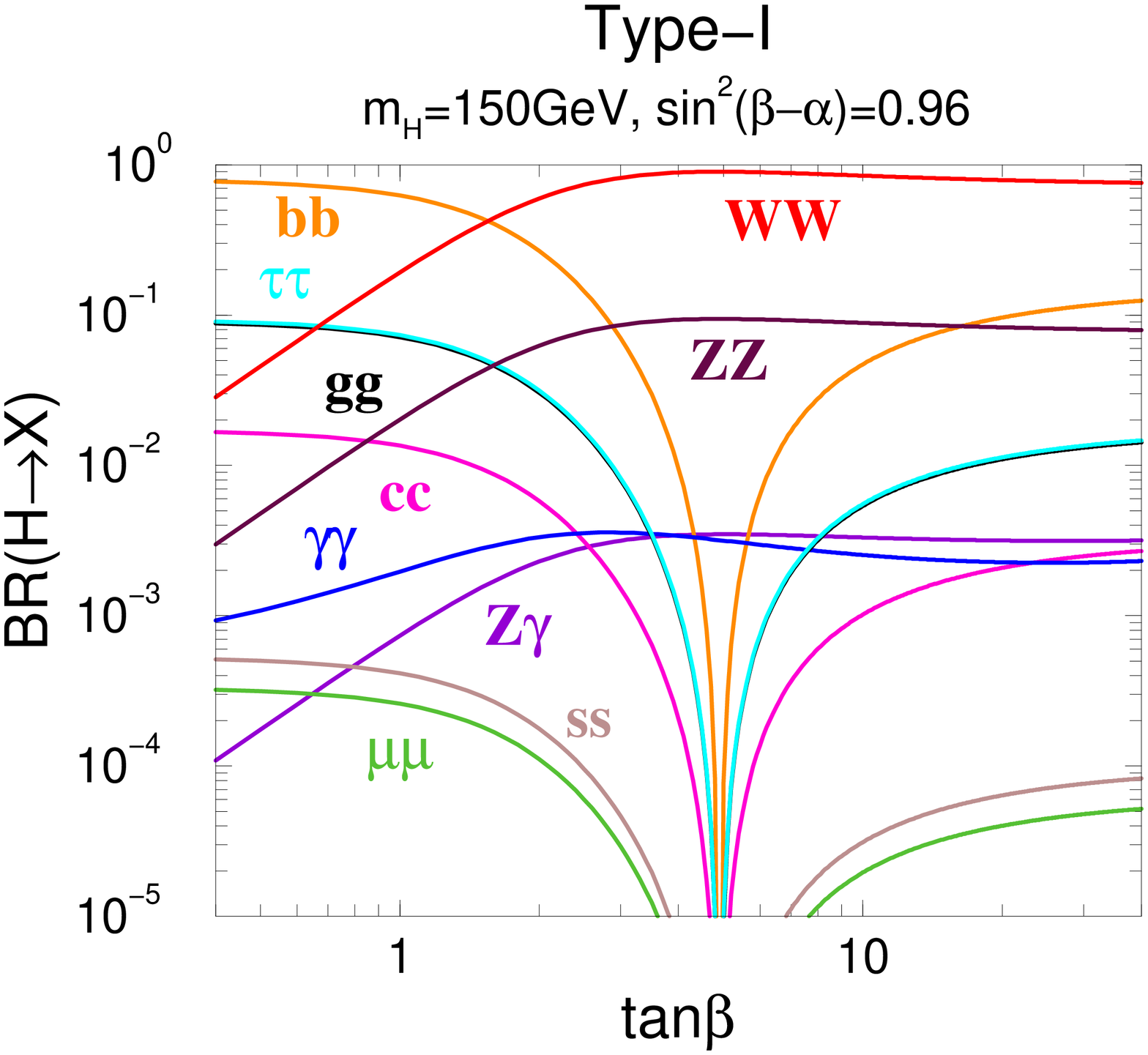}
\end{minipage}
\begin{minipage}{0.23\hsize}
\includegraphics[width=4.4cm,angle=-90]{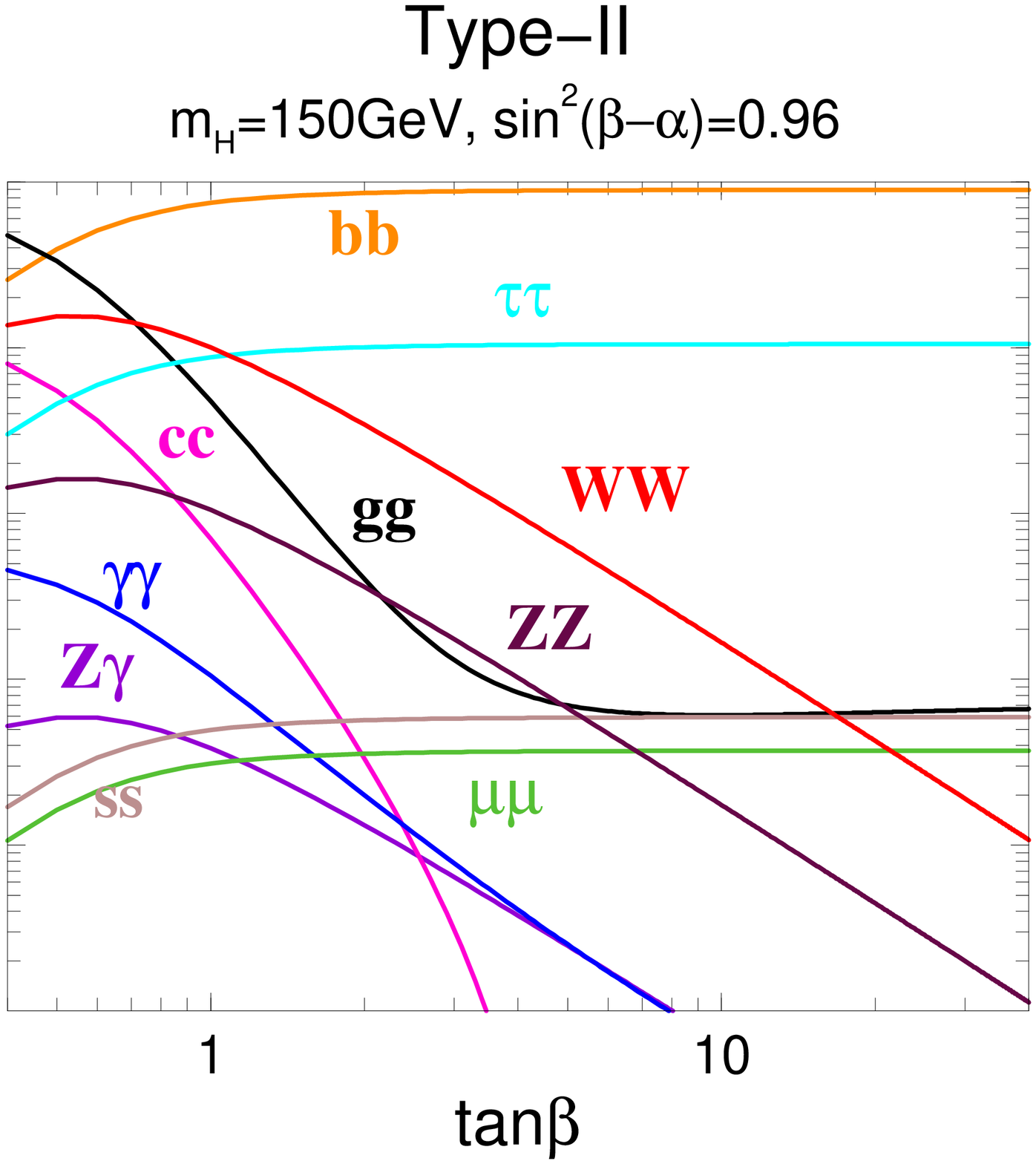}
\end{minipage}
\begin{minipage}{0.23\hsize}
\includegraphics[width=4.4cm,angle=-90]{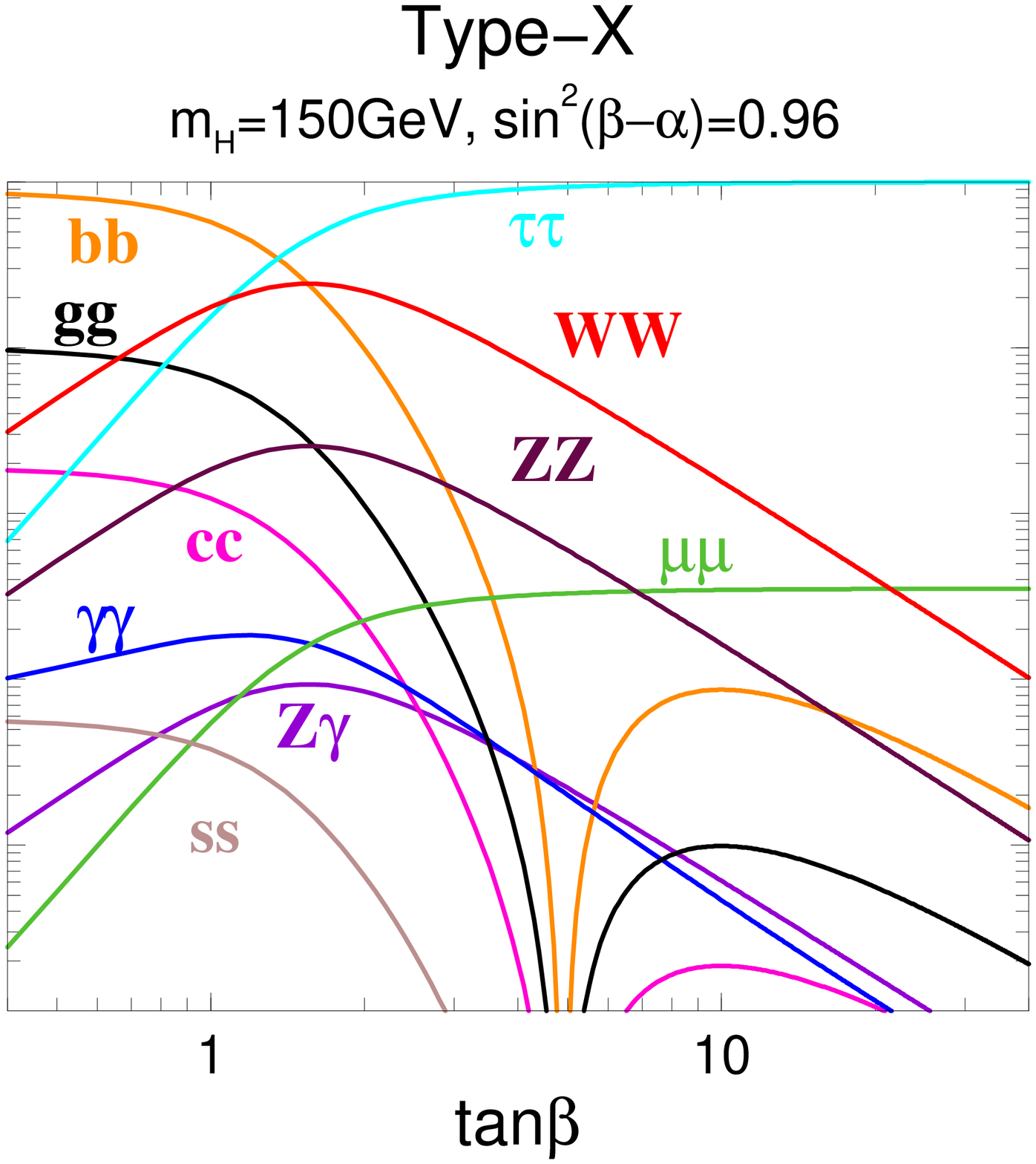}
\end{minipage}
\begin{minipage}{0.23\hsize}
\includegraphics[width=4.4cm,angle=-90]{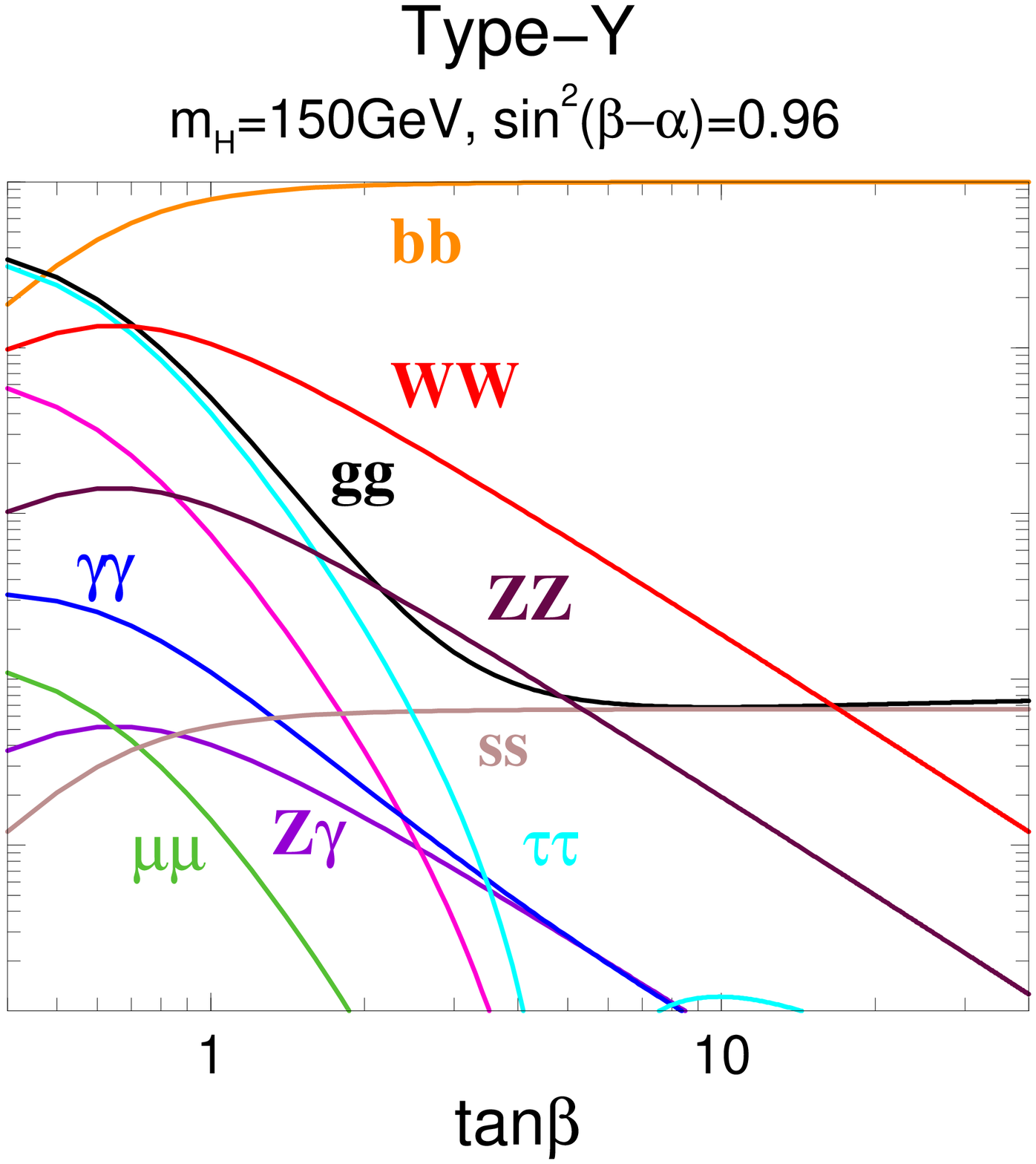}
\end{minipage}
\caption{Decay branching ratios of $h$ and $H$
in the four different types of THDM as a function of $\tan\beta$
for $m_h^{}=120$ GeV, $m_H^{}=150$ GeV, $M=148$ GeV and
$\sin^2(\beta-\alpha) =0.96$.}
\label{FIG:br_150'}
\end{center}
\end{figure}

In this section, we discuss the difference in decays of the Higgs bosons
for the types of Yukawa interactions in the THDM.
We calculate the decay rates of the Higgs bosons
and evaluate the total widths and the branching ratios.
In particular, we show the result with $\sin(\beta-\alpha) \simeq 1$~\cite{Ref:SMlike,Ref:KOSY},
where $h$ is the SM-like Higgs
boson while the VEV of $H$ is very small or negligible.
The decay pattern of $h$ is almost the same as that of the SM
Higgs boson with the same mass at the leading order except
for the loop-induced channels when $\sin(\beta-\alpha)=1$.
In this case, $H$ does not decay into the gauge
boson pair at tree level, so it mainly decays into fermion
pairs\footnote{In the case with a more complicated mass spectrum a heavy Higgs boson
can decay into the states which contain lighter Higgs
bosons~\cite{Ref:Yagyu}.}.
We note that $A$ and $H^\pm$ do not decay into the gauge boson pair at the
tree level for all
values of $\sin(\beta-\alpha)$.

The decay patterns are therefore completely
different among the different types of Yukawa interactions~\cite{Ref:Barger,Ref:Grossmann}.
For the decays of $H$ and $A$, we take into account the decay channels of
$q\bar q$, $\ell^+\ell^-$, ($WW^{(\ast)}$, $ZZ^{(\ast)}$) at the tree level,
and $gg$, $\gamma\gamma$, $Z\gamma$ at the leading order, where $q$ represents
$s$, $c$, $b$ (and $t$), and $\ell$ represents $\mu$ and $\tau$.
Running masses for $b$, $c$, and $s$ quarks are fixed as 
$\overline{m_b}=3.0$ GeV, $\overline{m_c}=0.81$ GeV and
$\overline{m_s}=0.046$ GeV, respectively.
For the decay of the charged Higgs boson, the modes into
$tb$, $cb$, $cs$, $\tau\nu$, and $\mu\nu$
are taken
into account as long as they are kinematically allowed.
The analytic formulas of each decay rate are given in the Appendix for completeness.
They are consistent with the previous results for the type-II THDM~\cite{Ref:HHG}.

In FIG.~\ref{FIG:width_mass}, the total widths of $H$, $A$ and $H^\pm$
are shown as a function of the mass of decaying Higgs bosons for
several values of $\tan\beta$ in the four different types of Yukawa
interactions. We assume $\sin(\beta-\alpha)=1$ and
$m_\Phi^{}=m_H^{}=m_A^{}=m_{H^\pm}^{}$. To keep perturbative unitarity
for a wide region of $\tan\beta$,
the soft breaking parameter is taken to be $M=m_\Phi^{}-1$ GeV~\cite{Ref:PU}.
The widths strongly depend on the types of Yukawa interactions for
each $\tan\beta$ value before and after the
threshold of the $t\bar t$ ($tb$) decay mode opens.

In FIG.~\ref{FIG:br_150}, the decay branching ratios of $H$, $A$ and
$H^\pm$ are shown for $m_\Phi^{}=150$ GeV, $\sin(\beta-\alpha)=1$,
and $M=m_\Phi-1$ GeV as a function of $\tan\beta$.
In the type-I THDM, the decay of $H$ into a gauge boson pair
$\gamma\gamma$ or $Z\gamma$ can increase for large $\tan\beta$ values,
because all the other fermionic decays (including the $gg$ mode)
are suppressed but the charged scalar loop contribution to $\gamma\gamma$
and $Z\gamma$ decay modes is not always suppressed
for large $\tan\beta$.
Such an enhancement of the bosonic decay modes cannot be seen
in the decay of $A$ since there is no $AH^+H^-$ coupling.
In the type-X THDM, the main decay mode of $H$ and $A$ is
$\tau^+\tau^-$ for $\tan\beta \gsim 2$, and the leptonic decays
$\tau^+\tau^-$ and $\mu^+\mu^-$ become almost $99\%$ and $0.35\%$ for
$\tan\beta \gtrsim 10$,
 while the $b\bar b$ (or $gg$) mode is always the main decay mode
 in all other cases.
In the type-Y THDM, the leptonic decay modes of $H$ and $A$ are rapidly
suppressed for large $\tan\beta$ values, and only the branching ratios
of $b\bar b$ and $gg$ modes are sizable for $\tan\beta \gsim 10$.
In charged Higgs boson decays with $m_{H^\pm}^{}=150$ GeV,
the decay into $\tau\nu$ is dominant in the type-I THDM, the type-II THDM and
the type-X THDMs for $\tan\beta \gtrsim 1$,
while hadronic decay modes can also be dominant in the type-Y THDM~\cite{Ref:Barger}.

In FIG.~\ref{FIG:br_150'}, we show the decay branching ratios of CP-even Higgs
bosons, where $m_h=120$ GeV, $m_\Phi^{}(=m_H^{}=m_A^{}=m_{H^\pm}^{})=150$ GeV,
$M=148$ GeV, and $\sin^2(\beta-\alpha)=0.96$ are taken.
Because of the CP-even Higgs boson mixing, the lightest Higgs boson $h$ is no longer
purely SM-like. Instead, $H$ can decay into massive gauge bosons via
off-shell modes such as $WW^\ast$ and $ZZ^\ast$.
Decay patterns of $h$ and $H$ depend on $\tan\beta$ and also on the type
of Yukawa interaction.
When $\tan\beta\sim 5$, the angle $\alpha$ is nearly zero.
In such case coupling constants become small, so that some of fermionic decay
modes are suppressed. In order to satisfy the unitarity constraints
for the large $\tan\beta$ region, the soft breaking mass scale $M$ must be
degenerate to the mass of the decaying bosons~\cite{Ref:PU}.

\section{Constraints from the current experimental data on THDMs}

One of the direct signal of the THDM is the discovery of extra Higgs bosons,
which have been searched at the LEP and Tevatron~\cite{Ref:LEP,Ref:Tevatron}.
Indirect contributions of Higgs bosons to precisely measurable observables
can also be used to constrain Higgs boson parameters.
In this section, we summarize these experimental bounds.

A direct mass bound is given from the LEP direct search results as
$m_{H^0}^{}\gtrsim92.8$ GeV for CP-even Higgs bosons and
$m_A^{}\gtrsim93.4$ GeV for CP-odd Higgs bosons in supersymmetric models.
The bound for charged Higgs boson has also been set as $m_{H^\pm}^{}\gtrsim79.3$ GeV~\cite{Ref:LEP}.

In the THDM, one-loop contributions of scalar loop diagrams
to the rho parameter are expressed as~\cite{Ref:HHG}
\begin{align}
&\delta\rho_\text{THDM}^{}= \sqrt2G_F\frac1{(4\pi)^2}
\left\{F_\Delta(m_A^2,m_{H^\pm}^2)\right.\nonumber\\
&\quad\left.-c_{\alpha-\beta}^2\left[F_\Delta(m_h^2,m_A^2)
-F_\Delta(m_h^2,m_{H^\pm}^2)\right]-s_{\alpha-\beta}^2\left[F_\Delta(m_H^2,m_A^2)
-F_\Delta(m_H^2,m_{H^\pm}^2)\right]\right\},
\end{align}
where $F_\Delta(x,y)=\tfrac12(x+y)-\tfrac{xy}{x-y}\ln\tfrac{x}{y}$ with
$F_\Delta(x,x)=0$\footnote{
There are other (relatively smaller in most of the parameter space)
contributions to the rho parameter in the THDM, i.e.,
those from the diagrams where the gauge boson (as well as Nambu-Goldstone
boson) and the Higgs boson are running together in the
loop\cite{Ref:HHG}. We have included these effect in our numerical analysis.}.
These quadratic mass contributions can cancel out when
Higgs boson masses satisfy the following relation: (i) $m_A^{}\simeq m_{H^\pm}^{}$,
(ii) $m_H^{}\simeq m_{H^\pm}^{}$ with $\sin(\beta-\alpha)\simeq1$,
and (iii) $m_h^{}\simeq m_{H^\pm}^{}$ with $\sin(\beta-\alpha)\simeq0$.
These relations correspond to the custodial symmetry
invariance~\cite{Ref:Csym,Ref:Csym2}.
This constraint is independent of the type of Yukawa interaction.

When $m_{H^\pm}^{}\lesssim m_t-m_b$, the top quark can decay into
the charged Higgs boson.
The decay mode $t\to H^+b$ has been studied by using the Tevatron
data~\cite{Ref:mH+TeVatron}.
The partial decay width is calculated by using the factors $\xi_A^f$ in
TABLE~\ref{Tab:MixFactor} as
\begin{align}
\Gamma(t\to H^+b) =&
\frac{G_F\left|V_{tb}\right|^2}{8\sqrt2\pi m_t}
\lambda\left(\frac{m_b^2}{m_t^2},\frac{m_{H^\pm}^2}{m_t^2}\right)^{1/2}
\nonumber\\
&\times
\left\{m_t^2\left[m_t^2{\xi_A^u}^2\left(1+\frac{m_b^2}{m_t^2}-\frac{m_{H^\pm}^2}{m_t^2}\right)+m_b^2{\xi_A^d}^2\right]
+4m_t^2m_b^2\xi_A^u\xi_A^d\right\},
\end{align}
where $\lambda(x,y)$ is defined in Eq.~\eqref{Eq:lamb}.
The higher order corrections can be found in Ref.~\cite{Ref:topQCD}.
The decay branching fraction strongly depends on the type of Yukawa interaction.
In the type-I (type-X) THDM, the decay mode can be sizable only
for small $\tan\beta$. On the other hand, it can also
be substantial for the large $\tan\beta$ region in the type-II (type-Y) THDM
since the bottom quark Yukawa coupling receives considerable enhancement.
This fact gives a lower bound $1\lesssim\tan\beta$ for THDMs
and an upper bound $\tan\beta\lsim 60$ for the type-II (type-Y) THDM below
$m_{H^\pm}^{}\lsim 130$GeV~\cite{Ref:mH+TeVatron}.

\begin{figure}[tb]
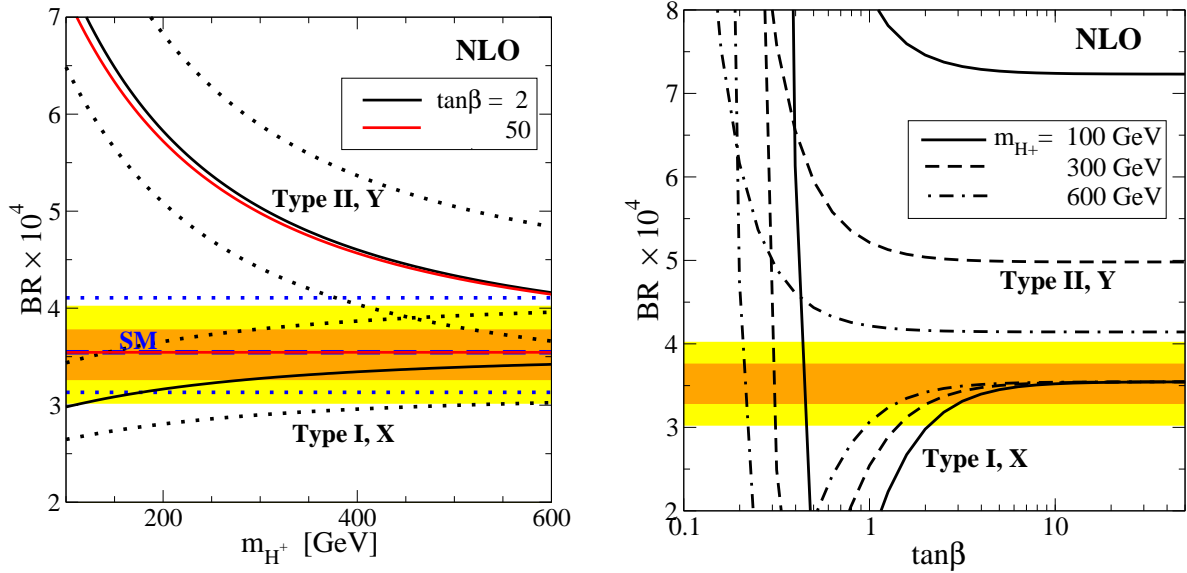

\begin{minipage}{0.49\hsize}
\includegraphics[width=7.5cm,angle=0]{BR_mh.eps}
\end{minipage}
\begin{minipage}{0.49\hsize}
\includegraphics[width=7.5cm,angle=0]{BR_tanb.eps}
\end{minipage}
\caption{Predictions of the decay branching ratio for $b\to s\gamma$
are shown at the NLO approximation as a function of $m_{H^\pm}^{}$ and $\tan\beta$.
The dark (light) shaded band represents $1\sigma$ $(2\sigma)$ allowed region
of current experimental data. In the left panel, solid (dashed) curves denote
the prediction for $\tan\beta=2$ $(50)$ in various THDMs. In the right panel, solid,
dashed and dot-dashed curves are those for $m_{H^\pm}^{}=100, 300$ and $600$ GeV,
respectively.}
\label{FIG:bsg}
\end{figure}

It has been known that the charged Higgs boson mass in the type-II THDM
is stringently constrained by the precision measurements of the
radiative decay of $b\to s\gamma$ at Belle~\cite{Ref:BELLE} and
BABAR~\cite{Ref:BABAR} as well as CLEO~\cite{Ref:CLEO}.
The process $b\to s\gamma$ receives contributions from the $W$ boson loop
and the charged Higgs boson loop in the THDM.
A notable point is that these two contributions
always work constructively in the type-II (type-Y) THDM,
while this is not the case in the type-I (type-X) THDM~\cite{Ref:Barger}.
In FIG.~\ref{FIG:bsg}, we show the branching ratio of $B\to X_s\gamma$ for
each type of THDM as a function of $m_{H^\pm}^{}$
(left-panel) and $\tan\beta$ (right-panel), which are evaluated at the next-to-leading 
order (NLO) following the formulas in Ref.~\cite{Ref:bsgNLO}.
The SM prediction at the NLO is also shown for comparison. 
The theoretical uncertainty is about $15\%$\footnote{In Ref.~\cite{Ref:bsgNLO}, 
the theoretical uncertainty is smaller because the value for the error
in $m_c^\text{pole}/m_b^\text{pole}$ is taken to be $7\%$, which gives main 
uncertainty in the branching ratio.} 
in the branching ratio ( as indicated by dotted curves in FIG.~\ref{FIG:bsg}), 
which mainly comes from the pole mass of charm quark $m_c^\text{pole}=1.65\pm 0.18$ 
GeV ~\cite{Ref:PDG}.
The experimental bounds of the branching ratio are also indicated, where
the current world average value is given by 
${\mathcal B}(B\to X_s\gamma)=(3.52\pm 0.23\pm0.09)\times 10^{-4}$~\cite{Ref:HFAG}.
It is seen in FIG.~\ref{FIG:bsg} that the branching ratio in
the type-I (type-X) THDM lies within the 2 $\sigma$ experimental error 
in all the regions of $m_{H^{\pm}}$ indicated for $\tan\beta\gtrsim 2$, 
while that in the type-II (type-Y) THDM is far from the value indicated by 
the data for a light charged Higgs boson region $(m_{H^\pm}^{}\lesssim 200$ 
GeV$)$.
In the right figure, a cancellation occurs in the type-I (type-X) THDM
since there are destructive interferences between the $W$ boson and
the $H^\pm$ contributions.
It is emphasized that the charged Higgs boson could be light
in the type-I (type-X) THDM under the constraint from $B\to X_s\gamma$ results.
We note that in the MSSM the chargino contribution can compensate
the charged Higgs boson contribution~\cite{Ref:bsgMSSM}.
This cancellation weakens the limit on $m_{H^\pm}^{}$ from $b\to s\gamma$ 
in the type-II THDM,
and allows a light charged Higgs boson
as in the type-I (type-X) THDM.

We give some comments on the NNLO analysis, 
although it is basically out of the scope of this paper.
At the NNLO, the branching ratio for $b\to s\gamma$ 
has been evaluated in the SM in
Ref.~\cite{Ref:bsgNNLO,Ref:bsgNNLO_THDM}.
The predicted value at the NNLO is less than that at the NLO approximation 
in a wide range of renormalization scale. 
In Ref.~\cite{Ref:bsgNNLO}, the SM branching ratio is $(3.15\pm 0.23)\times
10^{-4}$, and the lower bound of the $m_{H^\pm}^{}$, after adding the
NLO charged Higgs contribution, is estimated as 
$m_{H^\pm}^{} \gtrsim 295$ GeV ($95\%$ CL) in the type-II (type-Y)
THDM~\cite{Ref:bsgNNLO}\footnote{In Ref.~\cite{Ref:bsgNNLO_THDM} 
the NNLO branching ratio in the SM is calculated as $(2.98\pm 0.26)\times
10^{-4}$, and the mass bound is a little bit relaxed.}.
On the other hand, in the type-I (type-X) THDM, although the branching ratio 
becomes smaller as compared to the NLO evaluation, no serious bound on 
$m_{H^\pm}^{}$ can be found for $\tan\beta \gtrsim 2$. 
Therefore, charged Higgs boson mass is not expected to be strongly constrained in 
the type-I (type-X) THDM even at the NNLO, and our main conclusion that 
the type-I (type-X) THDM 
is favoured for $m_{H^\pm}^{}\lesssim 200$ GeV based on the NLO
analysis should not be changed.

\begin{figure}[tb]
\begin{minipage}{0.49\hsize}
\includegraphics[width=7.5cm]{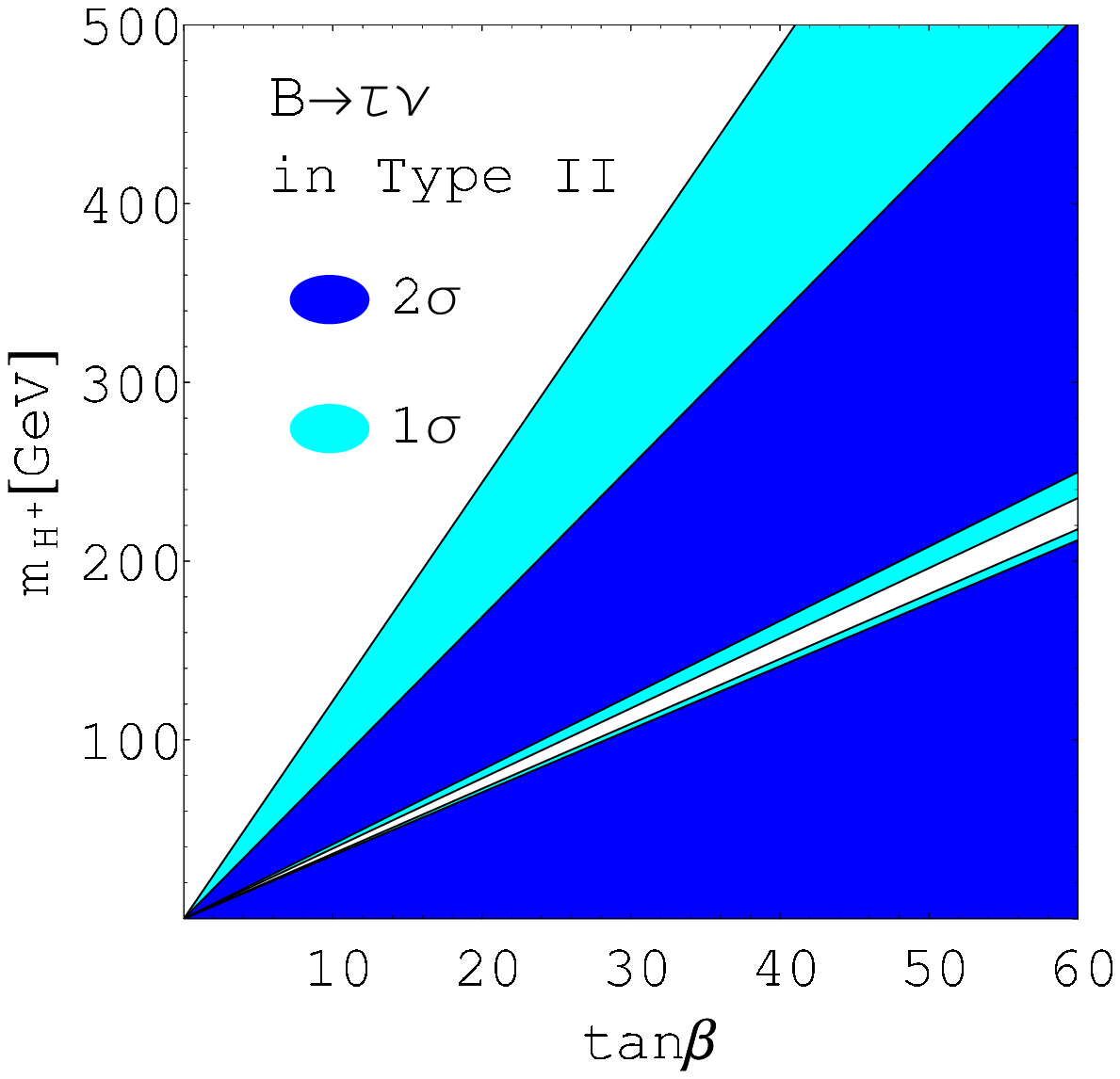}
\end{minipage}
\begin{minipage}{0.49\hsize}
\includegraphics[width=7.5cm]{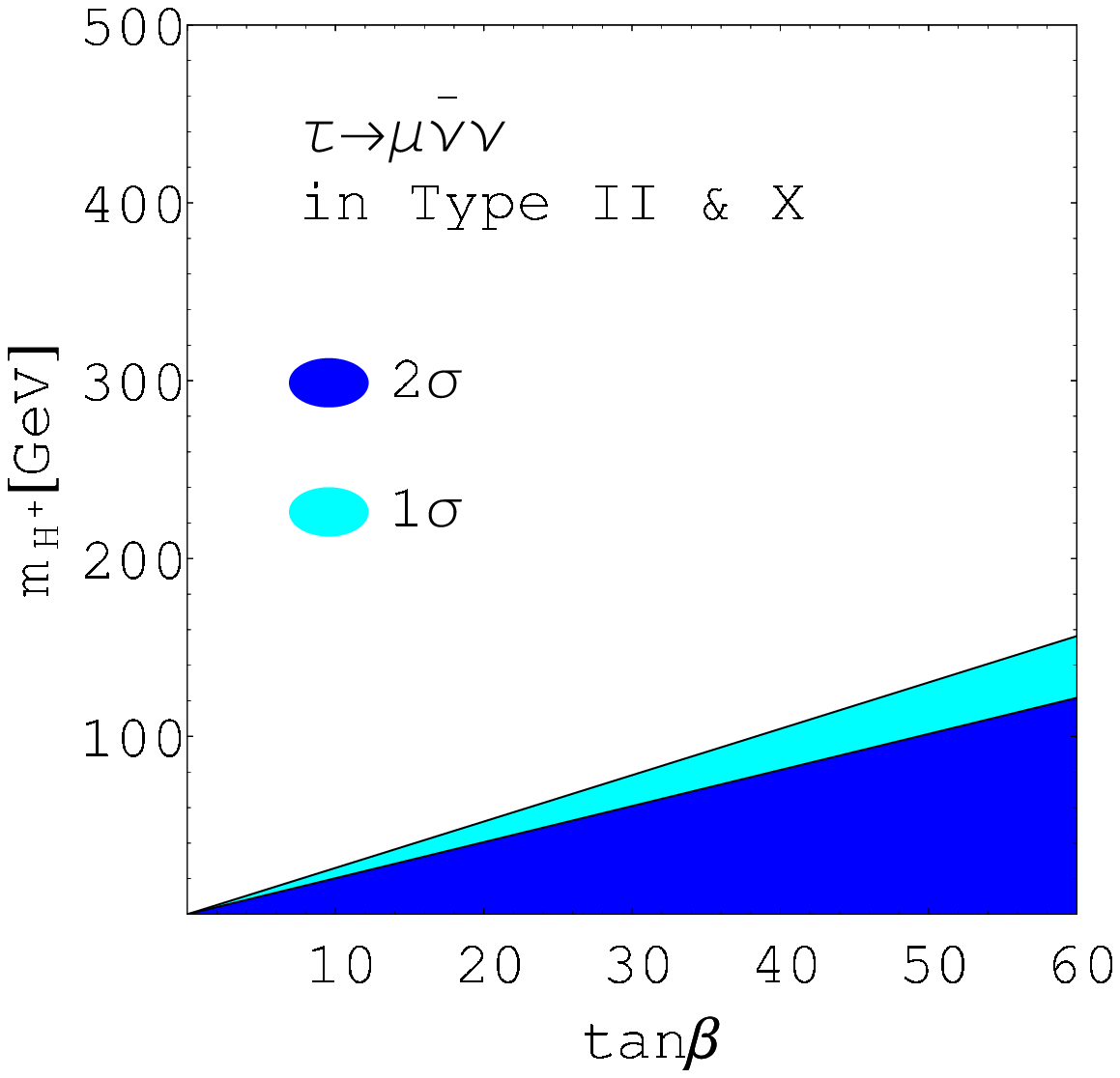}
\end{minipage}
\caption{Bounds from $B\to\tau\nu$ (left panel) and tau leptonic decay (right panel) on $m_{H^\pm}^{}$ as a function of $\tan\beta$ are shown. The dark (light) shaded region
corresponds to the $2\sigma$ $(1\sigma)$ exclusion of these experimental results.
In the type-II THDM the wide parameter space is constrained by $B\to\tau\nu$,
while only the tau leptonic decays are important for the type-X THDM.}
\label{FIG:mH+tanb}
\end{figure}

The decay $B\to\tau\nu$ has been discussed in the type-II THDM~\cite{Ref:MK,Ref:BTauNu}.
The data for ${\mathcal B}(B^+\to\tau^+\nu_\tau)=(1.4\pm0.4)\times 10^{-4}$
are obtained at the $B$ factories~\cite{Ref:PDG,Ref:BtaunuExp}.
The decay branching ratio can be written as~\cite{Ref:MK,Ref:IP}
\begin{align}
\frac{{\mathcal B}(B^+\to\tau^+\nu_\tau)_\text{THDM}}{{\mathcal B}(B^+\to\tau^+\nu_\tau)_\text{SM}}
\simeq\left(1-\frac{m_B^2}{m_{H^\pm}^2}\xi_A^d\xi_A^\ell\right)^2.
\end{align}
In FIG.~\ref{FIG:mH+tanb}, the allowed region from the $B\to\tau\nu$ results
is shown in the type-II THDM. The dark (light) shaded region denotes
the $2\sigma$ $(1\sigma)$ exclusion from the $B\to\tau\nu$ measurements.
The process is important only in the type-II THDM with large $\tan\beta$ values.
The other types of Yukawa interactions do not receive constraints form this process.

The rate for the leptonic decay of the tau lepton $\tau\to\mu\overline{\nu}\nu$
can be deviated from the SM value by the presence of a light charged Higgs boson~\cite{Ref:TauMuNuNu}.
The partial decay rate is approximately expressed as~\cite{Ref:MK}
\begin{align}
\frac{\Gamma_{\tau\to\mu\overline{\nu}\nu}^\text{THDM}}{\Gamma_{\tau\to\mu\overline{\nu}\nu}^\text{SM}}
&\simeq 1-\frac{2m_\mu^2}{m_{H^\pm}^2}{\xi_A^\ell}^2 \kappa\left(\frac{m_\mu^2}{m_\tau^2}\right)+\frac{m_\mu^2m_\tau^2}{4m_{H^\pm}^4}{\xi_A^\ell}^4,
\end{align}
where $\kappa(x)=g(x)/f(x)$ is defined by
$f(x)=1-8x-12x^2\ln x+8x^3-x^4,$ and $g(x)=1+9x-9x^2-x^3+6x(1+x)\ln x$.
In the type-II (type-X) THDM, the leptonic Yukawa interaction
can be enhanced in the large $\tan\beta$ region. Hence, both the models are weakly
constrained by tau decay data, as in FIG.~\ref{FIG:mH+tanb}.

The precision measurement for the muon anomalous magnetic moment
can give mass bound on the Higgs boson in the SM~\cite{Ref:g-2MuSM}.
This constraint can be applied for more
general interaction, including THDMs~\cite{Ref:g-2MuTHDM}.
At the one-loop level, the contribution is given by
\begin{align}
\delta a_\mu^{1-\text{loop}}
&\simeq \frac{G_Fm_\mu^4}{4\pi^2\sqrt2}
\left[\sum_{\phi^0=h,H}\frac{{\xi_{\phi^0}^\ell}^2}{m_{\phi^0}^2}
\left(\ln\frac{m_{\phi^0}^2}{m_\mu^2}-\frac76\right)
+\frac{{\xi_A^\ell}^2}{m_A^2}
\left(-\ln\frac{m_A^2}{m_\mu^2}+\frac{11}6\right)
-\frac{{\xi_A^\ell}^2}{6m_{H^\pm}^2}\right].
\end{align}
This process is also purely leptonic and only gives milder bounds
on the Higgs boson masses for very large $\tan\beta$ values
in the type-II (type-X) THDM. It gives no effective bound on the type-I (type-Y) THDM. 
It is also known that the two-loop (Barr-Zee type) diagram  
can significantly affect $a_\mu$~\cite{Ref:BZ,Ref:King}. 
The contribution can be large because of the enhancement factors 
of $m_f^2/m_\mu^2$ and also of the mixing factors $\xi_\phi^f$ as
~\cite{Ref:King}
\begin{align}
\delta a_\mu^\text{BZ}
&\simeq \frac{N_c^fQ_f^2G_F\alpha m_\mu^2}{4\pi^3v^2}
\left[-\sum_{\phi^0=h,H} 
\xi_{\phi^0}^\ell\xi_{\phi^0}^ff\left(\frac{m_f^2}{m_{\phi^0}^2}\right) 
+\xi_A^\ell\xi_A^fg\left(\frac{m_f^2}{m_A^2}\right) \right],
\end{align}
where 
\begin{align}
f(z)&=\frac{z}2\int_0^1dx\frac{1-2x(1-x)}{x(1-x)-z}\ln\frac{x(1-x)}{z},\\
g(z)&=\frac{z}2\int_0^1dx\frac1{x(1-x)-z}\ln\frac{x(1-x)}{z}.
\end{align}
The contribution from this kind of diagram is only important 
for large $\tan\beta$ values with smaller Higgs boson masses in the type-II THDM. 
For the other types of THDM, it would give a much less effective bound 
on the parameter space. 

\section{Collider signals in the Type-X THDM
at the LHC and the ILC
}

We discuss the collider phenomenology of the models at the LHC and the ILC.
There have already been many studies on the production and decays
of the Higgs bosons in the type-II THDM, especially in the context of
the MSSM, while the phenomenology of the other types of THDMs has not
yet been studied sufficiently.
Recently, the type-X THDM has been introduced in the model to explain
phenomena such as neutrino masses, dark matter, and baryogenesis at the
TeV scale~\cite{Ref:AKS}.
We therefore concentrate on the collider signals in the type-X THDM, and
discuss how we can distinguish the model from the type-II THDM (the
MSSM), mainly in scenarios with a light charged Higgs boson
($100$ GeV $\lesssim m_{H^\pm}^{} \lesssim 300$ GeV).
(Such a light charged Higgs boson is severely constrained by the $b\to s
\gamma$ result in the non-supersymmetric type-II THDM and the type-Y THDM,
while it can be allowed in the MSSM and the type-X (type-I) THDM.)
As we are interested in the differences in the types of the Yukawa
interactions,
we focus here on the case of $m_{H^\pm}^{} \simeq m_A^{} \simeq m_H^{}$ with
$\sin(\beta-\alpha) \simeq 1$ for definiteness.

\subsection{Charged Higgs boson searches at the LHC}

A light charged Higgs boson with $m_{H^\pm}^{} \lesssim m_t -m_b$ can be
produced in the decay of top quarks at the LHC.
The discovery potential for the charged Higgs boson via
the $t\overline{t}$ production has been studied in the MSSM~\cite{Ref:mH+TeVatron}.
Assuming an integrated luminosity of $30$ fb${}^{-1}$, the expected signal
significance of the event $t\overline{t}\to H^\pm W^\mp b\overline{b}\to
\ell\nu\tau\nu_\tau b\overline{b}$ is greater than $5 \sigma$ for
$m_{H^\pm}^{}\lsim 130$ GeV for $\tan\beta\lesssim2$
and $\tan\beta\gtrsim 20$~\cite{Ref:mH+TeVatron}.
The same analysis can also be applied for the type-X THDM, in which
a similar number of $H^\pm$ can be produced when $\tan\beta\sim{\cal O}(1)$.
The main decay mode ($\tau\nu$) is common in the type-II THDM and the type-X
THDM, except for very low $\tan\beta$ values.
Thus, searching for neutral Higgs bosons is important
to distinguish the difference in the types of Yukawa interactions.

For $m_{H^\pm}^{}\gtrsim m_t$, charged Higgs bosons can be produced in
$q\bar q/gg\to t\overline{b}H^-$, $gb\to tH^-$~\cite{Ref:tbH,Ref:ttH},
$gg$ $(q\bar q)\to H^+H^-$~\cite{Ref:H+H-,Ref:GGH2}
and $gg$ $(b\bar b)\to H^\pm W^\mp$~\cite{Ref:WH}.
These processes, except for the $H^+H^-$ production, 
are via the Yukawa coupling of $t \bar b H^-$, so
that the cross sections are significant for $\tan\beta\sim{\mathcal O}(1)$
or $\tan\beta \gtrsim 10$--$20$ in the type-II THDM
and only for $\tan\beta\sim{\mathcal O}(1)$ in the type-X THDM.
The type of Yukawa interaction in the THDM
can then be discriminated by measuring the difference in decay
branching ratios of $H^\pm$.
In the type-II THDM $H^\pm$ mainly decay into $tb$, while
$\tau\nu$ is dominant for $\tan\beta \gtrsim 10$ in the type-X THDM.

\subsection{Neutral Higgs boson $(A$ and $H)$ production at the LHC
}

At the LHC, the type of the Yukawa interaction may be determined
in the search for neutral Higgs bosons through the direct production
via gluon fusion $gg\to A/H$~\cite{Ref:GGH2,Ref:GGH},
vector boson fusion $V^\ast V^\ast \to H$, $V=W,Z$~\cite{Ref:VBF1,Ref:VBF2},
and also via associated production $pp \to b\bar b A$
$(b\bar b H)$~\cite{Ref:bbH,Ref:QQH}.
The production process $pp \to t\bar t A$ $(t\bar t H)$~\cite{Ref:ttH,Ref:QQH,Ref:ttH2}
can also be useful for $\tan\beta\sim 1$.
We discuss the possibility of discriminating the type of the Yukawa
interaction by using the production and decay processes of $A$ and $H$
for $\sin(\beta-\alpha)=1$.
Additional neutral Higgs bosons $A$ and $H$ are directly produced
by the gluon fusion mechanism at the one-loop level.
When $\sin(\beta-\alpha)\simeq 1$, the production rate can be significant
due to the top quark loop contribution for $\tan\beta \sim 1$
and, in the MSSM (the type-II THDM), also
for large $\tan\beta$ via the bottom quark loop contributions~\cite{Ref:GGH2}.
Notice that there is no rate of $V^\ast V^\ast \to A$ because there is no 
$VVA$ coupling, and that the production of $H$ from the vector boson
fusion $V^\ast V^\ast \to H$ is relatively unimportant when
$\sin(\beta-\alpha)\simeq 1$.
The associate production process $pp\to b\overline{b}A$
$(b\overline{b}H)$ can be significant for large
$\tan\beta$ values in the MSSM (the type-II THDM)~\cite{Ref:bbH}.

In the MSSM (the type-II THDM),
the produced $A$ and $H$ in these processes decay mainly into $b\bar b$
when $\sin(\beta-\alpha) \simeq 1$,
which would be challenging to detect because of huge QCD backgrounds.
Instead, the decays into a lepton pair $\tau^+\tau^-$ ($\mu^+\mu^-$)
would be useful for searches of $A$ (and $H$).
However, the decay branching ratios of $A \to
\tau^+\tau^-$ ($\mu^+\mu^-$) are less than $0.1$ ($0.0004$).
A simulation study~\cite{Ref:atlasTDR} shows that the Higgs boson search via the
associate production $b\bar b A$ ($b\bar b H$) is better than
that via the direct production from gluon fusion to see  both
$\tau^+\tau^-$ and $\mu^+\mu^-$ modes, especially in the large $\tan\beta$ area.
The largest background is the Drell-Yan process from
$\gamma^\ast/Z^\ast \to \tau^+\tau^-$ (and $\mu^+\mu^-$).
The other ones, such as $t\bar t$, $b\bar b$ and $W+jet$, also contribute to
the backgrounds.
The rate of the $\tau^+\tau^-$ process from the signal is much larger than
that of the $\mu^+\mu^-$ one.  However, the resolution for tau leptons is much
broader than that for muons, so that for relatively small $m_A$ ($m_H$)
the $\mu^+\mu^-$ mode can be more useful than the $\tau^+\tau^-$
mode~\cite{Ref:HanMuMu}.

In the type-X THDM, signals from the associate production $pp \to b\bar bA$
are very difficult to detect. The production cross section
is at most $150$ fb for $m_A^{}=150$ GeV at $\tan\beta=1$~\cite{Ref:atlasTDR}, 
where the branching ratio
$A\to \tau^+\tau^-$ and $A\to \mu^+\mu^-$ are small,
and the produced signals are less for larger values of $\tan\beta$.
On the other hand, the direct production from $gg\to A/H$ can be
used to see the signal.
The cross sections are significant for $\tan\beta \sim 1$, and
they decrease for larger values of $\tan\beta$ by a factor of $1/\tan^2\beta$.
However, the branching ratios of $A/H \to \tau^+\tau^-$
dominate over those of $A/H\to b\bar b$ for $\tan\beta
\gtrsim 2$ and become almost $100\%$ for $\tan\beta \gtrsim 4$ (see FIG.~\ref{FIG:br_150}).
Therefore, large significances can be expected around $\tan\beta \sim 2$
in the type-X THDM.
\begin{figure}[tb]
\begin{minipage}{0.49\hsize}
\includegraphics[width=7.5cm,angle=-90]{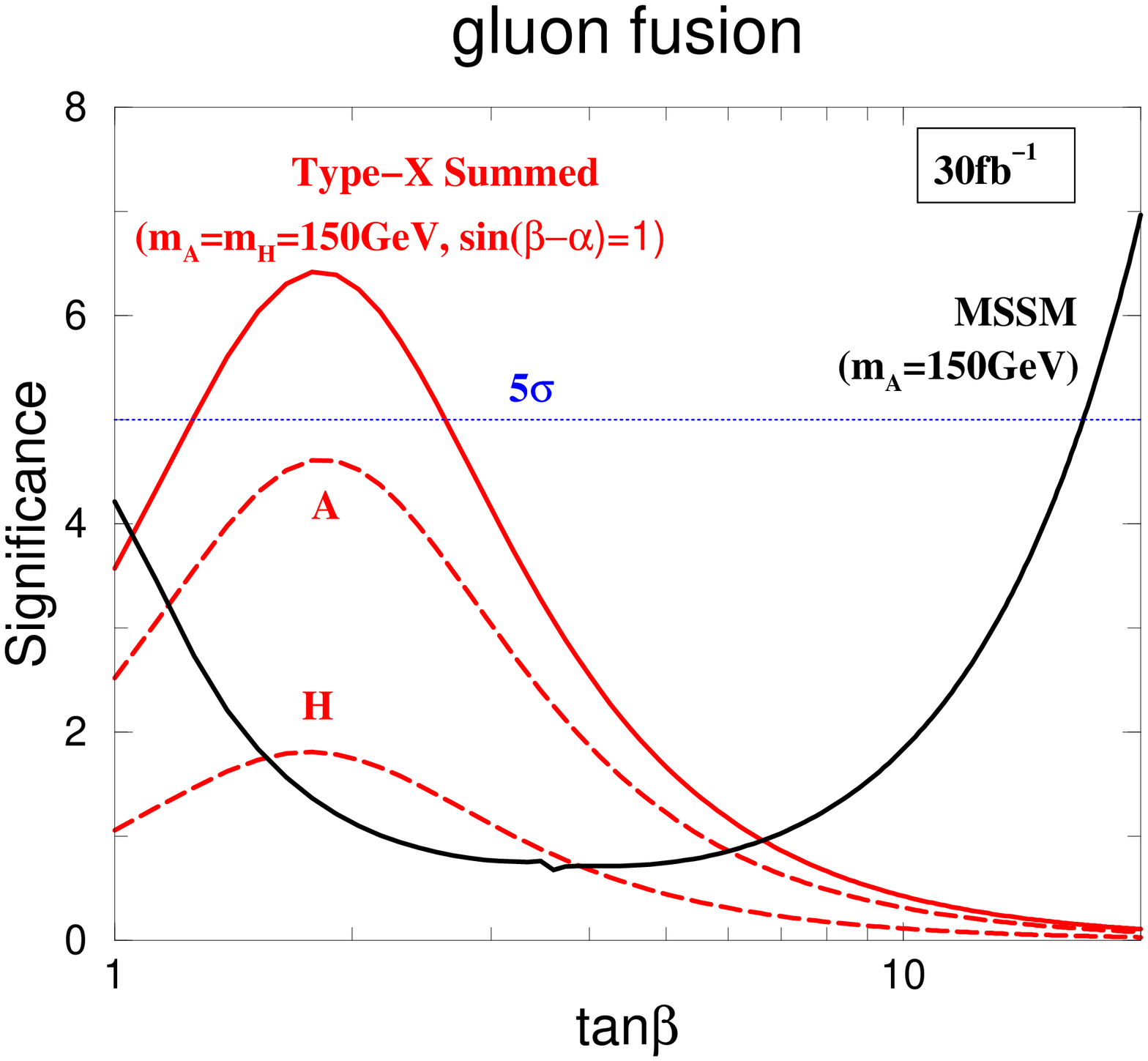}
\end{minipage}
\begin{minipage}{0.49\hsize}
\includegraphics[width=7.5cm,angle=-90]{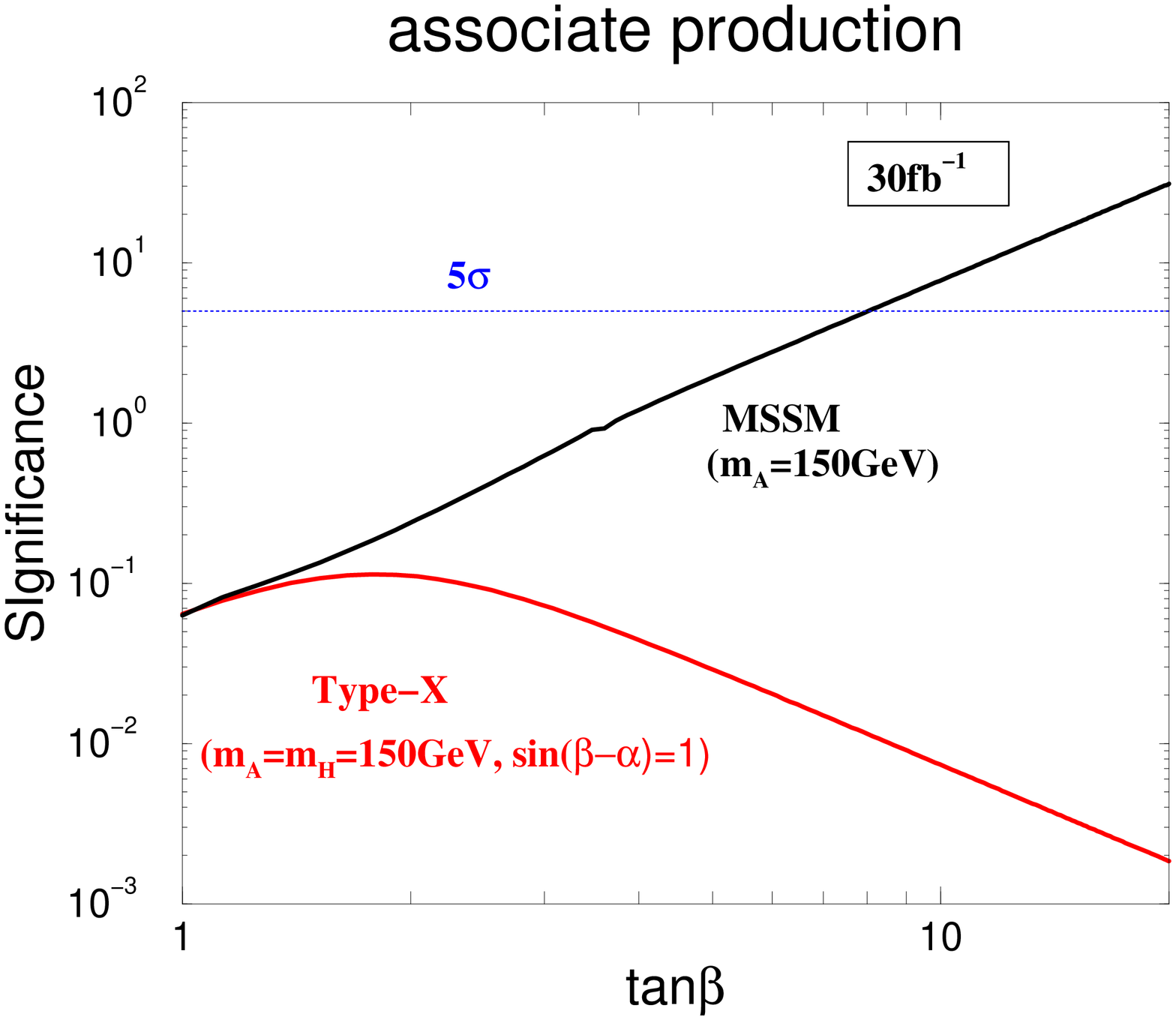}
\end{minipage}
\caption{Signal significance ($S/\sqrt{B}$) for gluon fusion $gg\to A/H$ (left panel)
and the associated production $pp\to b\bar b A$ $(b\bar b H)$ (right panel)
with the $\tau^+\tau^-$ final state in the type-X THDM and the MSSM.
In both figures, the dashed and the solid curves represents
the expected values of signal significance for $A/H$ and summed over $A$ and $H$.
The red (thin) curves denote the significance in the type-X THDM, while
the black (thick) solid curves denote that in the MSSM.}
\label{FIG:ggA}
\end{figure}

In FIG.~\ref{FIG:ggA}, we show the expected signal significance of the
direct production $pp (gg)\to A/H \to \tau^+\tau^-$ in the type-X THDM and
the MSSM at the LHC, assuming an integrated luminosity of $30$ fb$^{-1}$.
The mass of the CP-odd Higgs boson $m_A^{}$ is taken to be $150$ GeV
in both models, while $m_H^{}$ is $150$ GeV for the results
of the type-X THDM and that in the MSSM is deduced using the MSSM mass relation.
For the detailed analysis of background simulation, we
employed the one shown in the ATLAS TDR~\cite{Ref:atlasTDR}.
The basic cuts, such as the high $p_T^{}$ cut and the standard $A/H\to\tau^+\tau^-$ reconstruction, are assumed~\cite{Ref:atlasTDR}.
We can see that, for the search of the direct production, 
the signal significance in the type-X THDM can be larger than
that in the MSSM for $\tan\beta \lesssim 5$.
In particular, the signal in the type-X THDM can be expected to be
detectable $(S/\sqrt{B}>5)$ when $\tan\beta \sim 2$
for the luminosity of $30$ fb$^{-1}$.
For smaller values of $m_A$ ($m_H$), the production cross section becomes large
so that the signal rate is more significant, but the separation
from the Drell-Yan background becomes more difficult
because the resolution of the tau lepton is broad. 
Therefore, the significance becomes worse for $m_A^{} (m_H^{}) \lesssim 130$ GeV.

\begin{figure}[tb]
\includegraphics[width=7.5cm,angle=-90]{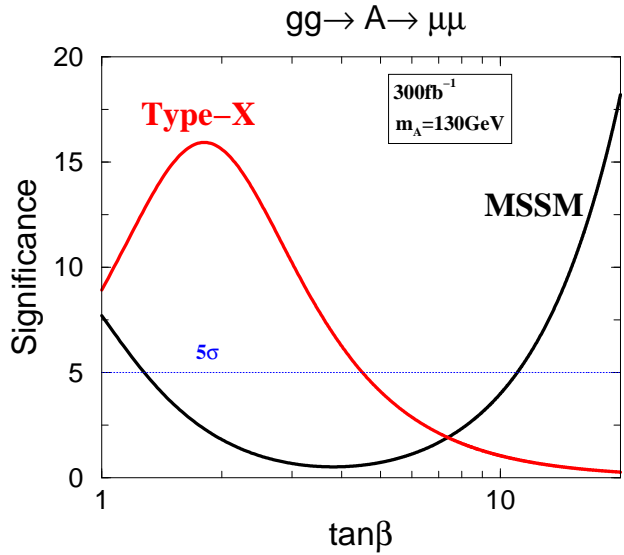}
\caption{Signal significance ($S/\sqrt{B}$) for the $\mu^+\mu^-$ final state 
from gluon fusion $gg\to A$ in the type-X THDM and the MSSM.
The red (thin) curves denote significance in the type-X THDM, while
the black (thick) solid curves denote that in the MSSM.}
\label{FIG:mumu}
\end{figure}

When $A$ and $H$ are lighter than $130$ GeV, the $\mu^+\mu^-$ mode can be more
useful than the $\tau^+\tau^-$ mode. The resolution of muons is much better
than that of tau leptons, so that the invariant mass cut is very effective
in reducing the background from $\gamma^\ast/Z^\ast \to \mu^+\mu^-$.
The feasibility of the process $gg\to A/H \to \mu^+\mu^-$
has been studied in the SM and the MSSM in Refs.~\cite{Ref:atlasTDR,Ref:HanMuMu}.
We evaluate the signal significances of $gg\to A/H \to \mu^+\mu^-$  in
the type-X THDM by using the result in Ref.~\cite{Ref:HanMuMu}.
In TABLE~\ref{Tab:signif_mumu},
we list the results of the significance in the SM and the
type-X THDM. According to Tao Han's paper, the
basic kinematic cuts of $p_T^{}>20$ GeV, $|\eta|<2.5$ and
the invariant mass cuts as
$m_{A/H}^{}-2.24$ GeV $ < M_{\mu\mu}^{} < m_{A/H}^{}+2.24$ GeV
are used~\footnote{This choice for the invariant mass cut is rather severe; i.e., 
it requires the precise determination of $m_A^{}$ and $m_H^{}$. 
If the range of the invariant mass cut is taken to be double, 
roughly speaking, background events also become double. 
This would suppress the signal significance in TABLE \ref{Tab:signif_mumu} by 
a factor of $\sim 1/\sqrt2$.}. 
The integrated luminosity is assumed to be $300$ fb$^{-1}$.
For the results in the type-X THDM, $\tan\beta=2$ and
 $\sin(\beta-\alpha)=1$ are taken. 
The results show that the significance can be substantial for
$m_A^{}\gtrsim 115$ GeV when $\tan\beta=2$.
For smaller masses of the extra Higgs bosons, the cross section for the
signal processes can be larger, but the invariant mass cut becomes
less effective in the reduction of the Drell-Yan background
because of the smaller mass difference between $m_Z^{}$ and
$m_{A}^{}$; hence, the signal significance becomes worse.
The $\tan\beta$ dependence in the signal significance for the muon final states 
is also shown in FIG.~\ref{FIG:mumu}. The shape of the curves is similar to that 
for the tau lepton final state in FIG~\ref{FIG:ggA}. 
\begin{table}[t]
\begin{center}
\begin{tabular}{|c||c|c|c|c|c|}
\hline  $m_\phi$ [GeV] & SM ($H_\text{SM}$) & MSSM($A$) & Type-X ($H$) & Type-X ($A$) & Type-X(Sum) ($m_A^{} \simeq m_H^{}$) \\  \hline
115  & 2.41 & 1.31 & 4.31 & 12.0 & 16.3 \\
120  & 2.51 & 1.49 & 4.89 & 13.4 & 18.3 \\
130  & 2.25 & 1.81 & 5.78 & 15.6 & 21.4 \\
140  & 1.61 & 2.11 & 6.60 & 17.5 & 24.1 \\
\hline
\end{tabular}
\end{center}
\caption{Expected signal significances for
$gg\to \phi \to \mu^+\mu^- (\phi=H_\text{SM}, H, A)$
 in the SM, the MSSM, and the type-X THDM.
For the results in the MSSM $\tan\beta=2$ is taken, 
and for that in the type-X THDM $\tan\beta=2$  
and $\sin(\beta-\alpha)=1$ are taken.
 The integrated luminosity is assumed to be $300$ fb$^{-1}$.
 }\label{Tab:signif_mumu}
\end{table}

In summary, we would be able to distinguish the type-X THDM from the MSSM by
measuring the leptonic decays of the additional Higgs bosons produced via
the direct search processes $gg \to A/H \to \tau^+\tau^-$ ($\mu^+\mu^-$) and
the associated processes $pp\to b\bar b A$ $(b\bar b H)$.
First, if a light scalar boson is found via $gg \to h \to \gamma\gamma$ or
$W^+W^- \to h \to \gamma\gamma$ ($\tau^+\tau^-$) and if
the event number is consistent with the prediction in the SM,
then we know that the scalar boson is of the SM or at least
SM-like: in the THDM framework this means $\sin(\beta-\alpha)\simeq 1$,
assuming that it is the lightest one.
Second, under the situation above, 
when the associated production $pp\to b\bar b A$ $(b\bar b H)$
is detected at a different invariant mass than the mass of the
SM-like one and no $gg\to A/H \to \tau^+\tau^-$ ($\mu^+\mu^-$) is found
at that point, we would be able to identify the MSSM Higgs sector
(or the type-II THDM) with high $\tan\beta$ values.
On the other hand, the type-X THDM with low $\tan\beta$ would be
identified by finding the signal from the gluon fusion process without
that from the $b\bar b \tau^+\tau^-$.
If signals from the direct production processes are found 
but the number is not sufficient, then the value of $\tan\beta$
would be around $6$--$10$ ($m_t \cot\beta \sim m_b \tan\beta$), 
where the rates in the MSSM and the type-X THDM are similar. In this case, it
would be difficult to distinguish these models from the above
processes. 
As we discussed in the next subsection, Higgs pair production processes
$pp\to AH^\pm, HH^\pm$, and $AH$ can be useful to measure
the Yukawa interaction through branching fractions, because these production
mechanism do not depend on $\tan\beta$ in such a situation.

We have concentrated on $\sin(\beta-\alpha)\simeq 1$ in this analysis because  
the parameter is motivated in Ref.~\cite{Ref:AKS}. 
Here we comment on the situation without the condition $\sin(\beta-\alpha)=1$. 
If $\sin(\beta-\alpha)$ is not close to unity, our conclusion can be modified. 
The production cross sections of $gg\to A/H$ and $pp\to b{\bar b}A(b{\bar b}H)$ 
can be enhanced in the type-X THDM for $\tan\beta\gtrsim 1$ since the 
factor $(\sin\alpha/\sin\beta)$ of quark-Higgs couplings can be large in a 
specific region of the parameter space. The signal of 
the CP odd Higgs boson $A$ can then be significant. On the other hand, 
the CP even Higgs boson $H$ can decay into $WW^*$ when $\sin(\beta-\alpha)$ 
is not unity. This would reduce leptonic branching fractions. 
The signal can be enhanced only for large $\tan\beta$ regions 
because leptonic decays are significant only for such a parameter space.
We also note that $H$ can be produced significantly 
by the vector boson fusion mechanism in a mixing case. 

\subsection{Pair production of extra Higgs bosons at the LHC}

\begin{figure}[tb]
\includegraphics[width=8cm]{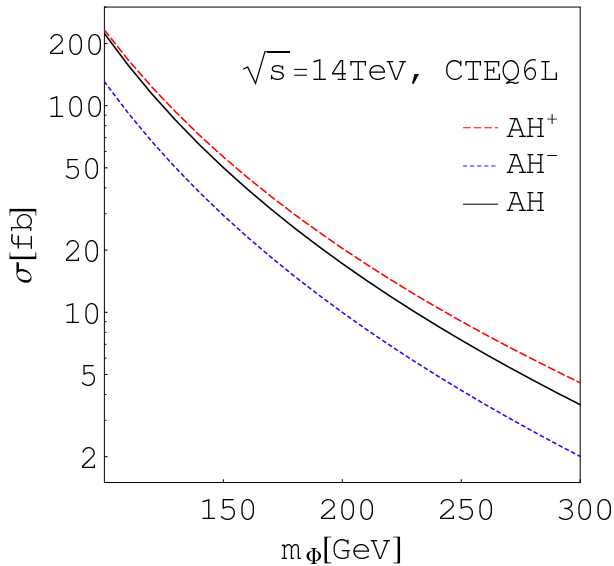}
\caption{Production cross sections of $pp\to AH^\pm$ and $AH$ are
shown at the leading order as a function of scalar boson masses in the THDM
where $m_\Phi=m_H^{}=m_A^{}=m_{H^\pm}^{}$ are chosen.
The long-dashed, dashed and solid curves denote the pair production
of $AH^+$, $AH^-$ and $AH$, respectively. The cross section of
$pp\to HH^\pm$ is the same as those of $pp\to AH^\pm$ when $\sin(\beta-\alpha)=1$. }
\label{FIG:AH}
\end{figure}

The types of the Yukawa interactions can be studied using the process of
$q\bar q' \to {W^\pm}^\ast \to AH^\pm$ ($HH^\pm$)
~\cite{Ref:GGH2,Ref:KY,Akeroyd:2003jp,Ref:CKY,Ref:TOBE} and
$q\bar q \to Z^\ast \to AH$~\cite{Ref:GGH2},
unless the extra Higgs bosons $H$, $A$ and/or $H^\pm$ are too
heavy~\footnote{When the mixing between $h$ and $H$ is large, the $hH^\pm$
production can also be important~\cite{Ref:TOBE}.}.
Hadronic cross sections for these processes are shown at the leading
order in
FIG.~\ref{FIG:AH} as a function of the mass of the produced scalar boson
$\Phi$, where $m_\Phi^{}=m_H^{}=m_A^{}=m_{H^\pm}^{}$.
Expected rates of $AH^\pm$(sum) and $AH$ are about $143$ fb 
and $85$ fb for $m_{H^\pm}^{} = 130$ GeV and about $85$ fb
and $50$ fb for $m_{H^\pm}^{} = 150$ GeV, respectively.
The NLO QCD corrections are expected to enhance these rates 
typically by about $20\%$~\cite{Ref:KY,Ref:CKY}.
The production rates are common in all the types of THDMs, because
the cross sections are determined only by the masses of the produced scalar bosons.
Therefore, they are very sensitive to the difference in the decay branching
ratios of the produced Higgs bosons.
In the MSSM, the $b\bar b \tau^\pm \nu$ ($b\bar b\tau^+\tau^-$)
events can be
the main signal of the $AH^\pm$ and $HH^\pm$ ($AH$) processes,
while in the type-X THDM ($\tan\beta \gtrsim 2$),
$\tau^+\tau^-\tau^\pm\nu$ ($\tau^+\tau^-\tau^+\tau^-$) would be the main signal events.

In the MSSM, the parton level 
background analysis for the $AH^\pm$ ($HH^\pm$) production process has been
performed in Ref.~\cite{Ref:CKY,Ref:TOBE} by using the $b\overline{b}\tau\nu$ 
final state.
The largest background comes from $q\overline{q'}\to Wg^*\to
Wb\overline{b}$, which can be reduced by basic kinematic cuts and
the invariant mass cut of $b\bar b$, as well as by 
the kinematic cut to select hard hadrons from the parent $\tau^\pm$ from $H^\pm$.
It has been shown that a sufficient signal significance can be
obtained for smaller masses of Higgs bosons~\cite{Ref:CKY}.

In the type-X THDM, the produced $AH^\pm$ ($HH^\pm$) and $AH$ pairs
can be studied via the leptonic decays.
Hence these channels can be useful to discriminate the type-X THDM from
the MSSM.
Assuming an integrated luminosity of $300$ fb$^{-1}$,
$8.6 \times 10^4$ and $5.1 \times 10^4$ of the
signal events are produced from both $AH^\pm$ and $HH^\pm$ production
for $m_\Phi^{}=130$ GeV and $150$ GeV, respectively, where
$m_\Phi^{} = m_H^{} = m_A^{} = m_{H^\pm}^{}$.
$A$ and $H$ ($H^\pm$) decay into $\tau^+\tau^-$ ($\tau\nu$)
by more than $95$\% and $95$\% ($99$\%) for $\tan\beta\gtrsim 4$, respectively.
The purely leptonic signal would have an advantage in the signal to
background ratio because the background from the intermediate state
$q\overline{q'}\to Wg^*$ would be negligible.
For $\tan\beta=7$, the produced $AH^\pm$ and $HH^\pm$ pairs almost all 
(more than $99\%$)
go to $\tau^+\tau^-\tau\nu$ final states.
The numbers of signal and background are summarized in TABLE~\ref{Tab:AH+}. 
The signal to background ratio for $\tau^+\tau^-\tau\nu$ final state 
is not so small ${\mathcal O}(0.1$--$1)$,  
before cuts~\footnote{The $\gamma W^\pm$ production may give a much larger 
cross section for background events. It may also be reduced considerably 
by kinematic cuts.}.
The backgrounds are expected to be reduced by using
high-$p_T^{}$ cuts, hard hadrons from the parent tau leptons from $H^\pm$,
and invariant mass cuts for $\tau^+\tau^-$ from $A$ and $H$, 
in addition to the basic cuts. 
However, the signal significance strongly depend on the rate of
miss identification of hadrons as tau leptons, so that
a realistic simulation is necessary.

\begin{table}[t]
\begin{center}
\begin{tabular}{|c||c|c|c|c|c|}
\hline & $AH^\pm, HH^\pm (m_\Phi^{}=130 \text{ GeV})$ & $AH^\pm, HH^\pm(m_\Phi^{}=150 \text{ GeV})$ & $ZW^\pm$\\
\hline $\tau^+\tau^-\tau\nu$ & $8.4 \times 10^4$ & $5.0 \times 10^4$ & 
$3.2\times 10^{4}$\\
\hline $\mu^+\mu^-\tau\nu$ & $3.0\times 10^2$ & $1.8\times 10^2$ & $3.1\times 10^{4}$\\
\hline
\end{tabular}
\end{center}
\caption{Events for the $\tau^+\tau^-\tau\nu$ 
and $\mu^+\mu^-\tau\nu$ final states 
from the Higgs boson pair production and $ZW^\pm$ background. 
The signal events are summed over $AH^\pm$ and $HH^\pm$. 
The integrated luminosity is taken to be $300$ fb$^{-1}$. 
Values for the decay branching ratios are taken to be 
${\mathcal B}(A/H\to \tau^+\tau^-)=0.99$, 
${\mathcal B}(A/H\to \mu^+\mu^-)=0.0035$, and 
${\mathcal B}(H^\pm\to \tau\nu)=0.99$, 
which correspond to the values for $\tan\beta\gtrsim 7$.   
The cross section of $pp\to ZW^\pm$ is evaluated 
as $\sigma_{ZW}=27.7$ pb by CalcHEP~\cite{Ref:CalcHEP}. 
}\label{Tab:AH+}
\end{table}

We also would be able to use the $\mu^+\mu^-\tau^+\nu$
events to identify the $AH^+$ and $HH^+$ production in the
type-X THDM, by using the much better resolution of $\mu^+\mu^-$
in performing the invariant mass cuts.
For $300$ fb$^{-1}$, the $AH^+$ and $HH^+$ process
can produce about
${\mathcal O}(100)$ 
of $\mu^+\mu^-\tau^+ \nu$ events for $m_A^{}=m_H^{}=130$ GeV.
The number of background events 
is about $3.1 \times 10^5$ of $\mu^+\mu^-\tau^+ \nu$ from $ZW^\pm$ production. 
Signals and background for $\mu^+\mu^-\tau^+\nu$ events are 
also summarized in TABLE~\ref{Tab:AH+}. 
The background can be expected to be reduced by imposing
a selection of the events around the invariant mass of $m_A^{} \sim
M_{\mu\mu}$ and the high $p_T^{}$ cuts. Hard hadrons from the
decay of $\tau$'s from $H^+$ can also be used to reduce the
background. In the MSSM, much smaller signals are expected,
so that this process can be a useful probe of the type-X THDM.

In a similar way, we may use $AH$ production~\cite{Ref:GGH2}
to identify the type-X THDM.
For the $\tau^+\tau^-\tau^+\tau^-$ decay mode, the signal is evaluated approximately as
$2.5\times 10^4$ events, assuming $300$ fb$^{-1}$ for $m_A^{}=m_H^{}=130$ GeV
and $\tan\beta=7$.
The main background may come from the $q\bar q\to ZZ$ process.
We also consider the $\mu^+\mu^-\tau^+\tau^-$ decay mode.
The number of signal event is ${\mathcal O}(100)$
for an integrated luminosity $300$ fb$^{-1}$. 
The numbers of signal and background event are listed in 
TABLE~\ref{Tab:AH}. 
It would be valuable to use the detailed background simulation.

\begin{table}[t]
\begin{center}
\begin{tabular}{|c||c|c|c|c|c|}
\hline & $AH (m_\Phi^{}=130 \text{ GeV})$ & $AH (m_\Phi^{}=150 \text{ GeV})$ & $ZZ$\\
\hline $\tau^+\tau^-\tau^+\tau^-$ & $2.5\times 10^5$ & $1.5\times 10^5$ & 
$3.6\times 10^{3}$\\
\hline $\tau^+\tau^-\mu^+\mu^-$ & $1.8\times 10^2$ & $1.0\times 10^2$ & 
$7.1\times 10^{3}$\\
\hline
\end{tabular}
\end{center}
\caption{Events for the $\tau^+\tau^-\tau^+\tau^-$ 
and $\tau^+\tau^-\mu^+\mu^-$ final states 
from the Higgs boson pair production and $ZZ$ background. 
The integrated luminosity is taken to be $300$ fb$^{-1}$. 
The cross section of $pp\to ZZ$ is evaluated 
as $\sigma_{ZZ}=10.5$ pb by CalcHEP~\cite{Ref:CalcHEP}. 
}\label{Tab:AH}
\end{table}

\subsection{Pair production of extra Higgs bosons at the ILC}

At the ILC, we would be able to test the types of the Yukawa interactions
via the pair production of the additional Higgs bosons
$e^+e^- \to AH$~\cite{Ref:GGH2,Ref:eeAH}.
\begin{figure}[t]
\includegraphics[width=7.5cm,angle=-90]{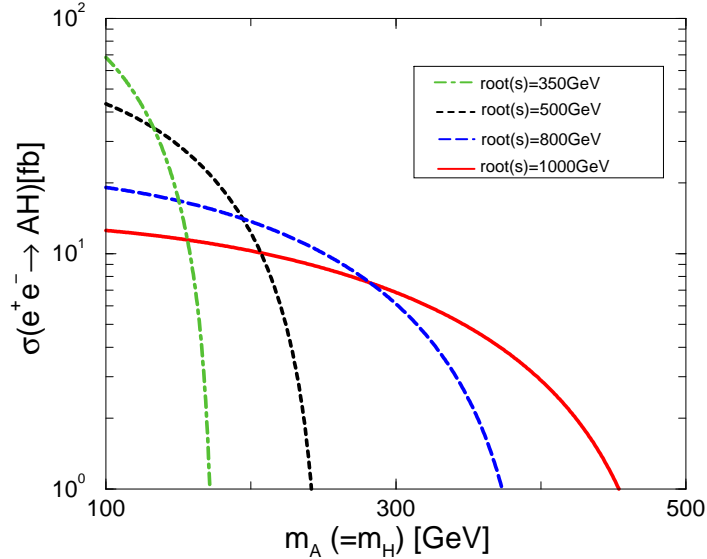}
\caption{The production cross section of $e^+e^- \to AH$ is shown as a
function of the Higgs boson mass.
The dot-dashed, dashed, long-dashed, and solid curves correspond to 
$\sqrt{s}=350, 500, 800$, and $1000$ GeV, respectively.}
\label{FIG:eeAH}
\end{figure}
In Fig.~\ref{FIG:eeAH}, the production cross section is shown for
$\sqrt{s}=350, 500$, $800$, and $1000$ GeV as a function of $m_A^{}$
assuming $m_{A}^{}=m_H^{}$ in the THDM. The production rate
is determined only by $m_{A}^{}$, and $m_H^{}$ at the leading order,
and  is independent of $\tan\beta$.
(In the MSSM, it depends indirectly on $\tan\beta$
via the mass spectrum.)
The signal of the type-X THDM can be identified by searching
for the events of $\tau^+\tau^-\tau^+\tau^-$ and
$\mu^+\mu^-\tau^+\tau^-$.
When $\sqrt{s}=500$ GeV, assuming an integrated luminosity of
$500$ fb$^{-1}$, the event number is estimated
as $1.6\times 10^4$ $(1.8\times 10^2)$
in the type-X (type-II) THDM for
$\tau^+\tau^-\tau^+\tau^-$, and $1.1\times 10^2$ $(0.6)$
for $\mu^+\mu^-\tau^+\tau^-$
assuming  $m_H^{}=m_A^{}=m_{H^\pm}^{}=130$ GeV, $\sin(\beta-\alpha)=1$
and $\tan\beta=10$. This number does not change much for $\tan\beta\gtrsim 3$.
The main background comes from the $Z$ pair production, whose rate is
about $4\times 10^2$ fb. The event numbers from the background are then
$2.3\times 10^2$ for $\tau^+\tau^-\tau^+\tau^-$ and $4.6\times 10^2$
for $\mu^+\mu^-\tau^+\tau^-$.
Therefore, the signal should be easily detected in the type-X THDM, by
which we would be able to distinguish the type-X from the type-II (the
MSSM).

The detailed measurement of the masses of additional
Higgs bosons and Yukawa coupling constants will make it possible to
determine the scenario of physics beyond the SM through the Higgs physics.

\section{Discussions and Conclusions}

We have discussed the phenomenology of the four possible types of
Yukawa interactions in the THDM with softly broken discrete  $Z_2$
symmetry. Although the particle contents are the same in these models,
their phenomenologies are completely different from each other.
In some new physics models, two Higgs doublets (plus singlets)
are introduced with a definite type of Yukawa interaction.
Therefore, when additional Higgs bosons are discovered, 
we may be able to distinguish new physics models via
the differences between the types of Yukawa interactions.

The differences between the types of the Yukawa interactions
largely affect the decay of the Higgs bosons.
We have evaluated the decay branching ratios in each type of THDM.
We have then summarized the current experimental bounds on the models from the
data of $b\to s\gamma$, $B\to\tau\nu$, $\tau\to\mu\overline{\nu}\nu$
and measurements of muon anomalous magnetic moment.
The charged Higgs boson mass can be light
($m_{H^\pm}^{}\lesssim m_t$) in the type-I and type-X THDMs
under these experimental limits. Such a light charged Higgs boson
is also possible in the MSSM.

We have discussed the collider phenomenology of the THDMs.
For definiteness, we have concentrated on phenomenological differences
between the MSSM Higgs sector and the type-X THDM in the case
of a light charged Higgs boson.
The type-X Yukawa interaction is used in TeV-scale models for neutrino
masses, dark matter, and baryon asymmetry.
At the LHC, the type-X THDM can be discriminated from the MSSM by
searching for the production and decays of the extra Higgs bosons $A$, $H$, 
and $H^\pm$, such as $gg\to A/H\to \ell^+\ell^-$,
$pp \to b\bar b A$ $(b\bar b H) \to b\bar b\ell^+\ell^-$,
where $\ell$ is $\tau$ or $\mu$ when $\sin(\beta-\alpha) \simeq 1$.
We also discussed the pair production processes
$pp \to AH^\pm$, $HH^\pm$, and $AH$ to test
the type-X THDM.
These processes would provide distinctive four lepton final states
$\ell^+\ell^-\tau^\pm\nu$ and $\ell^+\ell^-\tau^+\tau^-$ in the type-X THDM,
while the MSSM Higgs sector can be tested by $b\bar b\tau^\pm\nu$ and
$b\bar b\tau^+\tau^-$.
Although the realistic simulation study is necessary, the type-X
THDM can be identified at the LHC with an integrated luminosity of $300$ fb$^{-1}$.
At the ILC, the type-X THDM is expected to be studied very well
by the pair production $e^+e^-\to AH$. The signal should be four leptons
($\ell^+\ell^-\tau^+\tau^-$).

We have discussed phenomenological discrimination of
the types of Yukawa interactions in the THDM at future experiments.
In particular, we have mainly discussed the separation at the LHC and the ILC 
between the MSSM Higgs sector and the type-X THDM in the relatively light 
charged Higgs boson scenario. 
By extending this study to include other various cases,
we can expect that the types of the Yukawa interactions in extended Higgs models can be
 completely determined at current and future collider experiments.
Such information may be useful to select a true model
from many proposed new physics models at TeV scales.\\

\noindent
{\bf Acknowledgments}~~~\\[2mm]
We would like to thank Andrew Akeroyd, Yasuhiro Okada, Kaoru Hagiwara, 
and Jun-ichi Kanzaki for useful comments.
The work of S. K. was supported in part by Japan Society for Promotion of
Science (JSPS), No. 18034004. \\

\noindent
{\it Note added:}~~~\\[2mm]
After this paper was completed,
Refs.~\cite{Ref:Logan,Ref:Hall,Ref:2HDMC}
appeared in which similar Yukawa interaction is discussed
in a different context.

\newpage

\appendix
\section{Higgs boson decay rates in THDMs}

In the THDM, the partial decay widths of Higgs bosons decaying
into a fermion pair are computed at the leading order as
\begin{align}
&\Gamma(\varphi\to q{\bar q}) = N_C \frac{G_Fm_\varphi
m_q^2}{4\sqrt2\pi}{\xi^q_\varphi}^2 \times
\begin{cases}
\beta_q^3 \text{ for } \varphi=h,H\\
\beta_q \text{ for } \varphi=A
\end{cases},\\
&\Gamma(\varphi\to\ell^+\ell^-) = \frac{G_Fm_\varphi
m_\ell^2}{4\sqrt2\pi} {\xi^\ell_\varphi}^2 \times
\begin{cases}
\beta_\ell^3 \text{ for } \varphi=h,H\\
\beta_\ell \text{ for } \varphi=A
\end{cases},\\
&\Gamma(H^+\to u{\bar d}) = N_C
\frac{G_Fm_{H^\pm}^{}\left|V_{ud}\right|^2}{4\sqrt2\pi}\beta_{ud}^{}
\left\{\left(m_u^2{\xi^u_A}^2+m_d^2{\xi^d_A}^2\right)
\left(1-\frac{m_u^2+m_d^2}{m_{H^\pm}^2}\right)-\frac{4m_u^2m_d^2
{\xi^u_A}{\xi^d_A}}{m_{H^\pm}^2}\right\},\label{Eq:H+ud}\\
&\Gamma(H^+\to \ell^+\nu) =
\frac{G_Fm_{H^\pm}^{}m_\ell^2}{4\sqrt2\pi}{\xi^\ell_A}^2
\left(1-\frac{m_\ell^2}{m_{H^\pm}^2}\right)^2,\label{Eq:H+nul}
\end{align}
where the factors $\xi_\varphi^f$ are defined in Eq.~\eqref{Eq:Yukawa}, 
$q=u,d,s,c,t,b$; $\ell=e,\mu,\tau$; $N_C(=3)$ is the color factor; 
$V_{ud}$ is the Kobayashi-Maskawa matrix; and
\begin{align}
\lambda(x,y) &=
1+x^2+y^2-2x-2y-2xy,\label{Eq:lamb}\\
\beta_X^{} &= \lambda^{1/2}\left(\frac{m_X^2}{m_\varphi^2},
\frac{m_X^2}{m_\varphi^2}\right) =
\sqrt{1-\frac{4m_X^2}{m_\varphi^2}},\\
\beta_{XY}^{} &= \lambda^{1/2}\left(\frac{m_X^2}{m_\varphi^2},
\frac{m_Y^2}{m_\varphi^2}\right).
\end{align}

Formulas for the decay widths for loop induced decays are given by
\begin{align}
&\Gamma(\varphi\to\gamma\gamma) =
\frac{G_F\alpha_\text{EM}^2m_\varphi^3}{128\sqrt2\pi^3}
\left|\sum_fQ_f^2I_f^\varphi(m_f,N_C)+I_W^\varphi+I_{H^\pm}^\varphi\right|^2,\\
&\Gamma(\varphi\to Z\gamma) =
\frac{G_F\alpha_\text{EM}^2m_\varphi^3}{64\sqrt2\pi^3}
\left(1-\frac{m_Z^2}{m_\varphi^2}\right)^3
\left|\sum_fQ_fJ_f^\varphi(m_f,N_C)+J_W^\varphi+J_{H^\pm}^\varphi\right|^2,\\
&\Gamma(\varphi\to gg) =
\frac{G_F\alpha_S^2m_\varphi^3}{64\sqrt2\pi^3}
\left|\sum_{f=q}I_f^\varphi(m_f,1)\right|^2,
\end{align}
where fermionic loop functions are given by
\begin{align}
&I_f^\varphi(m_f,N_C) = \xi_\varphi^f \times
\begin{cases}
-N_C\frac{4m_f^2}{m_\varphi^2}
\left[2-\beta_f^2m_\varphi^2 C_0(0,0,m_\varphi^2,m_f^2,m_f^2,m_f^2)\right]
&\text{ for } \varphi=h,H\\
+4N_Cm_f^2C_0(0,0,m_\varphi^2,m_f^2,m_f^2,m_f^2) &\text{ for }
\varphi=A
\end{cases},\\
&J_f^\varphi(m_f,N_C) = \xi_\varphi^f \times
\begin{cases}
-4N_Cc_V^f\left[J_1(m_f)-J_2(m_f)\right]
&\text{ for } \varphi=h,H\\
-4N_Cc_V^f\left[-J_2(m_f)\right] &\text{ for } \varphi=A
\end{cases},
\end{align}
and 
\begin{align}
J_1(m) = & \frac{2m^2}{m_\varphi^2-m_Z^2}
\left\{1+2m^2C_0(0,m_Z^2,m_\varphi^2,m^2,m^2,m^2)
\right.\nonumber\\
&\qquad\left.+\frac{m_Z^2}{m_\varphi^2-m_Z^2}
\left[B_0(m_\varphi^2,m^2,m^2)-B_0(m_Z^2,m^2,m^2) \right]\right\},\\
J_2(m) = & m^2C_0(0,m_Z^2,m_\varphi^2,m^2,m^2,m^2),
\end{align}
with $c_V^f = \frac1{2s_Wc_W}\left(T_{3L}^f-2Q_fs_W^2\right)$, where 
$T_{3L}^f$ is the third component of the isospin. 
The functions $B_0$ and $C_0$ are the Passarino-Veltman functions~\cite{Ref:PV}, 
which can be expressed by analytic functions as 
\begin{align}
B_0(m_\varphi^2,m^2,m^2)&=\Delta-2g\left(\frac{4m^2}{m_\varphi^2}\right)\\
B_0(m_Z^2,m^2,m^2)&=\Delta-2g\left(\frac{4m^2}{m_Z^2}\right)\\
C_0(0,0,m_\varphi^2,m^2,m^2,m^2)&=\frac{-2}{m_\varphi^2}f\left(\frac{4m^2}{m_\varphi^2}\right)\\
C_0(0,m_Z^2,m_\varphi^2,m^2,m^2,m^2)&=\frac{-2}{m_\varphi^2-m_Z^2}[f\left(\frac{4m^2}{m_\varphi^2}\right)-f\left(\frac{4m^2}{m_Z^2}\right)]
\end{align}
where $\Delta$ denotes the regularized divergence part 
in the dimensional regularization, and 
\begin{align}
g(x)=\begin{cases}
(x-1)\arcsin\left(\sqrt{1/x}\right)&\text{ for } x\ge1\\
\frac{1}{2}(1-x)\left[\ln\left(\frac{1+\sqrt{1-x}}{1-\sqrt{1-x}}\right)
-i\pi\right]& \text{ for } x<1
\end{cases},\\
f(x)=\begin{cases}
\left[\arcsin\left(\sqrt{1/x}\right)\right]^2&\text{ for } x\ge1\\
-\frac{1}{4}\left[\ln\left(\frac{1+\sqrt{1-x}}{1-\sqrt{1-x}}\right)-i\pi\right]^2
&\text{ for } x<1
\end{cases}.
\end{align}
The loop functions $I_W, I_{H^\pm}, J_W$ and $J_{H^\pm}$ are common 
among all types of Yukawa interactions, which can be found in Ref.~\cite{Ref:HHG}.

\end{document}